\documentclass{emulateapj}
\usepackage{natbib}

\def\eq#1{\begin{equation} #1 \end{equation}}
\def\eqarray#1{\begin{eqnarray} #1 \end{eqnarray}}

\def\comm#1           {{\tt (COMMENT: #1)}}
\def\sm#1           {{\tt (MACRO: #1)}}

\def\scl#1            {}

\shorttitle{The Milky Way Stellar Number Density Distribution}
\shortauthors{Juri\'{c} et al.}

\begin{document}

\title{    The Milky Way Tomography with SDSS: I. Stellar Number Density Distribution     }

\author{
Mario Juri\'{c}\altaffilmark{1,2},
\v{Z}eljko Ivezi\'{c}\altaffilmark{3},
Alyson Brooks\altaffilmark{3},
Robert H. Lupton\altaffilmark{1},
David Schlegel\altaffilmark{1},
Douglas Finkbeiner\altaffilmark{1},
Nikhil Padmanabhan\altaffilmark{4},
Nicholas Bond\altaffilmark{1},
Branimir Sesar\altaffilmark{3},
Constance M. Rockosi\altaffilmark{3},
Gillian R. Knapp\altaffilmark{1},
James E. Gunn\altaffilmark{1},
Takahiro Sumi\altaffilmark{1,11},
Donald P. Schneider\altaffilmark{5},
J.C. Barentine\altaffilmark{6},
Howard J. Brewington\altaffilmark{6},
J. Brinkmann\altaffilmark{6},
Masataka Fukugita\altaffilmark{7},
Michael Harvanek\altaffilmark{6},
S.J. Kleinman\altaffilmark{6},
Jurek Krzesinski\altaffilmark{6,8},
Dan Long\altaffilmark{6},
Eric H. Neilsen, Jr.\altaffilmark{9},
Atsuko Nitta\altaffilmark{6},
Stephanie A. Snedden\altaffilmark{6},
Donald G. York\altaffilmark{10}
}

\altaffiltext{1}{Department of Astrophysical Sciences, Princeton University, Princeton, NJ 08544\label{Princeton}}
\altaffiltext{2}{School of Natural Sciences, Institute for Advanced Study, Princeton, NJ 08540\label{IAS}}
\altaffiltext{3}{University of Washington, Dept. of Astronomy, Box 351580, Seattle, WA 98195 
                 \label{Washington}}
\altaffiltext{4}{Princeton University, Dept. of Physics, Princeton, NJ 08544 \label{PrincetonPhy}}
\altaffiltext{5}{Department of Astronomy and Astrophysics, Pennsylvania
State University, University Park, PA 16802\label{PennState}}
\altaffiltext{6}{Apache Point Observatory, P.O. Box 59, Sunspot, NM 88349,
U.S.A.\label{APO}}
\altaffiltext{7}{University of Tokyo, Institute for Cosmic Ray
Research\label{Tokyo}}
\altaffiltext{8}{Mt. Suhora Observatory, Cracow Pedagogical University, ul.
Podchorazych 2, 30-084 Cracow, Poland\label{Poland}}
\altaffiltext{9}{Fermi National Accelerator Laboratory, P.O. Box 500,
Batavia, IL 60510, U.S.A.\label{FNAL}}
\altaffiltext{10}{Department of Astronomy and Astrophysics, and Enrico Femi Institute, The University of
Chicago, Chicago, IL 60037 USA \label{Chicago}}
\altaffiltext{11}{Solar-Terrestrial Environment Laboratory, Nagoya University, Furo-cho, Chikusa-ku,
Nagoya, 464-8601, Japan
\label{Nagoya}}


\begin{abstract}
Using the photometric parallax method we estimate the distances to 
$\sim$48 million stars detected by the Sloan Digital Sky Survey (SDSS) 
and map their three-dimensional
number density distribution in the Galaxy. The currently available
data sample the distance range from 100 pc to 20 kpc and cover 6,500 deg$^2$ of sky, mostly at high
galactic latitudes ($|b|>25$). These stellar number density maps
allow an investigation of the Galactic structure with no a priori assumptions about the
functional form of its components.
The data show strong evidence for a Galaxy consisting of an oblate halo, a disk component,
and a number of localized overdensities. The number density distribution of stars as traced by
M dwarfs in the Solar neighborhood ($D < 2$ kpc) is well fit by two exponential disks 
(the thin and thick disk) with scale heights and lengths, bias-corrected for an assumed 
35\% binary fraction, of  $H_1 = 300$ pc and $L_1 = 2600$ pc, and $H_2 = 900$ pc and 
$L_2 = 3600$ pc, and local thick-to-thin disk density normalization 
$\rho_{thick}(R_\odot)/\rho_{thin}(R_\odot) = 12$\%. 
We use the stars near main-sequence turnoff to measure the shape of the Galactic halo.
We find a strong preference for oblate halo models, with best-fit axis ratio $c/a = 0.64$, 
$\rho_H \propto r^{-2.8}$ power-law profile, and the local halo-to-thin disk normalization of 0.5\%. 
Based on a series of Monte-Carlo simulations, we estimate the errors of
derived model parameters 
not to be larger than $\sim 20$\% for the disk scales and $\sim10$\% for the density normalization, 
with largest contributions to error coming from the uncertainty in calibration of the 
photometric parallax relation and poorly constrained binary fraction.
While generally consistent with the above model, the measured
density distribution shows a number of statistically significant
localized deviations. In addition to known features, 
such as the Monoceros stream, we detect two overdensities in the thick disk region at 
cylindrical galactocentric radii and heights $(R, Z) \sim (6.5, 1.5)$~kpc 
and $(R, Z) \sim (9.5, 0.8)$~kpc,
and a remarkable density enhancement in the halo covering over a thousand square
degrees of sky towards the constellation of Virgo, at distances of $\sim$6-20 kpc. 
Compared to counts in a region symmetric with respect to the $l=0^\circ$ line and with the same 
Galactic latitude, the Virgo overdensity is responsible for a factor of 2
number density excess, and may be a nearby tidal stream or a low-surface
brightness dwarf galaxy merging with the Milky Way.
The $u-g$ color distribution of stars associated with it implies 
metallicity lower than that of thick disk stars, and
consistent with the halo metallicity distribution.
After removal of the resolved overdensities, the remaining data are consistent with a 
smooth density distribution; we detect no evidence of further unresolved clumpy 
substructure at scales ranging from $\sim 50$~pc in the disk, to 
$\sim1 - 2$~kpc in the halo.
\end{abstract}

\section{                       Introduction                     }

In the canonical model of Milky Way formation \citep*{ELS} the Galaxy began with a 
relatively rapid ($\sim 10^8$yr) radial collapse of the initial protogalactic cloud, 
followed by an equally rapid settling of gas into a rotating disk.
This model readily explained the origin and general structural, kinematic and
metallicity correlations of observationally identified populations of
field stars \citep{Baade44,OConnell58}: low metallicity
Population II stars formed during the initial collapse and populate the extended \emph{stellar halo};
younger Population I and Intermediate Population II stars formed after the gas has 
settled into the Galactic plane and constitute the \emph{disk}.

The observationally determined distribution of
disk stars is commonly described by exponential density laws
\citep{BahcallSoneira,Gilmore83,Gilmore89}, while power-laws or flattened
de Vaucouleurs spheroids are usually used to describe the halo
(e.g., \citealt{Wyse89}; \citealt{Larsen96b}; see also a review by \citealt{Majewski93}).
In both disk and the halo, the distribution of stars is expected to be a smooth 
function of position, perturbed only slightly by localized bursts of star formation
or spiral structure induced shocks.

However, for some time, starting with the pioneering work of \citet{SearleZinn}, continuing
with the studies of stellar counts and count asymmetries from Palomar Observatory Sky Survey (e.g.
\citealt{Larsen96b}, \citealt{Larsen96}, \citealt{Parker03}), and most recently with the data from
modern large-scale sky surveys (e.g., the Sloan Digital Sky Survey, \citealt{York00}; The 
Two Micron All Sky Survey, 2MASS, \citealt{Majewski03}; and the QUEST survey \citealt{Vivas01})
evidence has been mounting for a more complex picture of the
Galaxy and its formation. Unlike the smooth distribution easily captured by analytic 
density laws, new data argue for much more irregular 
substructure, especially in the stellar halo. Examples include the Sgr dwarf 
tidal stream in the halo \citep{Ivezic00,Yanny00,Vivas01,Majewski03}, or the Monoceros 
stream closer to the Galactic plane \citep{Newberg02,Rocha-Pinto03}. The existence of
ongoing merging points to a likely significant role of accretion events in the early formation 
of the Milky Way's components, making the understanding of both the distribution of merger remnants,
and of overall Milky Way's stellar content, of considerable theoretical interest.

\vspace{.5in}
The majority ($>90\%$) of Galactic stellar content resides in the form of
main-sequence (MS) stars. However, a direct measurement of their spatial
distribution requires accurate estimates of stellar distances to faint flux
levels, as even the most luminous main sequence stars have $V \sim 15-18$ for 
the 1--10 kpc distance range. This requirement, combined with the need to
cover a large sky area to sample a representative portion of the Galaxy, 
have historically made this type of measurement a formidable task.

A common workaround to the first part of the problem is to use bright
tracers for which reasonably accurate distance estimates 
are possible (e.g. RR Lyrae stars, A-type stars, M giants), and which are
thought to correlate with the overall stellar number density distribution. These tracers,
however, represent only a tiny fraction of stars on the sky, and their low
number density prevents tight constraints on the Galactic structure model parameters  
\citep{Reid96}. For the same reason, such tracers are unsuitable tools for 
finding localized overdensities with small contrast ratios over their surroundings.

Preferably, given precise enough multiband photometry, one would avoid
the use of tracers and estimate the distances to MS stars directly
using a color-absolute magnitude, or ``photometric parallax'', relation.
However, until now the lack of deep, large-area optical\footnote{For example,
near-IR colors measured by the all-sky 2MASS survey are not well suited 
for this purpose, because they only probe the Rayleigh-Jeans
tail of the stellar spectral energy distribution and thus are not very 
sensitive to the effective temperature.} surveys with sufficiently
accurate multiband photometry has prevented an efficient use of this
method.

Surveying a wide area is of particular importance.
For example, even the largest Galactic structure oriented data set to date to use accurate optical 
CCD photometry \citep{Siegel02} covered only $\sim15$ deg$^2$, with $\sim10^5$ stars.
To recover the overall Galactic density field their study, as others before it,
it has had to resort to model fitting and \emph{assume} a high degree of regularity in the 
density distribution and its functional form. This however, given that typical disk+halo models 
can have up to 10 free parameters, makes parameter estimation vulnerable to bias 
by unrecognized clumpy substructure.

Indeed, a significant spread in results coming from different studies has existed for 
quite some time (e.g., \citealt{Siegel02}, Table 1; \citealt{Bilir06}) indicating that 
either the unidentified substructures are confusing the model fits, that there are a multitude
of degenerate models that are impossible to differentiate from using a limited number of lines of sight,
or that the usual models provide an inappropriate description of the large-scale distribution 
of stars in the Galaxy. A direct model-free determination of the stellar 
number density distribution in a large volume of the Galaxy would shed light on, and 
possibly resolve, all these issues.

The large area covered by the SDSS, with accurate photometric measurements ($\sim$0.02 mag)
and faint flux limits ($r<22$), allows for a novel approach to studies of the
stellar distribution in the Galaxy: using a photometric parallax relation appropriate 
for main sequence stars, we estimate distances for a large number of 
stars and \emph{directly map the Galactic stellar number density} without the
need for an \emph{a-priori} model assumption\footnote{The use of photometric
parallax to determine Galactic model parameters is not particularly novel, having a long
history going back to at least \citet{Gilmore83}. The novelty in our approach is to use the
photometric parallax and wide area of SDSS to construct stellar density distribution maps first, 
and look for structure in the maps and fit analytic Galactic models second.}. In this paper,
we describe a study based on $\sim$48 million stars detected by the SDSS in
$\sim 6500$~deg$^2$ of sky. An advantage of this approach is that 
the number density of stars as a function of color and position in the Galaxy, $\rho(X, Y, Z, r-i)$
can be measured without assuming a particular Galactic model (e.g. the luminosity function and
functional forms that describe the density laws for disks and halo). Rather, with minimal
assumptions about the nature of the observed stellar population (that the large majority of the
observed stars are on the main sequence) and by using an adequate photometric parallax relation, 
the computed stellar number density maps can be used to get an overall picture about the 
distribution of stars first, and \emph{a-posteriori} constrain the density laws of 
Galactic components and look for deviations from them.

This is the first paper, in a series of three\footnote{We credit the late
J.R.R. Tolkien for demonstrating the virtues of this approach.}, that employs 
SDSS data and a photometric parallax relation to map the Galaxy. Here, we focus on the stellar
number density distribution. In Ivezi\'{c} et al. (2007, in prep., hereafter
Paper II) we discuss the distribution of photometric metallicity (calibrated 
using SDSS spectroscopic data), and in Bond et al. (2007, in prep., hereafter
Paper III) we analyze the stellar kinematics using radial velocity and proper
motion measurements. 
 
We begin by describing the SDSS data, the photometric parallax relations, 
and the construction of stellar number density maps in the following Section. 
Analysis of overall trends and identification of localized density features
(substructure) is described in Section~\ref{analysis}. In
Section~\ref{sec.galactic.model} we use the maps to derive best-fit parameters 
of density model for the Galactic disk and stellar halo. Section~\ref{vlgv} discusses the details of a 
remarkably large overdensity of stars identified in Section~\ref{analysis}.
Our results and their theoretical implications are summarized and
discussed in Section~\ref{Disc}.

\section{                 Data and Methodology                   }
\label{DM}

In this Section we list the basic characteristics of the SDSS imaging survey, 
discuss the adopted photometric parallax relation used to estimate 
the distance to each star, and describe a method for determining three-dimensional
number density distribution as a function of Galactic coordinates.

\subsection{  The Basic Characteristics of the SDSS Imaging Survey}

The SDSS is a digital photometric and spectroscopic survey which will cover up to one quarter 
of the celestial sphere in the North Galactic cap, and produce a smaller area ($\sim225$ deg$^{2}$)
 but much deeper survey in the Southern Galactic hemisphere\footnote{See also 
http://www.astro.princeton.edu/PBOOK/welcome.htm} \citep{York00,EDR,DR1,SDSSTelescope,SDSSMonitorTelescope}. 
The flux densities of detected objects are measured almost simultaneously in five bands ($u$, $g$, $r$, $i$, and $z$) with effective wavelengths of 3540 \AA, 
4760 \AA, 6280 \AA, 7690 \AA, and 9250 \AA\ \citep{Fukugita96,Gunn98,Smith02,Hogg01}. 
The completeness of SDSS catalogs for point sources is 
$\sim$99.3\% at the bright end ($r \sim 14$, where the SDSS CCDs saturate, \citealt{Ivezic01}), and drops to 95\% at  
magnitudes\footnote{These values are determined by comparing multiple scans of the same area 
obtained during the commissioning year. Typical seeing in these observations was 1.5$\pm$0.1
arcsec.} of 22.1, 22.4, 22.1, 21.2, and 20.3 in $u$, $g$, $r$, $i$ and $z$, respectively. 
All magnitudes are given on the AB$_{\nu}$ system (\citealt{Oke83}, for additional discussion 
regarding the SDSS photometric system see \citealt{Fukugita96} and \citealt{Fan99}). 
The final survey sky coverage of about 8,000 deg$^{2}$ will result in photometric 
measurements to the above detection limits for about 80 million stars and a similar number of 
galaxies. Astrometric positions are accurate to about 0.1 arcsec per coordinate for sources 
brighter than $r\sim$20.5$^{m}$ \citep{Pier03}, and the morphological information from the 
images allows robust point source-galaxy separation to $r\sim$ 21.5$^{m}$ \citep{Lupton02}. 
The SDSS photometric accuracy is $0.02$~mag (root-mean-square, at the bright end), with well
controlled tails of the error distribution \citep{Ivezic03a}. The absolute zero point
calibration of the SDSS photometry is accurate to within $\sim0.02$~mag \citep{Ivezic04}.
A compendium of technical details about SDSS can be found in \citet{EDR},
and on the SDSS web site (http://www.sdss.org).

\subsection{         The Photometric Parallax Method               }
\label{pp}

\begin{figure}                
\plotone{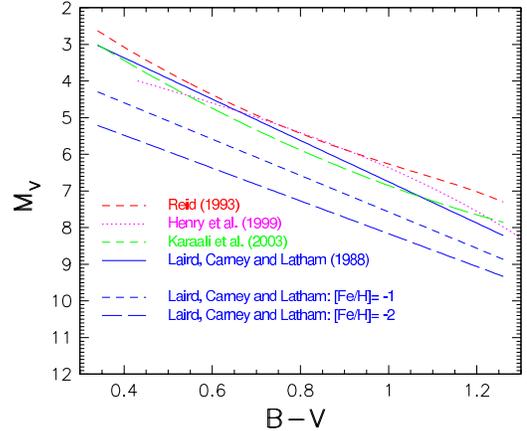}      
\caption{A comparison of photometric parallax relations, 
expressed in the Johnson system, from the literature. The relation 
from Henry et al. (1999) is valid for stars closer than 10 pc,
while other relations correspond to the Hyades main sequence. 
Note that the latter differ by a few tenths of a magnitude.
The relation from Laird, Carney \& Latham (1988) is also 
shown when corrected for two different metallicity values,
as marked in the legend. The gradient $dM_V/d[Fe/H]$ given
by their prescription is about 1 mag/dex at the blue end, and 
about half this
value at the red end. 
\label{pprel0}}
\end{figure}

\begin{figure}                
\plotone{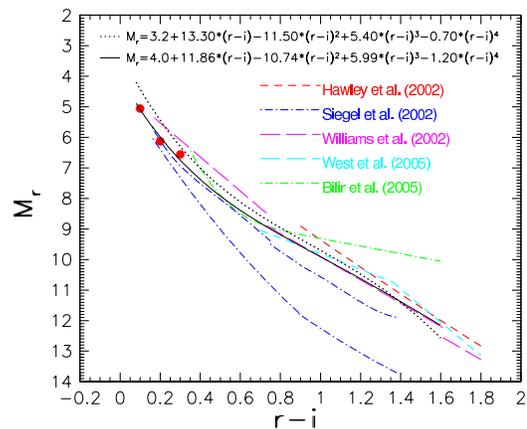}       
\caption{A comparison of photometric parallax relations in
the SDSS $ugriz$ system from the literature and adopted in
this work. The two relations adopted here are shown by the
dashed (``bright'' normalization) and solid (``faint'' 
normalization) lines. Other lines show  photometric parallax 
relations from the literature, as marked. The lower (thin) 
curve from Siegel et al. corresponds to low metallicity stars. 
The large symbols show SDSS observations of globular cluster M13.
\label{fig.Mr}}
\end{figure}

SDSS is superior to previous optical sky surveys because of its high catalog
completeness and  accurate multi-band CCD photometry to faint flux limits 
over a large sky area. The majority of stars detected
by SDSS are main-sequence stars ($\sim$98\%, \citealt{Finlator00}), which have a 
fairly well-defined color-luminosity relation\footnote{The uniqueness of color-luminosity
relation breaks down for stars at main sequence turn-off ($r-i \sim 0.11$~mag for disk, and
$r-i \sim 0.06$ for halo stars, \citealt{Chen01}). Those are outside of all but the bluest
bin of the $r-i$ range studied here.}. Thus, accurate SDSS colors can be 
used to estimate luminosity, and hence, distance, for each individual star. 
While these estimates are incorrect for a fraction of stars such as multiple
systems and non-main sequence stars, the overall contamination is small or controllable.

There are a number of proposed photometric parallax relations in the literature.
They differ in the methodology used to derive them (e.g., geometric parallax measurements,
fits to globular cluster color-magnitude sequences), photometric systems, 
and the absolute magnitude and metallicity range for which they are applicable.
Not all of them are mutually consistent, and most exhibit significant intrinsic
scatter of order a half a magnitude or more. Even the relations
corresponding to the same cluster, such as the Hyades, can differ by a few tenths
of a magnitude (see Fig.~\ref{pprel0}). 

In Fig.~\ref{fig.Mr} we compare several recent photometric parallax relations found
in the literature. They are all based on geometric parallax measurements, but the stellar
colors are measured in different photometric systems. In order to facilitate
comparison, we use photometric transformations between the Johnson and SDSS
systems derived for main-sequence stars by \citet{Ivezic07a}, and fits
to the stellar locus in SDSS color-color diagrams from \citet{Ivezic04}. 
As evident, different photometric parallax relations from the
literature are discrepant at the level of several tenths to a
magnitude. Furthermore, the relation proposed by \citet{Williams02}
is a piece-wise fit to restricted color ranges, and results in a discontinuous
relation. The behavior of Kurucz model atmospheres suggests that these
discontinuities are probably unphysical.

We constructed a fit, shown in Figure~\ref{fig.Mr}, that attempts to reconcile the 
differences between these relations. We require a low-order polynomial fit that 
is roughly consistent with the three relations at the red end, and properly reproduces the SDSS
observations of the position of the turn-off (median $M_r = 5$ at $r-i=0.10$) for 
globular cluster M13  (using a distance of 7.1 kpc, \citealt{Harris96}). The adopted relation
\eqarray{
\label{eq.Mr.faint}
 M_r = 4.0 + 11.86 \,(r-i) -10.74 \, (r-i)^2  \\ \nonumber
           + 5.99\, (r-i)^3 - 1.20\, (r-i)^4 
}
is very similar to the \citet{Williams02} relation at the red end, and agrees well
with the \citet{Siegel02} relation at the blue end.

In order to keep track of uncertainties in our results due to systematic
errors in photometric parallax relation, we adopt another relation. The
absolute magnitude difference between the two relations covers the plausible 
uncertainty range, and hence the final results are also expected to bracket 
the truth. While we could arbitrarily shift the normalization of eq.~\ref{eq.Mr.faint}
for this purpose, we instead use a relation that has an independent motivation.  

In Paper III, we propose a novel method to constrain the photometric parallax 
relation using kinematic data. The method relies on the large sky coverage by 
SDSS and simultaneous availability of both radial velocities and proper motion
data for a large number of stars. These data can be successfully modeled using 
simple models such as a non-rotating halo and a disk rotational lag that is 
dependent only on the height above the Galactic plane. The best-fit models
that are independently constrained using radial velocity and proper motion
measurements agree only if the photometric parallax relation is correct. 
That is, the tangential velocity components, that are proportional to distance
and measured proper motions, are tied to the radial velocity scale by adopting 
an appropriate distance scale. As a result of such kinematic analysis, 
we adopt a second photometric parallax relation
\eqarray{
\label{eq.Mr}
 M_r = 3.2 + 13.30 \,(r-i) -11.50 \, (r-i)^2  \\ \nonumber
      + 5.40\, (r-i)^3 - 0.70\, (r-i)^4. 
}
This relation is 0.66 mag brighter at the blue end ($r-i=0.1$), and matches 
eq.~\ref{eq.Mr.faint} at $r-i = 1.38$ (see Fig.~\ref{fig.Mr} for a
comparison). The normalization differences between
the two relations at the blue end correspond to a systematic distance scale
change of $\pm$18\%, relative to their mean.

To distinguish the two relations, we refer to the relation from
eq.~\ref{eq.Mr.faint} as the ``faint'' normalization, and to the relation from
eq.~\ref{eq.Mr} as the ``bright'' normalization. We note that, encouragingly, 
the {\it Hipparcos}-based $M_R$ vs. $R-I$ relation from \citet{Reid01} 
falls in between these two relations.

In sections to follow, we perform all the analysis separately for each relation, 
and discuss the differences in results when they are noticeable. For all figures,
we use the bright normalization, unless noted otherwise.

Equations~\ref{eq.Mr.faint}~and~\ref{eq.Mr} are quite steep, for example, 
$\Delta M_r / \Delta(r-i) \sim 10$~mag/mag at
the blue end ($r-i \sim 0.1$). Because of this steepness\footnote{This is not
an artifact of the SDSS photometric system, or the adopted photometric parallax relation.
For example, even for the linear $M_V$ vs. $B-V$ relation from \citet{Laird88}
$dM_V/d(B-V)=5.6$~mag/mag.}, very accurate photometry ($0.01$-$0.02$~mag) is
required to reach the intrinsic accuracy of the photometric relation (about
$0.2$~mag or better for individual globular clusters; for metallicity effects
see below). Older photographic surveys have photometric errors of
$\sim0.1$-$0.2$~mag \citep{Sesar06}, and inaccurate color measurements 
result in $M_r$ errors exceeding $\sim$1 mag. Hence, with the SDSS, the
intrinsic accuracy of the photometric parallax method can be approached to 
a faint flux limit and over a  large sky area for the first time.

\subsubsection{  Effects of Metallicity on the Photometric Parallax Relation  }
\label{sec.pp.metallicity}

\begin{figure}                
\plotone{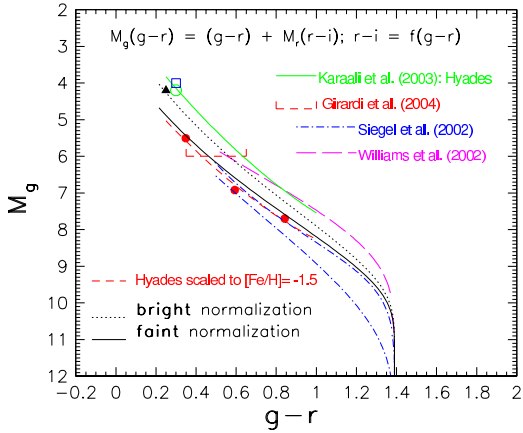}       
\caption{A comparison of photometric parallax relations
from the literature and adopted in this work, shown using
blue SDSS bands (stars with spectral types later than $\sim$M0 
have $g-r\sim1.4$). The two relations adopted here are shown by the
dotted (``bright'' normalization) and solid (``faint'' 
normalization) lines. These are the {\it same} relations as 
shown in Fig.~\ref{fig.Mr}, translated here into the $M_g(g-r)$ 
form using the $r-i=f(g-r)$ relation appropriate for main sequence
stars on the main stellar locus.
Other lines show  photometric parallax 
relations from the literature, as marked. The line marked 
Girardi et al. shows the range of model colors for $M_g=6$. 
The lower (thin) curve from Siegel et al. corresponds to 
low metallicity stars. The triangle, circle and square show
the SDSS observations of globular clusters Pal 5 ($[Fe/H]=-1.4$),
and the Hyades ($[Fe/H]=0.1$) and M48 ($[Fe/H]=-0.2$) open clusters, 
respectively. The three large dots show the SDSS observations of globular 
cluster M13 ($[Fe/H]=-1.5$). Note the good agreement between these
observations and the Hyades sequence scaled to M13's metallicity
using the prescription from Laird, Carney \& Latham (1988). 
For reference, $B-V = 0.95\,(g-r) + 0.20$ to within 0.05 mag. 
\label{pprel2}\vskip 1em}
\end{figure}

The main source of systematic errors in photometric parallax relation is its 
dependence on metallicity. For example, \citet{Siegel02} address this problem
by adopting different relations for low- and high-metallicity stars (c.f. 
Fig.~\ref{fig.Mr}). Another approach is to estimate metallicity, either from
a spectrum or using photometric methods such as a UV excess based $\delta$
method (e.g. \citealt{Carney79}),
and then apply a correction to the adopted photometric parallax relation 
that depends both on color and metallicity (e.g. \citealt{Laird88}),
as illustrated in Fig.~\ref{pprel0}. We have tested the \citeauthor*{Laird88}
metallicity correction by scaling the Hyades main sequence, as given 
by \citet{Karaali03}, using $[Fe/H]=-1.5$ appropriate for M13, and
comparing it to SDSS observations of that cluster. As shown in
Fig.~\ref{pprel2}, the agreement is very good ($\sim$0.1 mag). 

An application of $\delta$ method to SDSS photometric system was recently
attempted by \citet{Karaali05}. However, as they pointed out, 
their study was not based on SDSS data, and thus even small differences
between different photometric systems may have a significant effect on
derived metallicities (especially since the SDSS $u$ band photometry 
is not precisely on the AB system, see \citealt{Eisenstein06}). 

The expected correlation of metallicity and the SDSS $u-g$ and $g-r$ colors
was recently discussed by \citet{Ivezic07b}. Using SDSS photometry
and metallicity estimates derived from SDSS spectra \citep{AllendePrieto06}, they 
demonstrated a very strong dependence of the median metallicity on the position 
in the $g-r$ vs. $u-g$ color-color diagram. For example, for stars at the 
blue tip of the stellar locus ($u-g<1$, mostly F stars), the expression
\begin{equation}
        [Fe/H] = 5.11\,(u-g) - 6.33
\end{equation}
reproduces the spectroscopic metallicity with an rms of only 0.3 dex.
This relation shows that even in this favorable case (it is much harder 
to estimate metallicity for red stars), a 0.1 mag error of the $u-g$ color 
would introduce an error of $[Fe/H]$ as large as 0.5 dex, resulting in an 
error in the absolute magnitude of $\sim$0.5 mag.

We aim here to study the Galaxy to as large a distance limit as the SDSS
photometric sample of stars allows. While metallicity could be estimated
for bright blue stars using the above expression, for most stars in
the sample the SDSS $u$ band photometry is not sufficiently accurate to
do so reliably. For example, the random error of $u-g$ color becomes
0.1 mag at $u\sim20.5$ (\citealt{Ivezic03a}), which corresponds
to $g\sim19.5$ or brighter even for the bluest stars. Therefore, metallicity
estimates based on the $u-g$ color would come at the expense of a more
than 2 mag shallower sample. Hence, we choose not to correct the adopted 
photometric parallax relation for metallicity effects, and only utilize
the correlation between metallicity and $u-g$ color when constraining
the metallicity distribution of a large halo overdensity discussed in 
Section~\ref{vlgv}.

We point out that the adopted relations do account for metallicity effects 
to some extent. The metallicity distribution shows a much larger gradient
perpendicular to the Galactic plane than in the radial direction (see 
Fig.~3 in \citealt{Ivezic07b}). As we only consider high 
Galactic latitude data, the height above the plane is roughly proportional 
to distance. At the red end, the adopted relations are tied via geometric 
parallax to nearby metal-rich stars, and even the faintest M dwarfs in SDSS 
sample are only $\sim$1 kpc away. At the blue end, the adopted relations are
tied to globular clusters and halo kinematics, which is appropriate for 
the bluest stars in the sample, that are detected at distances from several
kpc to $\sim$10 kpc. Thus, in some loose ``mean'' sense, the adopted relation 
smoothly varies from a relation 
appropriate for nearby, red, high-metallicity stars to a relation appropriate 
for more distant, blue, low-metallicity stars\footnote{When the adopted photometric
parallax relation is applied to the Sun ($r-i=0.10$), the resulting
absolute magnitude is too faint by about 0.5~mag. This is an expected
result, because the relation is anchored to a low-metallicity globular
cluster at the blue end. For both relations, the predicted absolute magnitudes 
of low-temperature, low-metallicity stars are systematically too bright.
However, the majority of such stars (e.g., distant halo M-dwarfs) are 
faint, and well beyond the flux limit of the survey.}. 
Furthermore, \citet{Reid01} show that 
photometric parallax relations constructed using red photometric bands,
such as our $M_r$ vs. $r-i$ relation, are much less sensitive to metallicity 
than the traditional $M_V$ vs. $B-V$ relation (compare the top left and 
bottom right panel in their Fig.~15). 

Nevertheless, to further control metallicity and other systematic effects, 
we perform analysis in narrow color bins, as described in more detail in 
Section~\ref{sec.maps}.

\subsubsection{A Test of the Photometric Parallax Relation using Resolved Binary Stars }
\label{sec.widebinaries}

\begin{figure}
\scl{.45}
\plotone{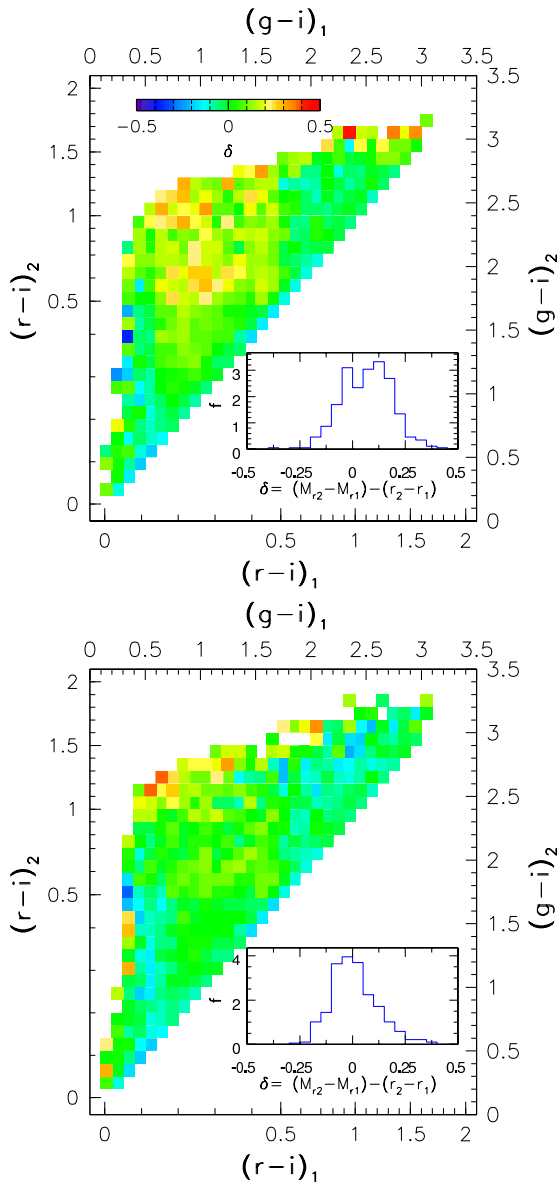}
\caption{
The distribution of the median $\delta$ for a sample of $\sim17,000$ candidate wide-angle binaries in 
the $(r-i)_1$ (color of brighter pair member; the primary) vs.~$(r-i)_2$ (color of 
fainter member; the secondary) color-color diagram. Here, $\delta=(M_{r,2}-M_{r,1})-(r_2-r_1)$,
is the difference of two estimates (one from the absolute, and the other from the apparent magnitudes)
of brightness difference between the two components. In the top panel, the
absolute magnitudes were estimated using eq.~\ref{eq.Mr}  (the ``bright''
paralax relation; the dotted line in Figure~\ref{fig.Mr}), and in the bottom panel
using eq.~\ref{eq.Mr.faint} (the ``faint''
paralax relation; the solid line in Figure~\ref{fig.Mr})
Inset histograms show the distribution of the median $\delta$ evaluated for
each color-color pixel. The distribution medians are 0.07 (top panel) and
-0.004 (bottom panel), and the dispersions (determined from the interquartile
range) are 0.13 and 0.10 mag, respectively.
\label{fig.plxbinaries}}
\end{figure}

The number of close stellar pairs in the SDSS survey with distances in the 
2--5 arcsec range shows an excess relative to the extrapolation
from larger distances (Sesar et al. 2007, accepted to ApJ). Statistically, they 
find that $\sim$70\% of such pairs are physically associated binary systems.
Since they typically have different colors, they also
have different absolute magnitudes. The difference in absolute
magnitudes, $\Delta M$, can be computed from an adopted photometric
parallax relation without the knowledge of the system's distance,
and should agree with the measured difference of their apparent
magnitudes, $\Delta m$. The distribution of the difference,
$\delta = \Delta m - \Delta M$ should be centered on zero and should
not be correlated with color if the shape of photometric parallax
relation is correct (the overall normalization is not constrained,
but this is not an issue since the relation can be anchored
at the red end using nearby stars with geometric parallaxes)\footnote{Note the
similarities of this method, and the method of reduced proper motions \cite{Luyten68}.
}.
The width of the $\delta$ distribution provides an upper limit
for the intrinsic error of the photometric parallax method 
(note, however, that $\delta$ is not sensitive to systematic errors 
due to metallicity since the binary components presumably have the
same metallicity). 

We have performed such a test of adopted parallax relations using a sample 
of 17,000 candidate binaries from SDSS Data Release 5. Pairs of stars with 
$14 < r < 20$ are selected as candidate wide binaries if their angular separation 
is in the 3--4 arcsec range. The brighter star (in the $r$ band) is designated 
as the primary (subscript 1), and the fainter one as the secondary (subscript
2). For each pair, we calculated $\delta$ twice -- once assuming the bright
photometric parallax relation (eq.~\ref{eq.Mr}), and once assuming the faint
relation (eq.~\ref{eq.Mr.faint}). We further remove from the sample all pairs
with $|\delta|>0.5$, those likely being the interlopers and not actual physical
pairs.

The results of this analysis are summarized in Figure~\ref{fig.plxbinaries}.
The color-coded diagrams show the dependence of $\delta$ on the $r-i$ colors 
of the primary and the secondary components. The median $\delta$ value in 
each $(ri_1, ri_2)$ pixel measures whether the absolute magnitude difference 
obtained using the parallax relation for stars of colors $ri_1$ and $ri_2$ is 
consistent with the difference of their apparent magnitudes (in each bin,
the $\delta$ distribution is much more peaked than for a random sample of 
stars, and is not much affected by the  $|\delta|<0.5$ cut). 
If the {\it shape} of the photometric parallax relation is correct, the 
median $\delta$ should be close to zero for all combinations of $ri_1$ and 
$ri_2$. 

The distributions of the median $\delta$ for each pixel are fairly narrow
($\sim 0.1$~mag), and centered close to zero (the medians are 0.07 mag for
the bright relation and $-0.004$~mag for the faint relation). Irrespective 
of color and the choice of photometric parallax relation, the deviations are 
confined to the $\sim \pm 0.25$mag range, thus placing a stringent upper 
limit on the errors in the shape of the adopted relations. 
The $\delta$ distributions root-mean-square width of $\sim 0.1$~mag 
implies average distance error of about 5\%. Nevertheless, the binary stars 
in a candidate pair are of presumably identical metallicities. As a large 
fraction of the intrinsic scatter of $M_r(r-i)$ comes the dependence of
absolute magnitude on metallicity, we adopt a conservative value of 
$\sigma_{M_r} = 0.3$. 

The coherent deviations seen in Figure~\ref{fig.plxbinaries} (e.g. around
$ri_1 \sim 0.3$ and $ri_2\sim 0.5$) indicate that the adopted parallax
relations could be improved. Given already satisfactory accuracy of
the adopted relations, such a study is presented separately (Sesar et al. 2007, accepted
to ApJ).

\subsubsection{   Contamination by Giants        }

\begin{figure}
\scl{.80}
\plotone{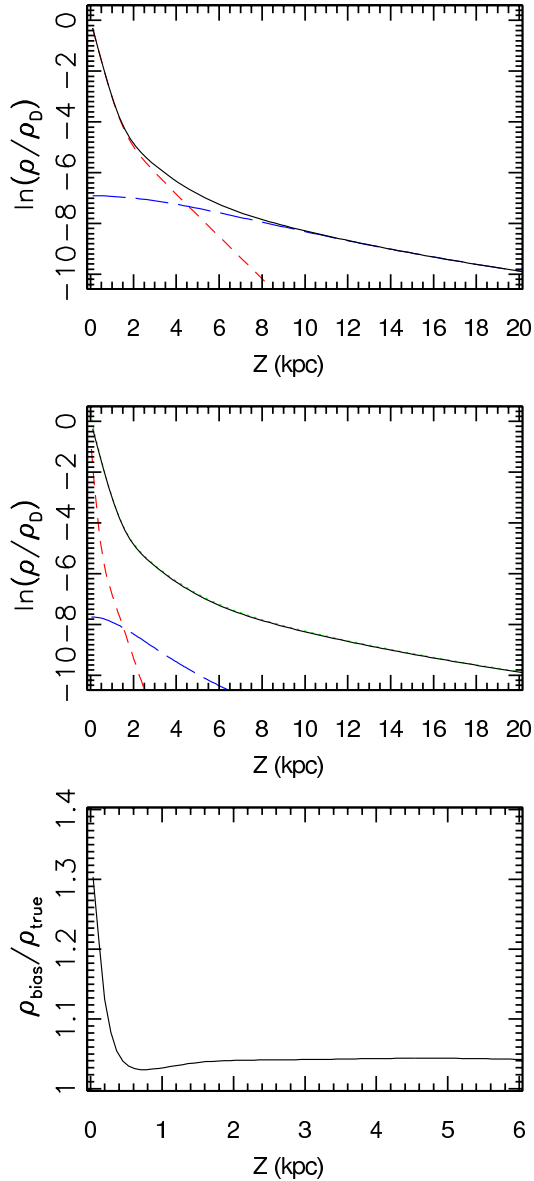}
\caption{An illustration of the effects of misidentifying giants
as main sequence stars. The top panel shows the $Z$ dependence of 
stellar density at $R$=8 kpc for a fiducial model consisting 
of two disks with scale heights of 300 pc and 1200 pc. The 
contribution of the disks is shown by the short-dashed line, and 
the long-dashed line shows the contribution of a power-law 
spherical halo with the power-law index of 3. The middle panel 
shows the contribution of misidentified giants from disks (short-dashed) 
and halo (long-dashed) for an assumed giant fraction of 5\%, and 
underestimated distances by a factor of 3. The ``contaminated'' model 
is shown by dotted line, just above the solid line, which is the same 
as the solid line in the top panel. The ratio of the ``contaminated'' 
and true density is shown in the bottom panel (note the different 
horizontal scale).
\label{models1}}
\end{figure}

The photometric parallax method is not without pitfalls, even when applied
to the SDSS data. Traditionally, the application of this method was prone
to significant errors due to sample contamination by evolved stars (subgiants and giants,
hereafter giants for simplicity), and their underestimated distances. This effect is 
also present in this study, but at a much less significant level because of the faint 
magnitudes probed by SDSS. At these flux levels, the distances corresponding to giants 
are large and sometimes even beyond the presumed edge of the Galaxy (up to $\sim$100 kpc). The stellar 
density at these distances is significantly smaller than at distances corresponding
to main sequence stars with the same apparent magnitude. The contamination with evolved stars
rapidly asymptotes (e.g., assuming a $\sim r^{-3}$ halo profile) and may decline when the edge
of the halo is reached.

A quantitative illustration of this effect is shown in Fig.~\ref{models1}
for a fiducial Galaxy model. The worst case scenario corresponds to G giants with 
$g-r\sim0.4-0.5$ and $r-i\sim0.15-0.20$, and their most probable fraction is about 5\%.
This color range and the fraction of giants was determined using the SDSS data 
for the globular cluster M13 (the data for the globular cluster Pal 5 imply similar 
behavior). To be conservative, we have also tested a model with a twice as large 
fraction of giants. This analysis (see bottom panel) shows that the 
effect of misidentifying giants as main sequence stars is an overall bias 
in estimated number density of $\sim$4\% ($\sim$8 \% when the fraction of
giants is 10\%), with little dependence on distance from the Galactic plane 
beyond 500 pc. This is the distance range probed by stars this blue, and thus
the worst effect of contamination by giants is a small overall overestimate
of the density normalization. Shorter distances are probed by redder stars, 
M dwarfs, for which the contamination by M giants is negligible because the 
luminosity difference between red giants and dwarfs is very large (e.g. there are 
tens of millions of M dwarfs in our sample, while the 2MASS survey revealed only 
a few thousand M giants in the same region of the sky, \citealt{Majewski03}). Hence, 
the misidentified giants are not expected to significantly impact our analysis. 

\subsubsection{   Unrecognized Multiplicity       }

Multiplicity may play a significant role by systematically making unresolved multiple systems, when
misidentified as a single star, appear closer then they truly are. The net effect of 
unrecognized multiplicity on derived distance scales, such as scale height and scale length, is
to underestimate them by up to $\sim$35\% (see Section~\ref{sec.binarity} here and \citealt{Siegel02}).
The magnitude of this bias is weakly dependent on the actual composition of the binaries 
(e.g. their color difference and luminosity ratio), but it is dependent on the fraction 
of multiple systems in the Galaxy. Since this fraction is not well constrained,
for the purpose of constructing
the number density maps (Section~\ref{mkmaps}) we assume all observed objects are single stars.
This biases the distance scales measured off the maps, making them effectively lower limits, and
we \emph{a-posteriori} correct for it, after making the Galactic model fits 
(Sections~\ref{sec.binarity}~and~\ref{sec.bestfit}). Note that this bias cannot affect the 
shapes of various density features seen in the maps, unless the properties of multiple systems 
varies greatly with the position in the Galaxy. 

\subsubsection{ Distance Range Accessible to SDSS Observations  of Main-Sequence Stars   }

A disadvantage of this method is its inability, when applied to main sequence stars, to probe distances
as large as those probed by RR Lyrae and M giants (20 kpc vs. 100 kpc). 
However, a significant advantage of using main sequence stars is the vastly 
larger number of stars (the number ratio of main sequence to RR Lyrae stars in the 
SDSS sample is $\sim$10,000, and even larger for M giants \citep{Ivezic03a,Ivezic03c,Ivezic05}.
This large number of main-sequence stars allows us to study
their number density distribution with a high spatial resolution, and without
being limited by Poisson noise in a large fraction of the observed volume.

\subsection {  The SDSS Stellar Sample      }
\label{sec.maps}

In this Section we describe the stellar sample utilized in this work, and 
the methods used to construct the three-dimensional number density maps.

\subsubsection{The Observations}

\begin{figure*}
\centering
\plotone{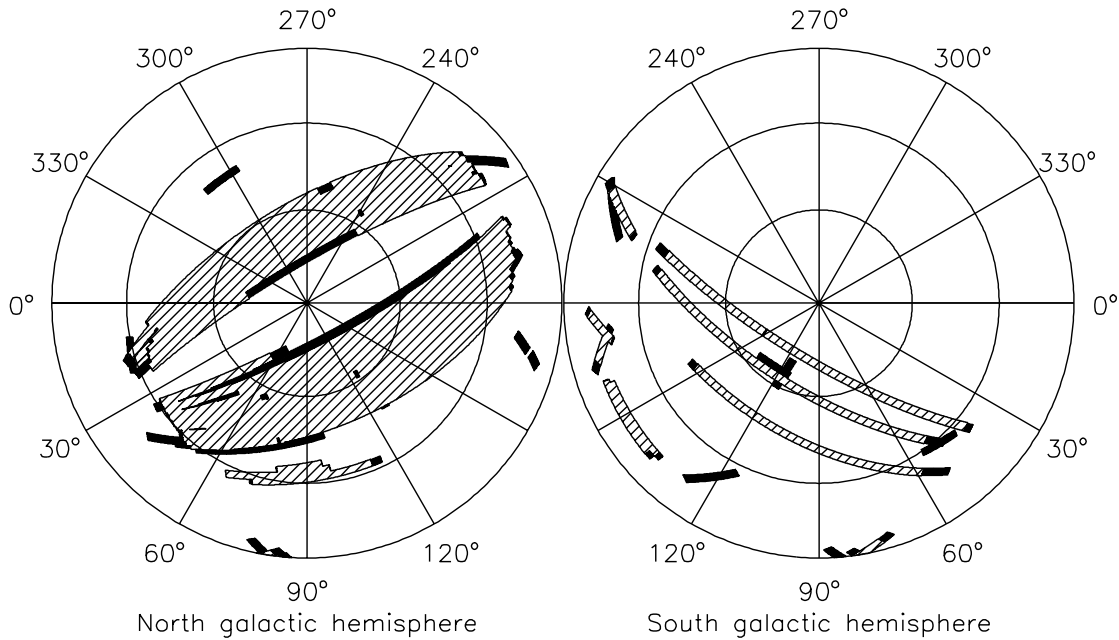}
\caption{Footprint on the sky of SDSS observations used in this work shown in Lambert equal area
projection (hatched region). The circles represent contours of constant Galactic latitude, with the
straight lines
showing the location of constant Galactic longitude. For this study, observations from 248 SDSS
imaging runs were used, obtained over the course of 5 years. The data cover $5450$~deg$^2$ of the north
Galactic hemisphere, and a smaller but more frequently sampled area of $1088$~deg$^2$ in the
southern Galactic hemisphere.
\label{fig.skymap}}
\end{figure*}

We utilize observations from 248 SDSS imaging runs obtained in a 5 year period through September
2003, which cover $6,538$~deg$^2$ of the sky. This is a superset of imaging runs described in SDSS Data
Release 3 \citep{DR3}, complemented by a number of runs from SDSS Data Release 4 \citep{DR4}
 and the so called ``Orion'' runs \citep{Finkbeiner04}. The sky coverage of
these 248 runs is shown in figure \ref{fig.skymap}. They cover $5450$~deg$^2$ in the northern
Galactic hemisphere, and $1088$~deg$^2$ in the south.

We start the sample selection with 122 million detections classified 
as point sources (stars) by the SDSS photometric pipeline, {\it Photo} \citep{Lupton02}. For a
star to be included in the starting sample, we require that $r < 22$, and that it is also 
detected (above 5$\sigma$) in at least the $g$ or $i$ band. The latter requirement is necessary 
to be able to compute either the $g-r$ or $r-i$ color. The two requirements reduce the sample to
$87$ million observations. For each magnitude measurement, {\it Photo} 
also provides a fairly reliable estimate of its accuracy \citep{Ivezic03a}, 
hereafter $\sigma_g$, $\sigma_r$ and $\sigma_i$. We correct all measurements for the 
interstellar dust extinction using the \citet{Schlegel98} (hereafter SFD) maps.

\subsubsection{The Effects of Errors in Interstellar Extinction Corrections }
\label{extinction}

The SFD maps are believed to be correct within 10\%, or better.
This uncertainty plays only a minor role in this work because the
interstellar extinction is fairly small at the high galactic latitudes
analyzed here ($|b|>25$): the median value of the extinction in
the $r$ band, $A_r$, is 0.08, with 95\% of the sample with $A_r < 0.23$
and 99\% of the sample with $A_r<0.38$. Thus, only about 5\% of stars
could have extinction correction uncertain by more than the photometric
accuracy of SDSS data ($\sim$0.02 mag).
The SFD maps do not provide the wavelength dependence of the interstellar
correction, only its magnitude. The extinction corrections in the five
SDSS photometric bands are computed from the SFD maps using conversion
coefficients derived from an $R_V=3.1$ dust model. Analysis of the position
of the stellar locus in the SDSS color-color diagrams suggests that these
coefficients are satisfactory at the level of accuracy and galactic
latitudes considered here \citep{Ivezic04}.

We apply full SFD extinction correction to all stars in the sample.
This is inappropriate for the nearest stars because they are not
beyond all the dust. Distances to the nearest stars in our sample,
those with $r-i=1.5$ (the red limit) and $r\sim 14$ (approximately
the SDSS $r$ band saturation limit), are $\sim$30 pc (distance determination
is described in the next two sections). Even when these
stars are observed at high galactic latitudes, it is likely that
they are over-corrected for the effects of interstellar extinction.
To estimate at what distances this effect becomes important, we have
examined the dependence of the $g-r$ color on apparent magnitude for
red stars, selected by the condition $r-i>0.9$, in the region defined
by $210 < l < 240$ and $25 < b < 30$. The distribution of the intrinsic
$g-r$ color for these stars is practically independent of their $r-i$
color (see Fig.~\ref{locusfit}), with a median of 1.40 and a standard deviation of only 0.06 mag
\citep{Ivezic04}. This independence allows us to test at what
magnitude (i.e. distance) the applied SFD extinction corrections become
an overestimate because, in such a case, they result in $g-r$ colors that are bluer than
the expected value of $\sim1.40$. We find that for $r>15$ the median
$g-r$ color is nearly constant -- it varies by less than 0.02 mag over
the $15 < r < 20$ range. On the other hand, for stars with $r< 15$ the
median $g-r$ color becomes much bluer -- at $r=14.5$ the median value
is 1.35. This demonstrates that stars at $r>15$ are already behind most
of the dust column. With the median $r-i$ color of 1.17, the implied
distance corresponding to $r=15$ is $\sim$80 pc. For the probed
galactic latitude range, this indicates that practically all the
dust is confined to a region within $\sim$70 pc from the galactic
midplane (here we define midplane as a plane parallel to the galactic
plane that has $Z=-25$ pc, because the Sun is offset from the midplane
towards the NGP by $\sim$25 pc; for more details see below). We arrive
to the same conclusion about the dust distribution when using an
analogous sample in the south galactic plane with $|b|\sim12$ (in this
case the median $g-r$ color is systematically bluer for $r<19$, due to
different projection effects and the Sun's offset from the midplane).  
Hence, in order to avoid the
effects of overestimated interstellar
extinction correction for the nearest stars, we exclude stars that are
within 100 pc from the galactic plane when fitting galaxy models
(described below). Only 0.05\% of stars in the sample are at such
distances. In summary, the effects of overestimated interstellar
extinction correction, just as the effects of sample contamination
by giants, are not very important due to the faint magnitude range
probed by SDSS.

\subsubsection{The Treatment of Repeated Observations}

\begin{figure}
\centering
\plotone{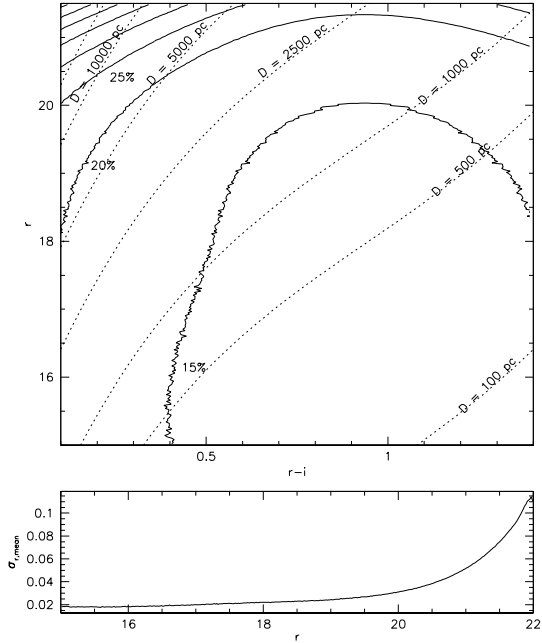}
\caption{The top panel shows the mean fractional distance error as a function of the $r-i$ 
color and $r$ band magnitude, assuming the intrinsic photometric parallax relation scatter of
$\sigma_{M_{r}} = 0.3$~mag. The solid lines are contours of constant fractional distance 
error, starting with $\sigma_D/D = 15$\% (lower right) and increasing in increments of $5$\% 
towards the top left corner. The dotted lines are contours of constant distance, and
can be used to quickly estimate the distance errors for an arbitrary combination of 
color and magnitude/distance. Fractional distance errors are typically smaller than $\sim 20$\%.
Note that the distance errors act as a $\sigma_{M_{r}}$ wide convolution kernel 
in magnitude space, and leave intact structures larger than the kernel scale. In particular,
they have little effect on the slowly varying Galactic density field and the determination 
of Galactic model parameters.
\label{magerr2}}
\end{figure}

\begin{deluxetable}{rrr}
\tablewidth{3.2in}
\tablecaption{Repeat Observation Statistics\label{tbl.starcat}}
\tablehead{
	\colhead{$N_{app}$} & \colhead{$N(r < 22)$} & 
	\colhead{$N(r < 21.5)$}
}
\startdata 
1            & 30543044 & 2418472 \\
2            & 11958311 & 1072235 \\
3            & 3779424  & 3471972 \\
4            & 856639   & 785711 \\
5            & 220577   & 199842 \\
6            & 105481   & 93950 \\
7            & 141017   & 132525 \\
8            & 43943    & 40065 \\ 
9            & 59037    & 57076 \\ 
10           & 15616    & 15002 \\
11           & 1522     & 1273 \\ 
12           & 2012     & 1772 \\ 
13           & 2563     & 2376 \\ 
14           & 1776     & 1644 \\ 
15           & 1864     & 1741 \\ 
16           & 3719     & 3653 \\ 
17           & 1281     & 1253 \\ 
             &          &  \\ 
$N_{stars}$  & 47737826 & 39716935 \\
$N_{obs}$    & 73194731 & 62858036 \\
\enddata
\tablecomments{Repeat observations in the stellar sample: Because of partial imaging scan overlaps and
the convergence of scans near survey poles, a significant fraction of observations are
repeated observations of the same stars. In columns $N(r<22)$ and $N(r < 21.5)$
we show the number of stars observed $N_{app}$ times for stars with average
magnitudes less than $r = 22$ and $r = 21.5$, respectively. The final two rows
list the total number of stars in the samples, and the total number
of observations.}
\end{deluxetable}

\begin{deluxetable*}{rccccrrrrr}
\tabletypesize{\scriptsize}
\tablecaption{Number Density Distribution Maps\label{tbl.bins}}
\tablewidth{6in}
\tablecolumns{9}
\tablehead{
	 &  &  & &  & 
	\multicolumn{2}{c}{Bright} & \multicolumn{2}{c}{Faint} \\
	\colhead{\#}
	& \colhead{$ri_0$ - $ri_1$} & \colhead{$N_{stars} (\times 10^{6})$}
	& \colhead{$<gr>$} &  \colhead{$SpT$}
	& \colhead{$\tilde{M_r}$} & \colhead{$D_0-D_1 (dx)$} 
	& \colhead{$\tilde{M_r}$} & \colhead{$D_0-D_1 (dx)$}
}
\startdata 
 1 & 0.10 - 0.15 & 4.2 & 0.36   & $\sim$F9&  4.69  & 1306 - 20379 (500) &  5.33 &  961 -  15438  (500) \\ 
 2 & 0.15 - 0.20 & 3.8 & 0.48   & F9-G6   &  5.20  & 1021 - 16277 (400) &  5.77 &  773 -  12656  (400) \\
 3 & 0.20 - 0.25 & 2.8 & 0.62   & G6-G9   &  5.67  & 816 -  13256 (400) &  6.18 &  634 -  10555  (400) \\
 4 & 0.25 - 0.30 & 2.0 & 0.75   & G9-K2   &  6.10  & 664 -  10989 (300) &  6.56 &  529 -   8939  (300) \\
 5 & 0.30 - 0.35 & 1.5 & 0.88   & K2-K3   &  6.49  & 551 -   9259 (200) &  6.91 &  448 -   7676  (200) \\
 6 & 0.35 - 0.40 & 1.3 & 1.00   & K3-K4   &  6.84  & 464 -   7915 (200) &  7.23 &  384 -   6673  (200) \\
 7 & 0.40 - 0.45 & 1.2 & 1.10   & K4-K5   &  7.17  & 397 -   6856 (200) &  7.52 &  334 -   5864  (200) \\
 8 & 0.45 - 0.50 & 1.1 & 1.18   & K5-K6   &  7.47  & 344 -   6008 (150) &  7.79 &  293 -   5202  (150) \\
 0 & 0.50 - 0.55 & 1.0 & 1.25   & K6      &  7.74  & 301 -   5320 (150) &  8.04 &  260 -   4653  (150) \\
10 & 0.55 - 0.60 & 0.9 & 1.30   & K6-K7   &  8.00  & 267 -   4752 (150) &  8.27 &  233 -   4191  (150) \\
11 & 0.60 - 0.65 & 0.8 & 1.33   & K7      &  8.23  & 238 -   4277 (100) &  8.49 &  210 -   3798  (100) \\
12 & 0.65 - 0.70 & 0.8 & 1.36   & K7      &  8.45  & 214 -   3874 (100) &  8.70 &  190 -   3458  (100) \\
13 & 0.70 - 0.80 & 1.4 & 1.38   & K7-M0   &  8.76  & 194 -   3224 (75)  &  9.00 &  173 -   2897  (100) \\
14 & 0.80 - 0.90 & 1.4 & 1.39   & M0-M1   &  9.15  & 162 -   2714 (60)  &  9.37 &  145 -   2450  (60)  \\
15 & 0.90 - 1.00 & 1.3 & 1.39   & M1      &  9.52  & 136 -   2291 (50)  &  9.73 &  122 -   2079  (50)  \\
16 & 1.00 - 1.10 & 1.3 & 1.39   & M1-M2   &  9.89  & 115 -   1925 (50)  & 10.09 &  104 -   1764  (50)  \\
17 & 1.10 - 1.20 & 1.3 & 1.39   & M2-M3   & 10.27  & 96 -    1600 (40)  & 10.45 &   88 -  1493   (40)  \\
18 & 1.20 - 1.30 & 1.1 & 1.39   & M3      & 10.69  & 80 -    1306 (30)  & 10.81 &   74 -  1258   (30)  \\
19 & 1.30 - 1.40 & 0.9 & 1.39   & M3      & 11.16  & 65 -    1043 (25)  & 11.18 &   63 -  1056   (25)  
\enddata
\tablecomments{The number density map parameters. Each of the 19 maps is a volume limited three-dimensional
density map of stars with $ri_0 < r-i < ri_1$, corresponding to MK spectral
types and mean $g-r$ column listed in columns $SpT$ and $<gr>$, respectively.
Median absolute magnitude $\tilde{M_r}$, distance limits $D_0-D_1$ (in parsecs) and binning pixel 
scale $dx$ (also in parsecs) are given in columns labeled ``Bright'' and ``Faint'', for the bright
(Equation~\ref{eq.Mr}) and faint (Equation~\ref{eq.Mr.faint}) photometric parallax relation.
The number of stars in each $r-i$ bin is given in $N_{stars}$ column (in millions).}
\end{deluxetable*}

SDSS imaging data are obtained by tracking the sky in six parallel scanlines, each 13.5 arcmin wide.
The six scanlines from two runs are then interleaved to make a filled stripe. Because of the scan
overlaps, and because of the convergence of the scans near the survey poles, about 40\%
of the northern survey is surveyed at least twice. Additionally, the southern survey areas will be
observed dozens of times to search for variable objects and, by stacking the frames, to push 
the flux limit deeper. For these reasons, a significant fraction of measurements are repeated 
observations of the same stars.

We positionally identify observations as corresponding to the same object if they are within 1 arcsec of
each other (the median SDSS seeing in the $r$ band is 1.4 arcsec). Out of the initial $\sim$122 
million observations, the magnitude cuts and positional matching produce
a catalog of 47.7 million unique stars (the ``star catalog'', Table~\ref{tbl.starcat}). 
They span the MK spectral types from $\sim$F9 to $\sim$M3 (Table \ref{tbl.bins}).
There are two or more observations for about 36\% (17.2 million) of observed stars. For stars 
with multiple observations we take the catalog magnitude of the star to be equal to the weighted mean of all 
observations. In this step there is a tacit assumption that the variability is not important, 
justified by the main-sequence nature of the stellar sample under consideration (for the
variability analysis of the SDSS stellar sample see \citealt{Sesar06}).

As discussed in Section~\ref{pp}, an accurate determination of stellar distances by photometric
parallax hinges on a good estimate of the stellar color and magnitude. In the bottom panel of
Fig.~\ref{magerr2} we show the mean $r$ magnitude error of stars in the catalog as a function 
of the $r$ band magnitude. The photometric errors are $\sim$0.02 mag for bright objects
(limited by errors in modeling the point spread function), and steadily increase towards the
faint end due to the photon noise. At the adopted sample limit, $r=22$, the $r$ band photometric 
errors are $\sim$0.15 mag. The $g$ and $i$ band magnitude errors display similar behavior as 
for the $r$ band.

\subsubsection {     Maximum Likelihood Estimates of True Stellar Colors       }

\begin{figure}
\centering
\plotone{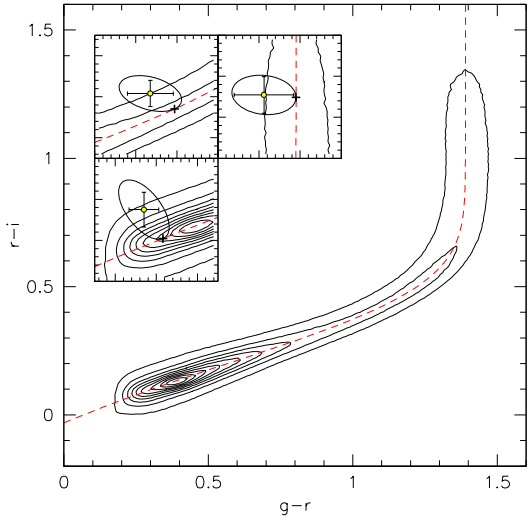}
\caption{The distribution of $\sim$48 million stars analyzed in this work in the  $r-i$ vs. $g-r$ 
color-color diagram, shown by isodensity contours. Most stars lie on a narrow 
locus, shown by the dashed line, whose width at the bright end is 0.02 mag for blue stars
($g-r\la1$)
and 0.06 mag for red stars ($g-r\sim1.4$). The inserts illustrate the maximum likelihood method
used to improve color estimates: the ellipses show measurement errors, and the crosses
are the color estimates obtained by requiring that a star lies exactly on the stellar
locus. Note that the principal axes of the error ellipses are not aligned with the axes of the
color-color diagram because both colors include the $r$ band magnitude.
\label{locusfit}}
\end{figure}

The photometric parallax relation (eq.~\ref{eq.Mr}) requires only the knowledge
of $r-i$ color to estimate the absolute magnitude. The accuracy of this estimate
deteriorates at the faint end due to increased $r-i$ measurement error. It also 
suffers for blue stars ($r-i < 0.2$) of all magnitudes because the slope of the photometric parallax 
relation, $\Delta M_r / \Delta(r-i)$, is quite large at the blue end -- for these stars 
it would be better to use the $g-r$ (or $u-g$) color to parametrize the photometric parallax 
relation. On the other hand, the $g-r$ color is constant for stars later than $\sim$M0 ($g-r
\sim 1.4$), and cannot be used for this purpose. These problems can be alleviated to some extent
by utilizing the fact that colors of main sequence stars form a very narrow, 
nearly one-dimensional locus.

The $r-i$ vs. $g-r$ color-color diagram of stars used in this work is shown in 
Fig.~\ref{locusfit}. We find that the stellar locus is well described by the 
following relation: 
\eqarray{
	g-r = 1.39 (1-\exp[-4.9(r-i)^3 \\ \nonumber
	- 2.45(r-i)^2 - 1.68(r-i) - 0.050] )  
\label{eq.locus} 
}
which is shown by the solid line in the figure.

The intrinsic width of the stellar locus is $0.02$~mag for blue stars and $0.06$~mag for red
stars \citep{Ivezic04}, which is significantly smaller than the measurement error 
at the faint end. To a very good approximation, any deviation of observed colors from the 
locus can be attributed to photometric errors. We use this assumption to improve 
estimates of true stellar colors and apparent magnitudes at the faint end, 
and thus {\it to increase the sample effective distance limit by nearly a
factor of two.}

As illustrated in Fig.~\ref{locusfit},
for each point and a given error probability ellipse, we find a point on the locus with the highest 
probability\footnote{This is effectively a Bayesian maximum likelihood (ML) procedure with the
assumption of a uniform prior along the one-dimensional locus. As seen from from Fig~\ref{locusfit},
the real prior is not uniform. We have tested the effects of non-uniform priors. Adopting an
observationally determined (from Fig~\ref{locusfit}) non-uniform prior would change the loci of
posterior maxima by only $\sim 0.005$~mag (worst case), while further complicating the ML
procedure. We therefore retain the assumption of uniform prior.}, and adopt the corresponding
$(g-r)_e$ and $(r-i)_e$ colors. The error ellipse
is not aligned with the $g-r$ and $r-i$ axes because the $g-r$ and $r-i$ errors are correlated
($\sigma^2_{g-r,r-i} = \sigma^2_{g,r} + \sigma^2_{g,-i} + \sigma^2_{-r,r} + 
\sigma^2_{-r,-i} = -\sigma_r^2$).

We exclude all points further than 0.3~mag from the locus, as such large deviations 
are inconsistent with measurement errors, and in most cases indicate the source is 
not a main-sequence star. This requirement effectively removes
hot white dwarfs \citep{Kleinman04}, low-redshift quasars ($z<2.2$, \citealt{Richards02}, and
white dwarf/red dwarf unresolved binaries \citep{Smolcic04}.

Using the maximum likelihood colors, we estimate the magnitudes ($g_e$, $r_e$, $i_e$) by 
minimizing:
\eq{
\chi^2 = \frac{(r-r_e)^2}{\sigma_r^2} + \frac{(g-g_e)^2}{\sigma_g^2} + \frac{(i-i_e)^2}{\sigma_i^2},
}
which results in 
\eqarray{
	r_e & = & \frac{w_r r + w_g (g-(g - r)_e) + w_i (i+(r - i)_e)}{w_r + w_g + w_i}	\\
	g_e & = & (g - r)_e + r_e	\\
	i_e & = & (r - i)_e - r_e
}
where $w_j = 1/\sigma_j^2$ for $j = g,r,i$.

The adopted $(r - i)_e$ color and $r_e$ magnitude uniquely determine (through
eqs.~\ref{eq.Mr.faint} and \ref{eq.Mr}) 
the absolute magnitude $M_r$ for each star in the catalog. 
We dub this procedure a ``locus projection'' method, and refer to the derived colors as 
``locus-projected colors''. In all subsequent calculations we use these ``locus-projected'' 
colors, unless explicitly stated otherwise. This method is the most natural
way to make use of all available color information, and performs well in cases where the measurement
of one color is substantially worse than the other (or even nonexistent).
It not only improves the color estimates at the faint end, but also helps with debiasing 
the estimate of density normalization in regions of high gradients in $(g-r, r-i)$ color-color
diagram (e.g., near turnoff). This and other aspects of locus projection are further 
discussed in Appendix A.

\subsubsection{ The Contamination of Stellar Counts by Quasars }

\begin{figure*}
\plotone{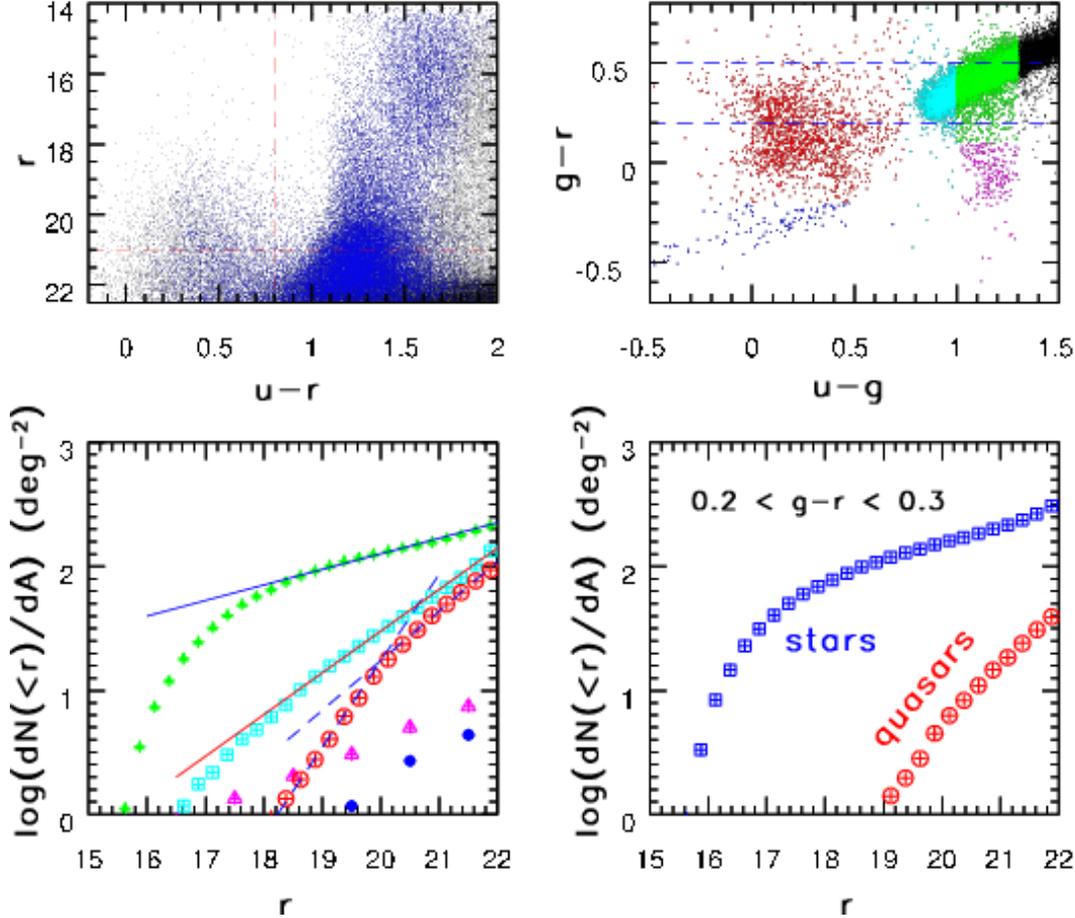}
\caption{The dots in the top left panel shows point sources from the
SDSS Stripe 82 catalog of coadded observations in the $r$ vs. $u-r$ 
color-magnitude diagram. Sources with $0.3 < g-r<0.5$
are marked blue. Sources with $u-r<0.8$ are dominated by low-redshift
quasars, those with $u-r\sim1.3$ by low-metallicity halo stars, and
the bright stars ($r<18$) with $u-r\sim1.6$ are dominated by thick
disk stars. Note the remarkable separation of halo and disk stars
both in magnitude (a distance effect) and color (a metallicity effect) 
directions. The top right panel shows a subset of sources with
$r<21$ in the $g-r$ vs. $u-g$ color-color diagram. Cumulative counts 
of sources from several regions of this diagram (blue: hot stars, dominated
by white dwarfs; red: quasars; magenta: blue horizontal branch stars;
cyan: halo stars; green: thick disk stars) are shown in the lower left 
panel, with the same color coding. The solid lines have slopes of
0.13 (blue) and 0.34 (red) for thick disk and halo stars, while the
quasar counts change slope at $r\sim20$ from $\sim$0.7 to $\sim$0.4, 
as indicated by the dashed lines. The bottom right panel compares 
cumulative counts of two subsets of sources with $0.2 < g-r< 0.3$ that 
are separated by the $u-r = 0.8$ condition. The fraction of $u-r<0.8$
sources is $\sim$10\% for $r<21.5$ and $\sim$34\% for $21.5<r<22$. 
\label{qsoFig}}
\end{figure*}

The stellar samples selected using the $g-r$ and $r-i$ colors, as described
above, are contaminated by low-redshift quasars. While easily recognizable
with the aid of $u-g$ color, a significant fraction of quasars detected by 
SDSS have the $g-r$ and $r-i$ colors similar to those of turn-off stars. 
The SDSS sample of spectroscopically confirmed quasars is flux-limited
at $i=19.1$ (\citealt{Richards02}, and references therein) and thus it is not 
deep enough to assess the contamination level at the faint end relevant
here. Instead, we follow analysis from \citep{Ivezic03e}, who were
interested in the contamination of quasar samples by stars, and obtain an 
approximate contamination level by comparing the counts of faint 
blue stars and photometrically selected quasar candidates. We use a catalog 
of coadded photometry based on about ten repeated SDSS observations recently 
constructed by \citep{Ivezic07c}. The catalog covers a 300 deg$^2$ 
large sky region at high galactic latitudes ($|b|\sim60^\circ$) and thus
the estimated contamination fraction represents an upper limit. With its 
significantly improved $u-g$ color measurements relative to single SDSS scans, 
this catalog allows efficient photometric selection of low-redshift quasar 
candidates to flux levels below $r=21$. 

As summarized in Fig.~\ref{qsoFig}, the largest contamination of stellar
sample by quasars is expected in blue bins. The bluest bin (0.10$<r-i<$ 0.15)
includes stars with 0.2$<g-r<$0.5, and $\sim$5\% of sources in the $r<21.5$
subsample have $u-r<0.8$, consistent with quasars. Even if we restrict the 
sample to 0.2$<g-r<$0.3, and thus maximize the sample contamination by quasars, 
the estimated fraction of quasars does not exceed 10\% for $r<21.5$ (see the 
bottom right panel).

\subsubsection{                   Estimation of Distances                         }
\label{sec.distance.estimates}

Given the photometric parallax relation (eq.\ref{eq.Mr}), the locus-projected maximum likelihood 
$r$ band magnitude, and $r-i$ color, it is straightforward to determine the distance $D$ to each
star in the catalog using
\eq{
	D = 10^{\frac{r - M_r}{5}+1} \,\, {\rm pc}, \label{eq.D}
}
Depending on color and the chosen photometric parallax relation, for the magnitude range probed 
by our sample ($r$=15--21.5) the distance varies from $\sim$100 pc to $\sim$20 kpc.

Due to photometric errors in color, magnitude, and the intrinsic scatter of the photometric
parallax relation, the distance estimate has an uncertainty, $\sigma_D$, given by:
\eqarray{
 \sigma_{M_r}^2 & = & (\frac{\partial M_r}{\partial (r-i)})^2 \sigma_{r-i}^2 + \sigma_{M_{r}}^2 \label{eq.MrErr}\\
 \sigma_D^2 & = & (\frac{\partial D}{\partial M_r})^2 \sigma_{M_r(r-i)}^2 + (\frac{\partial D}{\partial r})^2 \sigma_{r}^2
}
where $\sigma_{M_{r}}$ is the intrinsic scatter in the photometric parallax relation. With an
assumption of $\sigma_{r-i}^2 \approx 2 \sigma_{r}^2$, this reduces to a simpler form:
\eq{
\frac{\sigma_D}{D} = 0.46 \sqrt{(1 + 2\,(\frac{\partial M_r}{\partial (r-i)})^2)
\sigma_{r}^2 + \sigma_{M_{r}}^2 } 
\label{eq.disterr}
}

The fractional distance error, $\sigma_D/D$, is a function of color, apparent magnitude and magnitude 
error (which itself is a function of apparent magnitude). In the top panel in figure \ref{magerr2} we show the 
expected $\sigma_D/D$ as a function of $r$ and $r-i$ with an assumed intrinsic photometric
relation scatter of $\sigma_{M_{r}} = 0.3$~mag. This figure is a handy reference for estimating 
the distance accuracy at any location in the density maps we shall introduce in Section~\ref{mkmaps}. 
For example, a star with $r-i = 0.5$ and $r = 20$ (or, using eq.~\ref{eq.Mr}, at a distance of $D =
3$ kpc) has a $\sim$18\% distance uncertainty. Equivalently, when the stars are binned to
three-dimensional grids to produce density maps (Section~\ref{mkmaps}), this uncertainty gives rise
to a nearly Gaussian kernel smoothing the maps in radial direction, with color and distance dependent 
variance $\sigma^2_D$. Note that this convolution leaves intact structures larger than the kernel scale 
and, in particular, has little effect on the slowly varying Galactic density field and determination 
of Galactic model parameters (Section~\ref{sec.malmquist.effects}).

\vspace{5mm}
To summarize, due to measurement errors, and uncertainty in the absolute calibration of 
the adopted photometric parallax relations,
the derived density maps, described below, will differ from the true stellar distribution.
First, in the radial direction the spatial resolution is degraded due to the smoothing 
described above. A similar effect is produced by misidentification 
of binaries and multiple systems as single stars. Second, the distance scale may have systematic 
errors, probably color and metallicity dependent, that ``stretch or shrink'' the density maps. 
Third, for a small fraction of stars, the distance estimates may be grossly incorrect due to 
contamination by giants and multiple unresolved systems. Finally, stars with metallicities 
significantly different than assumed at a particular $r-i$ int the parallax relation may be 
systematically placed closer or farther away from the origin (the Sun).

However, all of these are either small (e.g., contamination by giants), have a small total effect on 
the underlying Galactic density field (radial smearing due to dispersion in distance estimates), 
or cause relative radial displacements of \emph{entire} clumps of stars with metallicities 
different than that of the background while not affecting their relative parallaxes, 
and thus allowing the discrimination of finer structure. Altogether, the maps fidelity will be 
fairly well preserved, making them a powerful tool for studying the Milky Way's stellar number 
density distribution.

\subsection{                 The Construction of the Density Maps           } 
\label{mkmaps}

The distance, estimated as described above, and the Galactic longitude and latitude, 
$(l,b)$, fully determine the three-dimensional coordinates of each star in the sample.
To better control the systematics, and study the dependence of density field on
spectral type, we divide and map the sample in 19 bins in $r-i$ color\footnote{To avoid excessive usage of
parenthesis, we sometimes drop the minus sign when referring to the colors (e.g. $g-r \equiv gr$ or 
$(r-i)_1 \equiv ri_1$).}:
\eq{
  	        ri_0 < r-i < ri_1
}
Typically, the width of the color bins, $\Delta_{ri} \equiv ri_1 - ri_0$, is $\Delta_{ri} = 0.1$
for bins redder than $r-i = 0.7$ and $\Delta_{ri} = 0.05$ otherwise. The bin limits
$ri_0$ and $ri_1$ for each color bin are given in the second column of table \ref{tbl.bins}. 
This color binning is roughly equivalent to a selection by MK spectral type
(Covey et al. 2005), or stellar mass. 
The range of spectral types corresponding to each $r-i$ bin is given in $SpT$ column of 
table \ref{tbl.bins}.

For each color bin we select a volume limited sample given by:
\eqarray{
	D_0 & = & 10^{\frac{r_{min} - M_r(ri_0)}{5}+1}\,\, {\rm pc}, 	\\
	D_1 & = & 10^{\frac{r_{max} - M_r(ri_1)}{5}+1}\,\, {\rm pc}, \nonumber
}
Here $r_{min}=15$ and $r_{max}=21.5$ are adopted as bright and faint magnitude limits 
(SDSS detectors saturate at $r\sim14$). In each color bin
$D_{1}/D_{0}\sim$15, and for the full sample $D_{1}/D_{0}\sim$300. 

We define the ``Cartesian Galactocentric coordinate system'' by the following 
set of coordinate transformations:
\eqarray{
	X & = & R_\odot - D \cos(l) \cos(b) \\ \label{eq.gc}
	Y & = & - D \sin(l) \cos(b) \\ \nonumber
	Z & = & D \sin(b) \nonumber
}
where $R_\odot = 8$ kpc is the adopted distance to the Galactic center \citep{Reid93a}.

The choice of the coordinate system is motivated by the expectation of cylindrical symmetry around
the axis of Galactic rotation $\hat{Z}$, and mirror symmetry of Galactic properties with respect to the
Galactic plane. Its $(X,Y)$ origin is at the Galactic center, the $\hat{X}$ axis points towards the
Earth,
and the $\hat{Z}$ axis points towards the north Galactic pole. The $\hat{Y} = \hat{Z} \times
\hat{X}$ axis is defined so as to keep the system right handed. The $\hat{X}-\hat{Y}$ plane is
parallel to the plane of the Galaxy, and the $Z=0$ plane contains the Sun. The Galaxy rotates
clockwise around the $\hat{Z}$ axis (the rotational velocity of the Sun is
in the direction of the $-\hat{Y}$ axis).

We bin the stars onto a three dimensional rectangular grid in these
coordinates. The choice of grid pixel size is driven by compromise between
two competing requirements: keeping the Poisson noise in each pixel at a
reasonable level, while simultaneously avoiding over-binning (and related
information loss) in high-density regions of the maps. By manual
trial-and-error of a few different pixel sizes, we come to a size (for each
color bin) which satisfies both requirements. The adopted pixel sizes are
listed in table \ref{tbl.bins}. For bins with $r-i > 0.3$ the median number
of stars per pixel is $\sim 10$, growing to $\sim 30$ for the bluest $r-i$
bin.

For each volume limited $(ri_0, ri_1)$ color bin sample, this binning
procedure results in a 
three-dimensional data cube, a \emph{map}, of observed stars with each $(X, Y, Z)$ pixel value 
equal to the number of stars observed in $(X-dx/2, X+dx/2)$, $(Y-dx/2, Y+dx/2)$,
$(Z-dx/2,Z+dx/2)$ interval.

Not all of the pixels in the maps have had their volume fully sampled by the SDSS survey. This
is especially true near the edges of the survey volume, and at places where there are holes in the
footprint of the survey (cf. figure \ref{fig.skymap}). In order to convert the number of stars
observed in a particular pixel $(X, Y, Z)$ to density, we must know the fraction of pixel
volume that was actually sampled by the survey. Although simple in principle, the problem of
accurately binning the surveyed volume becomes nontrivial due to overlap of observing runs,
complicated geometry of the survey, and the large survey area. We solve it by shooting a dense,
horizontal, rectangular grid of vertical $(X_r=const, Y_r=const)$ rays through the observed volume,
with horizontal spacing of rays $dx_r$ being much smaller than the pixel size $dx$ (typically,
$dx_r/dx = 0.1$). For each ray, we calculate the intervals in $Z$ coordinate in which it intersects
each imaging run ({\it "ray-run intersections"}). Since imaging runs are bounded by simple geometric
shapes (cones, spheres and planes), the ray-run intersection calculation can be done almost
entirely analytically, with the only numerical part being the computation of roots of a
$4^\mathrm{th}$~order polynomial. For each ray, the union of all ray-run intersections is the set of
$Z$ intervals ($[Z_0, Z_1), [Z_2, Z_3), [Z_4, Z_5), ...$) at a given column $(X_r, Y_r)$ which
were sampled by the survey. It is then a simple matter to bin such interval sets in $\hat{Z}$
direction, and assign their parts to pixels through which they passed. Then, by approximating that
the ray sweeps a small but finite area $dx_r^2$, the
survey volume swept by the ray contributing to pixel $(X, Y, Z)$ is simply $dx_r^2$ times the
length of the ray interval(s) within the pixel. By densely covering all of the $(X, Y)$ plane with
rays, we eventually sweep the complete volume of the survey and partition between all of the
$(X,Y,Z)$ pixels. This ray-tracing method is very general and can handle any survey geometry
in any orientation, as long as the survey geometry can be represented by a set of {\it runs} along
great circles. Using this approach, we compute the volume observed within each pixel with an accuracy of one
part in $10^3$.

In summary, for each of the 19 $r-i$ color bins, we finish with a three-dimensional map in which 
each $(X, Y, Z)$ pixel holds the number of observed stars ($N$) and the observed volume ($V$). 
We estimate the number density in the pixel by simply dividing the two:
\eq{
	\rho(X,Y,Z) = \frac{N(X,Y,Z)}{V(X,Y,Z)}.
}
with the error in density estimate due to shot noise being
\eq{
	  \sigma_{\rho}(X,Y,Z) = \frac{\sqrt{N(X,Y,Z)}}{V(X,Y,Z)}
}
For each pixel we also track additional auxiliary information (e.g. a list of all contributing
SDSS runs), mainly for quality assurance and detailed a posteriori analysis.

\section {                 Stellar Number Density Maps      }
\label{analysis}

\begin{figure*}
\scl{.70}
\plotone{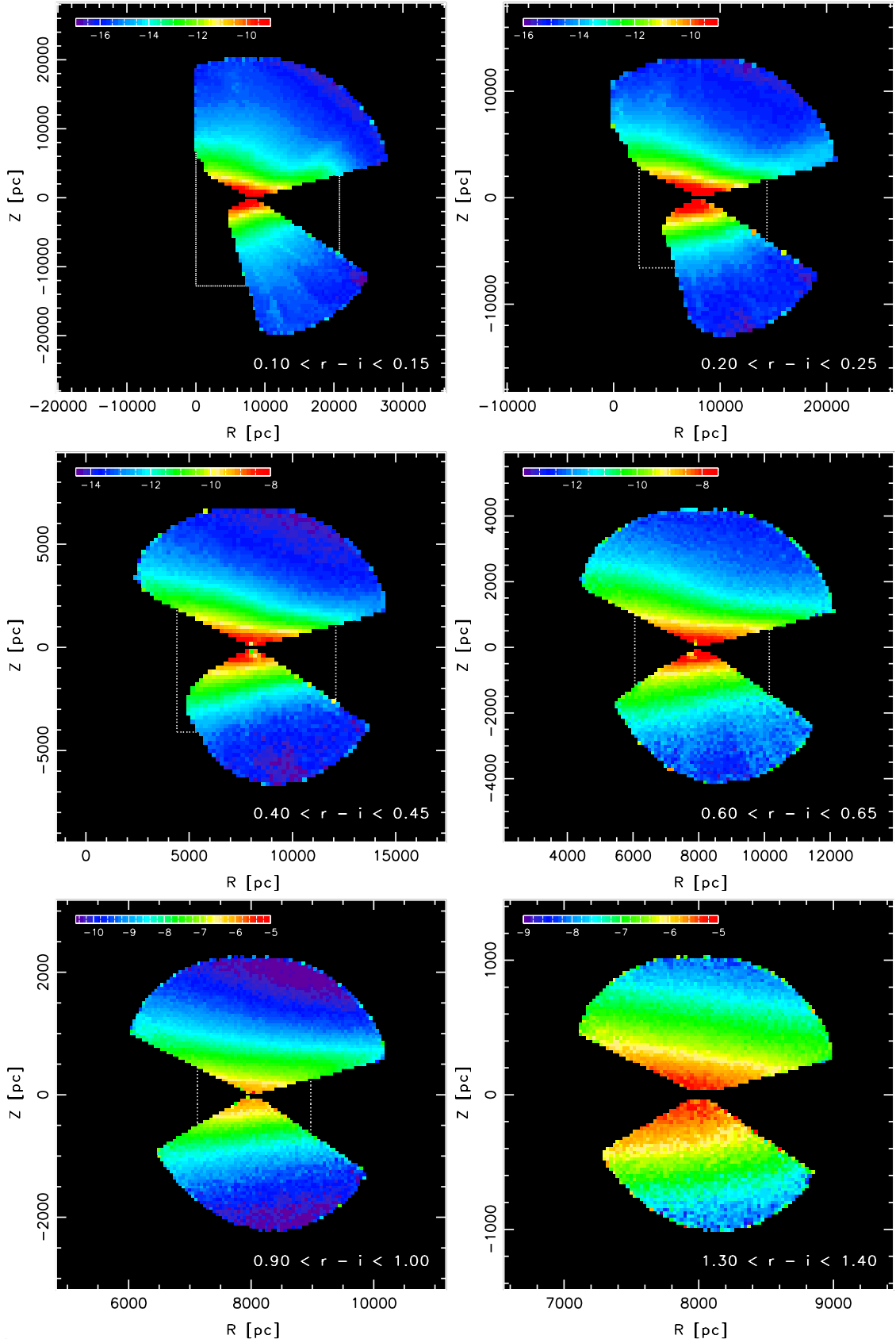}
\caption{The stellar number density as a function of Galactic cylindrical coordinates $R$ (distance 
from the axis of symmetry) and $Z$ (distance from the plane of the Sun), for different $r-i$ color bins,
as marked in each panel. Each pixel value is the mean for all polar angles $\phi$. The density is shown on
a natural log scale, and coded from blue to red (black pixels are regions without the data). Note 
that the distance scale greatly varies from the top left to the bottom right panel -- the size of 
the the bottom right panel is roughly equal to the size of four pixels in the top left panel. Each
white dotted rectangle denotes the bounding box of region containing the data on the subsequent
panel.
\label{RZmedians}}
\end{figure*}

In this Section we analyze the 19 stellar number density maps constructed as described above.
The $0.10 < r-i < 1.40$ color range spanned by our sample probes a large distance range --
as the bin color is varied from the reddest to the bluest, the maps cover distances 
from as close as 100 pc traced by M dwarfs ($r-i \sim 1.3$), to 20 kpc traced by stars 
near the main sequence turnoff ($r-i \sim 0.1$). We begin the analysis with a qualitative 
survey of the various map cross-sections, and then proceed to a quantitative description 
within the context of analytic models.

\subsection{   The Number Density Maps in the $R-Z$ Plane  }
\label{sec.rzmaps}

\begin{figure}
\plotone{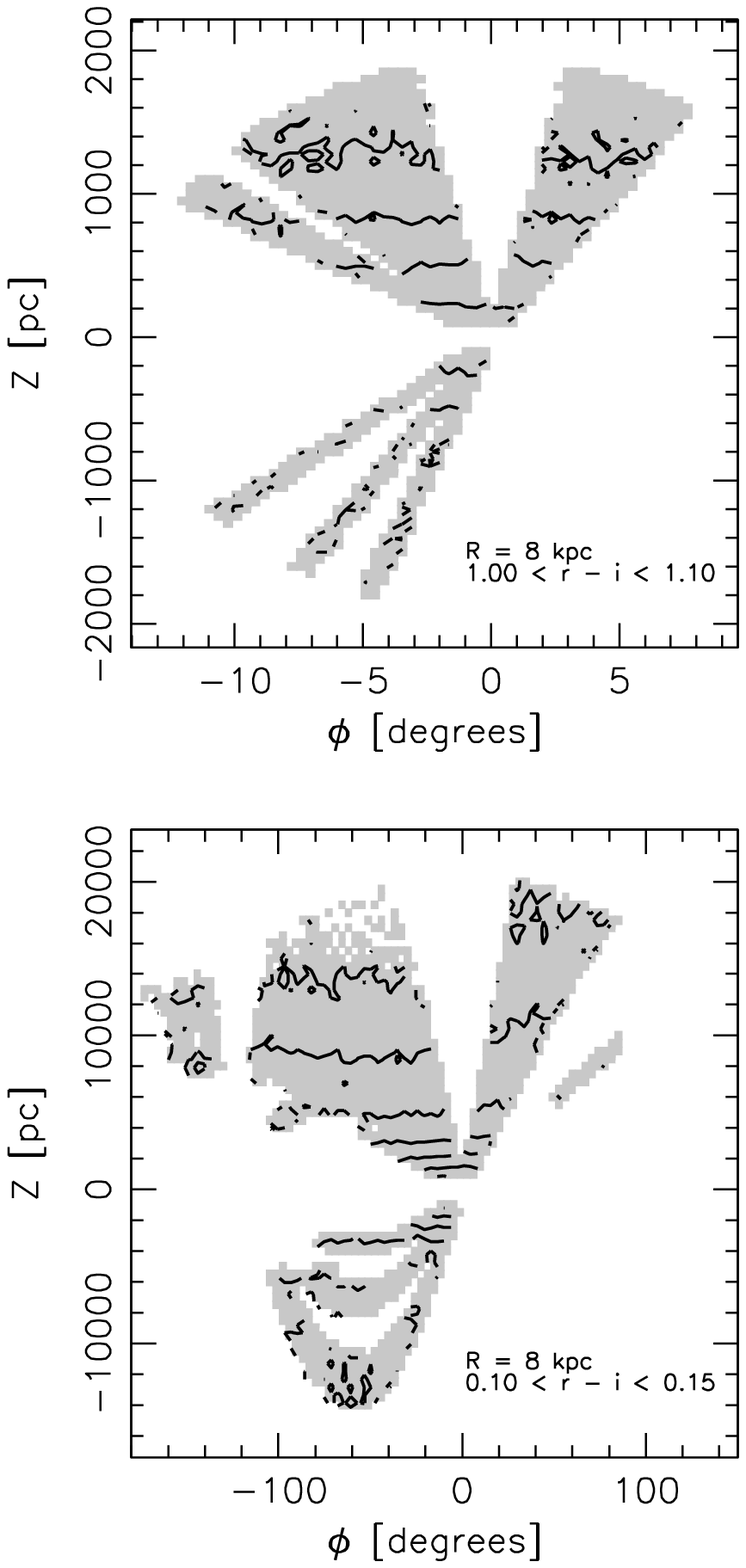}
\caption{The azimuthal dependence of the number density for $R=R_\odot$ cylinder around the 
Galactic center. The shaded region is the area covered by the SDSS survey, and the lines show
constant density contours for two color bins ($1.0 < r - i < 1.1$ in the top panel and 
$0.10 < r - i < 0.15$ in the bottom panel). 
The fact that isodensity contours are approximately horizontal supports 
the assumption that the stellar number density distribution is cylindrically symmetric 
around the Galactic center, and at the same time indicates that the assumed photometric 
parallax distribution is not grossly incorrect. Nevertheless, note that deviations from
cylindrical symmetry do exist, e.g. at $Z\sim10$~kpc and $\phi \sim 40^\circ$ in the bottom panel.
\label{figcyl}}
\end{figure}

We first analyze the behavior of two-dimensional maps in the $R-Z$ plane, where
$R=\sqrt{X^2+Y^2}$ and $Z$ are the galactocentric cylindrical coordinates. Assuming the 
Galaxy is circularly symmetric (we critically examine this assumption below), 
we construct these maps from the three-dimensional maps by taking a weighted mean of all 
the values for a given $Z-R$ pixel (i.e. we average over the galactocentric polar angle
$\phi=\arctan{\frac{Y}{X}}$). 

We show a subset of these maps in Fig.~\ref{RZmedians}. They bracket the
analyzed $r-i$ range; the remaining maps represent smooth interpolations
of the displayed behavior.

The bottom two panels in Fig.~\ref{RZmedians} correspond to the reddest bins, 
and thus to the Solar neighborhood within $\sim$2 kpc. They show a striking simplicity
in good agreement with a double exponential disk model:
\eq{
\label{oneD}
 \rho(R,Z) = \rho(R_\odot,0)\,e^\frac{R_\odot}{L}\,\exp\left(-\frac{R}{L}-\frac{Z+Z_\odot}{H}\right)
}.
Here $\rho$ is the number density of disk stars, $R_\odot$ and 
$Z_\odot$ are the cylindrical coordinates of the Sun, and $L$ and $H$ are the exponential scale 
length and scale height, respectively. 
This model predicts that the isodensity contours have the linear form 
\eq{
                 |Z+Z_\odot| = C - {H \over L} \, R,       
}
where $C$ is an arbitrary constant, a behavior that is in good agreement with the data.

As the bin color becomes bluer (the middle and top panels), and probed distances larger, 
the agreement with this simple model worsens. First, the isodensity contours become 
curved and it appears that the disk flares for $R>14$ kpc. Further, as we discuss
below, the $Z$ dependence deviates significantly from the single exponential given by 
eq.~\ref{oneD}, and additional components or a different functional form, are required 
to explain the observed behavior.

We test whether the number density maps are circularly symmetric by examining isodensity 
contours on a cylindrical surface at $R = R_\odot$ kpc. Fig.~\ref{figcyl} shows such
projections for two color bins, where we plot the dependence of isodensity
contours on galactocentric polar angle $\phi$, and distance from the plane $Z$. In case of
cylindrical symmetry, the contours would be horizontal. The top panel shows the isodensity contours
for the $1.0 < r-i < 1.1$ color bin and is representative of all bins redder
than $r-i \geq 0.35$~mag. The contours are horizontal, and the number density maps are indeed
approximately cylindrically symmetric.
However, for bins $r-i < 0.35$~mag, detectable deviations from cylindrical symmetry do exist,
especially at large distances from the Galactic plane (a few kpc and beyond). We show an example of this in
the bottom panel, where there is a slight upturn of the isodensity contour at $Z\sim$10,000 and
$\phi \sim 40^\circ$, indicating the presence of an overdensity. We will discuss such overdensities in
more detail in the following section.

\subsection{  The $X-Y$ Slices of the 3-dimensional Number Density Maps  }
\label{XYsection}

\begin{figure*}
\plotone{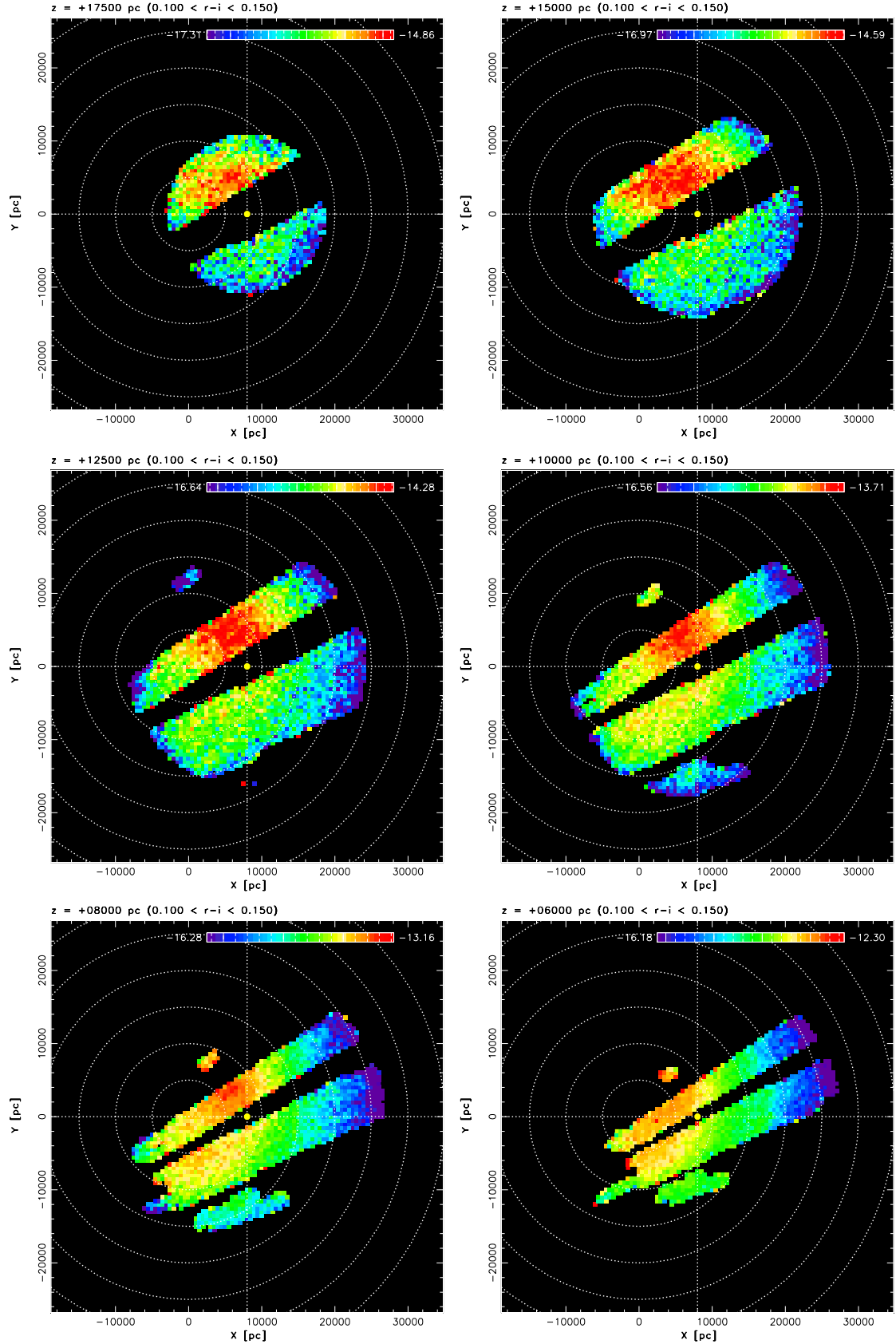}
\caption{The stellar number density for the same color bin as in the top left panel
in Fig.~\ref{RZmedians} ($0.10 < r-i < 0.15$), shown here in slices parallel to the 
Galactic plane, as a function of the distance from the plane. The distance from the plane 
varies from 17.5 kpc (top left) to 6 kpc (bottom right), in steps of 2 and 2.5 kpc. The 
circles visualize presumed axial symmetry of the Galaxy, and the origin marks the 
location of the Galactic center (the Sun is at $X=8, Y=0$~kpc). Note the strong asymmetry 
with respect to the $Y=0$ line.
\label{XYslices1}}
\end{figure*}

\begin{figure*}
\plotone{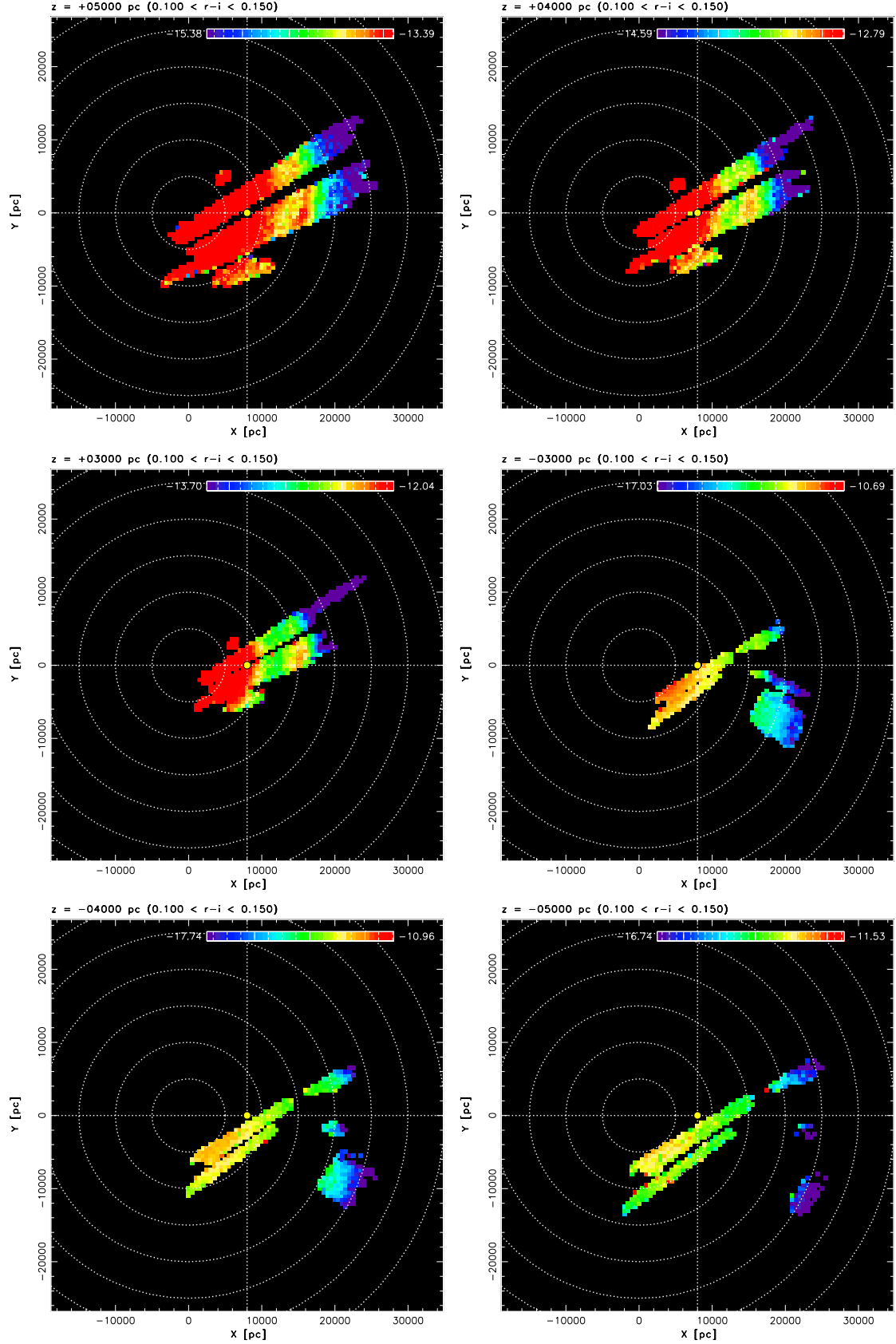}
\caption{Analogous to Fig.~\ref{XYslices1}, except that three symmetric slices at $Z$=3, 4 and 
5 kpc above and below the plane are shown. The color stretch in panels for $Z$=3, 4 and 5 kpc 
is optimized to bring out the Monoceros overdensity at $R\sim16$ kpc and $Y\sim0$.
\label{XYslices2a}}
\end{figure*}

\begin{figure*}
\plotone{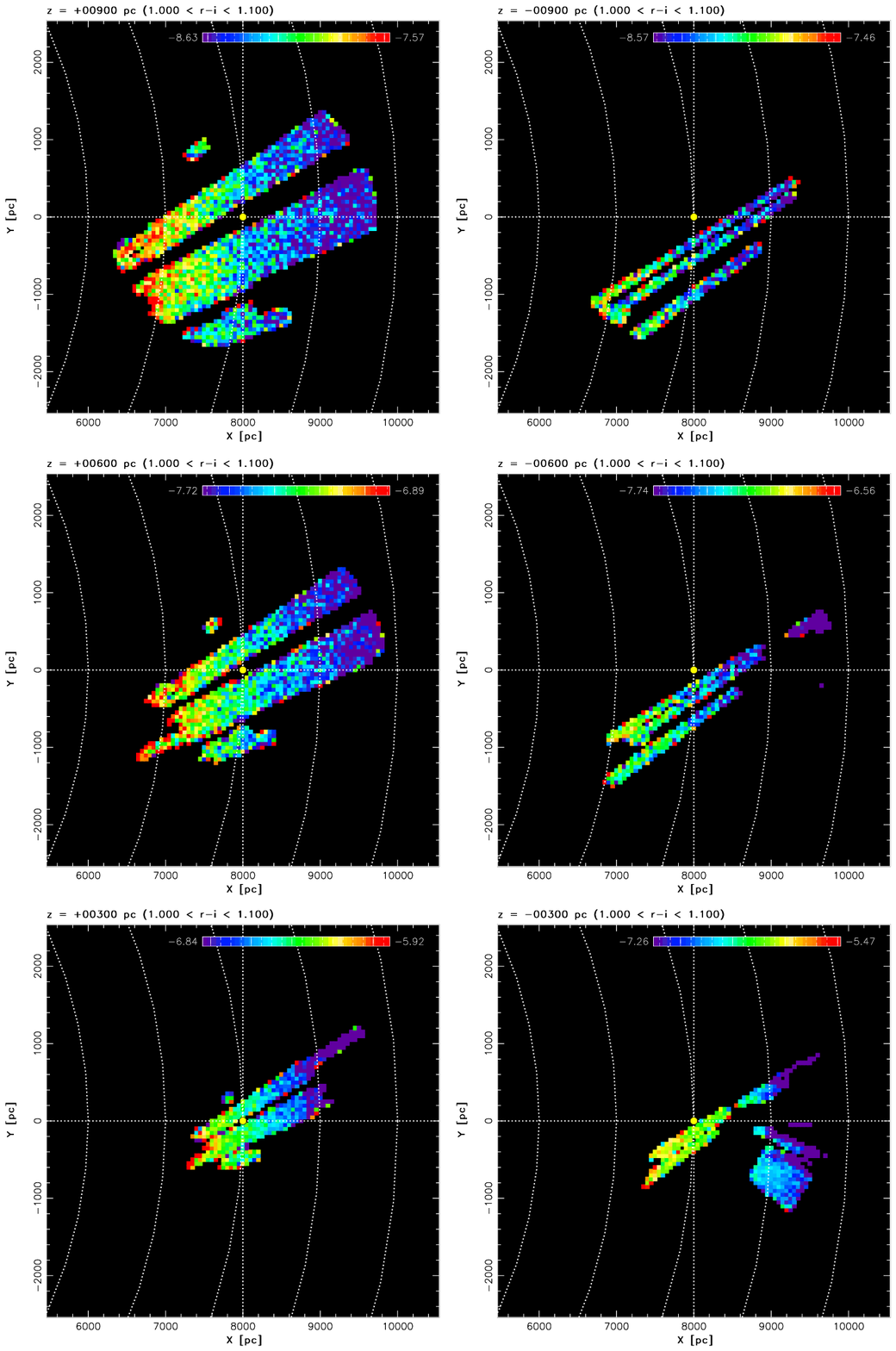}
\caption{Analogous to Fig.~\ref{XYslices1}, except that here three symmetric slices at $Z$=300, 600
and 900 pc above and below the plane are shown, for the $1.00 < r - i < 1.10$ color bin. Note that
at these distance scales there is no obvious discernible substructure in the density distribution.
\label{XYslices2b}}
\end{figure*}

Instead of contracting the three-dimensional maps by taking the mean of all $\phi$ values 
for a given $Z-R$ pixel, two-dimensional analysis can be based on simple cross-sections
parallel to an appropriately chosen plane. A convenient choice is to study the $X-Y$ 
cross-sections that are parallel to the Galactic plane. A series of such projections
for the bluest color bin is shown in Figs.~\ref{XYslices1}--\ref{XYslices2b}. Their
outlines are determined by the data availability. In particular, the gap between 
the two largest data regions will be eventually filled in as more SDSS imaging data becomes
available\footnote{This region of the sky has already been imaged, and will be a part of SDSS Data
Release 6 projected to be released in July 2007.}.

An unexpected large overdensity feature is easily discernible in five of the six panels in 
Fig.~\ref{XYslices1}. In all standard Galaxy models, the stellar density in the upper 
half ($Y > 0$) should mirror the bottom half ($Y < 0$), and in most models density depends 
only on the distance from the center of the Galaxy (each annulus enclosed by two successive 
circles should have roughly the same color). In contrast, the observed density map, 
with a strong local maximum offset from the center, is markedly different from these 
model predictions. This is the same feature that is responsible for the structure
visible at $Z\sim$10 kpc and $R\sim$5 kpc in the top left panel in Fig.~\ref{RZmedians},
and for the upturn of the isodensity contour at $Z\sim$10,000 and $\phi \sim 40^\circ$
in the bottom panel in Fig.~\ref{figcyl}. We discuss this remarkable feature in 
more detail in Section~\ref{vlgv}. 

The top three panels ($Z$=3-5 kpc) in Fig.~\ref{XYslices2a} clearly show another 
local overdensity at $R\sim16$ kpc and $Y\sim0$. This is the ``Monoceros Stream''
discovered by \citet{Newberg02} using a subset of the data analyzed here 
(this overdensity is also discernible in the top left panel in Fig.~\ref{RZmedians} at 
$R\sim 16$ kpc and $Z \sim 3$ kpc). The maps discussed here suggest that the 
stream is well localized in the radial direction with a width of $\sim 3$ kpc. 
This well-defined width rules out the hypothesis that this overdensity is due to disk 
flaring.

An alternative hypothesis, that of a ``ring'' around the Galaxy, was proposed by \citet{Ibata03},
but put question by observations of \citet{Rocha-Pinto03}. In particular, \citeauthor{Rocha-Pinto03} 
analyzed the distribution of 2MASS M giants in the Monoceros feature and concluded its 
morphology was inconsistent with a homogeneously dense ring surrounding the Milky Way.
Instead, a more likely explanation is a merging dwarf galaxy with tidal arms. The inhomogeneity 
of the stream apparent in top three panels of Fig.~\ref{XYslices2a}, as well 
as $R=const$. projections of these maps and a theoretical study by \citet{Penarrubia05}, 
support this conclusions as well.

Closer to the plane, at distances of less than about 1 kpc, the number density maps
become smoother and less asymmetric, with deviations from a simple exponential 
model given by eq.~\ref{oneD} not exceeding 30-40\% (measured upper limit). This is true of all
color bins for which region closer than $\sim 2$~kpc is well sampled, and is shown in
Fig.~\ref{XYslices2b} for $1.0 < r-i < 1.1$ color bin.

\subsection{   Overall Distribution of Stellar Number Density    }

Traditionally, the stellar distribution of the Milky Way has been decomposed
into several components: the thin and thick disks, the central bulge, and a much 
more extended and tenuous halo. While it is clear from the preceding discussion
that there are a number of overdensities that complicate this simple model,
the dynamic range of the number density variation in the Galaxy (orders of magnitude)
is large compared to the local density excess due to those features (a factor of few). 
Hence, it should still be possible to capture the overall density variation using analytic 
models.

Before attempting a full complex multi-parameter 
fits to the overall number density distribution, we first perform a simple qualitative exploration 
of density variation in the radial ($R$) and vertical ($Z$) directions. This type of
analysis serves as a starting point to understand of what types of models are at all 
compatible with the data, and to obtain reasonable initial values of model parameters 
for global multi-parameter fits (Section~\ref{sec.modelfit}).

\subsubsection{       The Z-dependence of the Number Density    }
\label{rhoZsec}

\begin{figure}
\plotone{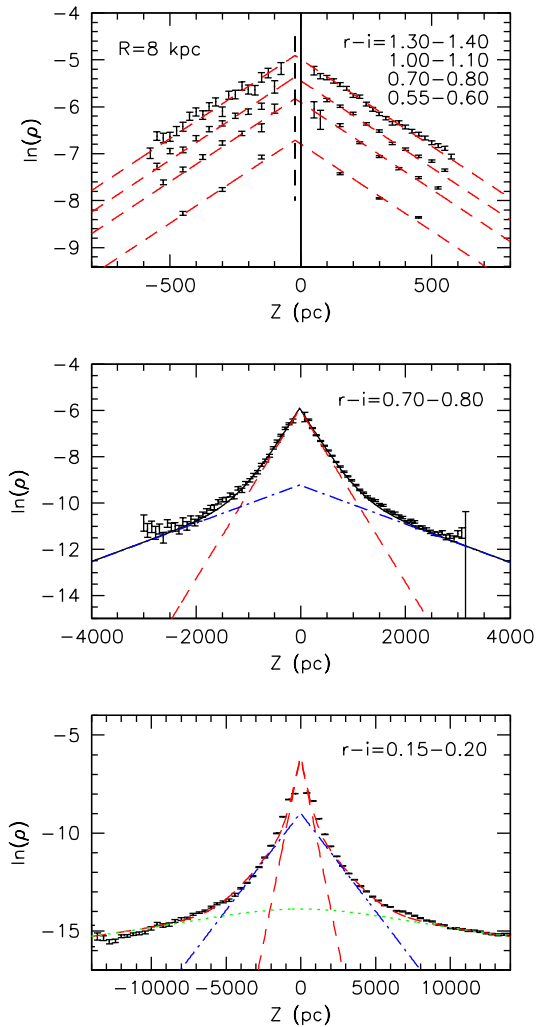}
\caption{The vertical ($Z$) distribution of SDSS stellar
counts for $R=8$ kpc, and different $r-i$ color bins, 
as marked. 
The lines are exponential models fitted to the points. The dashed lines
in the top panel correspond to a fit with a single, exponential
disk having a 270 pc scale height. The vertical dot-dashed line marks the position
of the density maximum, and implies a Solar offset from
the Galactic plane of $\sim 20$ pc. 
The dashed line in the middle panel correspond to 
a sum of two disks with scale heights of 270 pc and 1200 pc,
and a relative normalization of 0.04 (the ``thin'' and the ``thick'' disks).
The dot-dashed line is the contribution of the 1200 pc disk. 
Note that its contribution becomes important for $|Z|>1000$ pc. 
The dashed line in the bottom panel (closely following the data points) 
corresponds to a sum of two disks 
(with scale heights of 260 pc and 1000 pc, and the relative 
normalization of 0.06), and a power-law spherical halo with 
power-law index of 2, and a relative normalization with 
respect to the 260 pc disk of 4.5$\times10^{-4}$. 
The dashed line is the contribution of the 260 pc disk, 
the dot-dashed line is the contribution of the 1000 pc disk, and 
the halo contribution is shown by the dotted line.
Note that both the disk and halo models shown here are just the \emph{initial estimates}
of model parameters, based solely on this $Z$ crossection. As we discuss in
Section~\ref{sec.degeneracies} these are not the only combinations of model 
parameters fitting the data, and the true model parameters fitting \emph{all} of the 
data are in fact substantially different (Table~\ref{tbl.finalparams}).
\label{rhoZ}}
\end{figure}

Fig.~\ref{rhoZ} shows the stellar number density for several color bins
as a function of the distance $Z$ from the plane of the Sun at $R=R_\odot$. 
The behavior for red bins, which probe the heights from 50 pc to $\sim$2 kpc,
is shown in the top panel. They all appear to be well fit by an exponential 
profile\footnote{Motivated by theoretical reasoning (e.g., \citealt{GalacticDynamics}), 
sometimes the sech$^2$ function is used instead
of exponential dependence. However, the exponential provides a significantly 
better description of the data than sech$^2$. For example, the exponential
distribution is a good fit all the way towards the plane to 1/6 or so of 
the scale height, where the sech$^2$ function would exhibit significant curvature
in the $\ln(\rho)$ vs. $Z$ plot.}
with a scale height of $\sim270$~pc\footnote{Note that this is just an \emph{initial estimate}
for the scale height, based on a single effective line of sight (SGP -- NGP) and
limited $Z$ coverage. In Section~\ref{sec.modelfit}
we will derive the values of Galactic model parameters using the entire dataset.}. While 
the best-fit value of this scale height is uncertain up to 10--20\%, it is encouraging 
that the same value applies to all the bins. This indicates that the slope of the adopted 
photometric parallax relation is not greatly incorrect at the red end.

The extrapolations of the best exponential fits for $Z<0$ and $Z>0$ to small
values of $|Z|$ cross at $Z \sim -25$~pc. This is the well-known Solar
offset from the Galactic plane towards the north Galactic pole (e.g. \citealt{Reid93a}),
which is here determined essentially directly using a few
orders of magnitude greater number of stars (several hundred thousand) than in previous
work.

By selecting bluer bins, the $Z$ dependence of the number density can be 
studied beyond 1 kpc, as illustrated in the middle panel. At these distances 
the number density clearly deviates from a single exponential disk model. 
The excess of stars at distances beyond 1 kpc, compared to this model, is 
usually interpreted as evidence of another disk component, the thick disk. 
Indeed, the data shown in the middle panel in Fig.~\ref{rhoZ} can be modelled
using a double-exponential profile.

The need for yet another, presumably halo, component, is discernible
in the bottom panel in Fig.~\ref{rhoZ}, which shows the number density for 
the bluest color bin. The data show that beyond 3-4 kpc even the thick disk 
component underpredicts the observed counts. The observations can be explained
by adding a power-law halo component, such as described by eq.~\ref{haloModel}.

\subsubsection{  The R-dependence of the Number  Density     }
\label{Rdep}

\begin{figure}
\scl{0.8}
\plotone{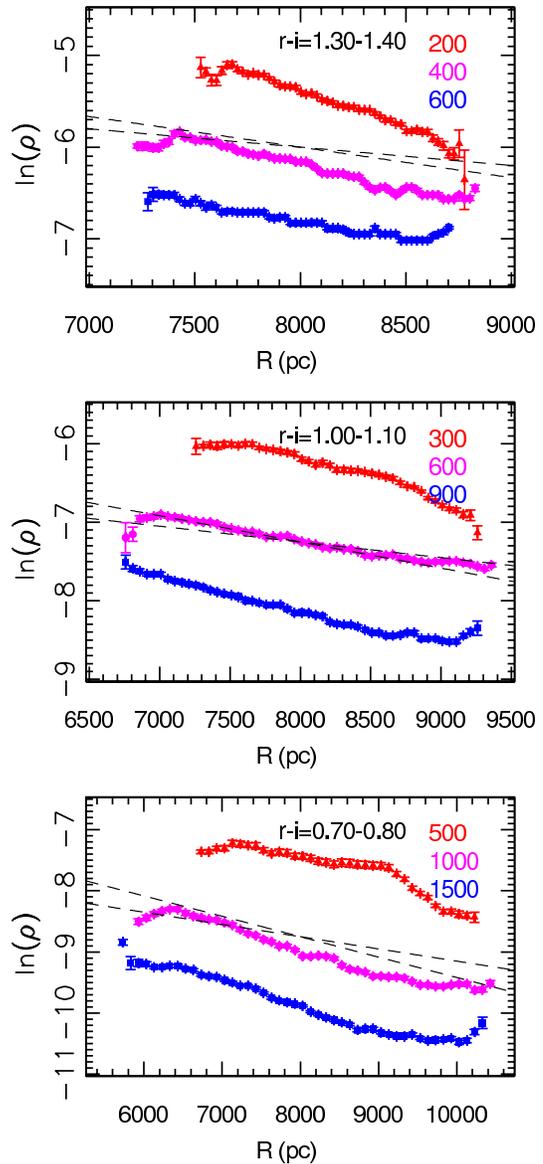}
\caption{The radial distribution of SDSS stellar
counts for different $r-i$ color bins, and at different 
heights above the plane, as marked in each panel (pc).
The two dashed lines show the exponential radial dependence
of density for scale lengths of 3000 and 5000 pc (with 
arbitrary normalization).
\label{rhoR1}}
\end{figure}

\begin{figure}
\plotone{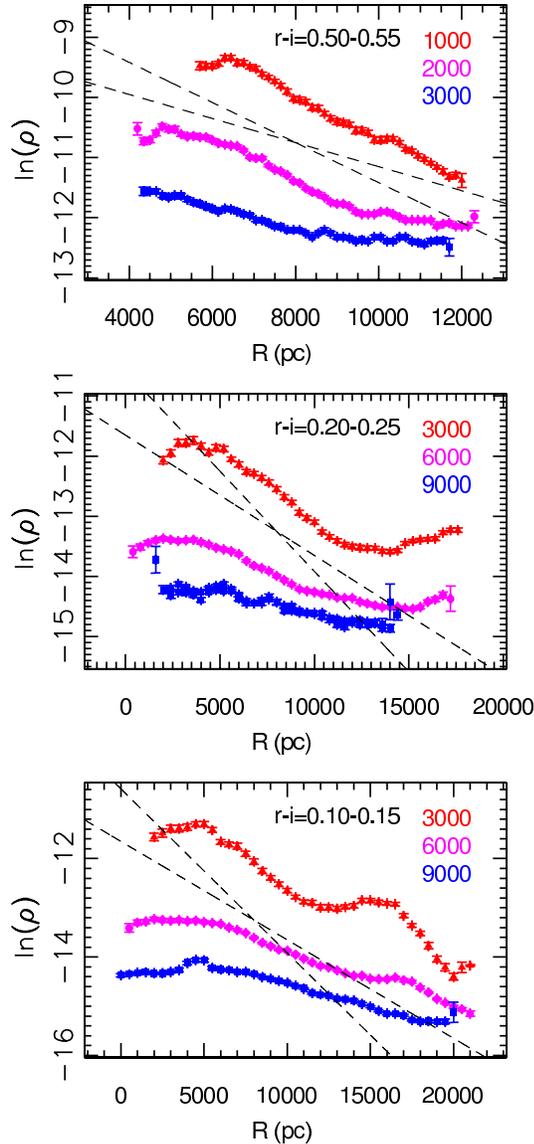}
\caption{Analogous to Fig.~\ref{rhoR1}, except for bluer
color bins, which probe larger distances. 
\label{rhoR2}\vskip 1em}
\end{figure}

\begin{figure}
\plotone{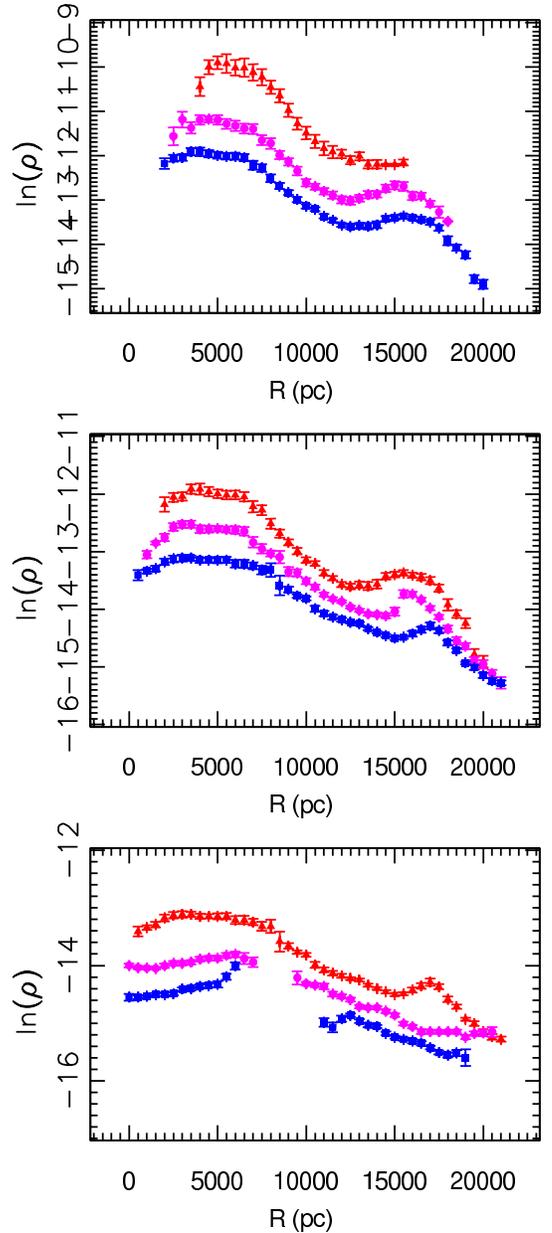}
\caption{The radial distribution of SDSS stellar counts for 
$0.10 < r-i < 0.15$ color bin, with the data restricted to $|y|<1$ kpc.
The selected heights are, from top to bottom, 
(2,3,4), (4,5,6) and (6,8,10) kpc. The Monoceros stream at
is easily visible as local maxima at $R=16-17$ kpc, and the Virgo overdensity 
as the wide bump at $R \sim6$ kpc.
\label{rhoRMon}}
\end{figure}

We examine the dependence of number density on the (cylindrical) distance
from the Galactic center in Figs.~\ref{rhoR1}, \ref{rhoR2} and \ref{rhoRMon}.
Each figure shows the number density as a function of $R$ for a given 
$r-i$ color bin at different heights above the Galactic plane. For 
red bins, which probe the Solar neighborhood within $\sim$2 kpc, the density
profiles are approximately exponential (i.e. straight lines in 
ln($\rho$) vs. $R$ plot, see Fig.~\ref{rhoR1}). 
The exponential scale length seems to increase with the distance from the Galactic plane,
or alternatively, requires the introduction of an additional exponential dependence with a
different scale. Due to the small baseline this variation
or the scale lengths are not strongly constrained with plausible values around 
$L \sim 3.5$~kpc and an uncertainty of at least 30\%. 

At distances from the Galactic plane exceeding 1-2 kpc, the exponential
radial dependence becomes a fairly poor fit to the observed density distribution
(Fig.~\ref{rhoR2}). The main source of discrepancy are
several overdensities noted in Section~\ref{XYsection}. In particular, the Monoceros
stream is prominent at $Z\sim$2-8 kpc, especially when the
density profiles are extracted only for $|Y|<1$ kpc slice (Fig.~\ref{rhoRMon}).

\section{       Galactic Model      }
\label{sec.galactic.model}

The qualitative exploration of the number density maps in the preceding
section, as well as the analysis of the density variation in the 
radial $R$ and vertical $Z$ directions, suggest that the gross behavior can be 
captured by analytic models. These typically model the number density
distribution with two exponential disks, and a power-law (or de Vaucouleurs 
spheroid) elliptical halo.

Following earlier work (e.g. \citealt{Majewski93}, \citealt{Siegel02}, \citealt{Chen01}), we 
decompose the overall number density into the sum of disk and halo contributions
\eq{
\label{galModel}
                \rho(R,Z) =  \rho_D(R,Z) + \rho_H(R,Z).
}
We ignore the bulge contribution because the maps analyzed here only
cover regions more than 3-4 kpc from the Galactic center, where the 
bulge contribution is negligible compared to the disk and halo contributions
(for plausible bulge parameters determined using IRAS data for asymptotic 
giant stars, see e.g. \citealt{Jackson02}).

Following \citet{BahcallSoneira} and \citet{Gilmore83}, we further decompose the 
disk into a sum of two exponential components (the ``thin'' and the ``thick'' disk), allowing 
for different scale lengths and heights of each component:
\eq {
\label{twoD}
 \rho_D(R,Z) = \rho_{D}(R,Z;L_1,H_1) + f\rho_{D}(R,Z;L_2,H_2)
}
where
\eq{
\label{diskZ}
  \rho_{D}(R,Z;L,H) = \rho_{D}(R_\odot,0)\,e^\frac{R_\odot}{L}\,\exp\left(-\frac{R}{L}-\frac{Z+Z_\odot}{H}\right)
}
Here $H_1$, $H_2$ and $L_1$ and $L_2$ are the scale heights and lengths for the thin and thick disk,
respectively, $f$ is the thick disk normalization relative to the thin disk at 
($R=R_\odot, Z=0$), and $Z_\odot$ is the Solar offset from the Galactic plane.
From previous work, typical best-fit values are $H_1\sim$300 pc, $H_2\sim$1-2 kpc, 
$f\sim$1-10\% and $Z_\odot\sim$10-50 pc (e.g. \citealt{Siegel02}, table 1).

We also briefly explored models where thin and thick disk had the same scale length 
that was allowed to vary linearly with distance distance from the Galactic plane 
($L = L_0 + k Z$), but found these to be unnecessary as the two-disk formalism
was able to adequately capture the behavior of the data.

We model the halo as a two-axial power-law ellipsoid\footnote{For the halo
component, $Z+Z_\odot \approx Z$ is a very good approximation.}
\eq{
\label{haloModel}
  \rho_H(R,Z) = \rho_D(R_\odot,0)\, f_H \, \left({R_\odot \over \sqrt{R^2 +
(Z/q_H)^2}}\right)^{n_H}. 
}
The parameter $q_H$ controls the halo ellipticity, with the ellipsoid described
by axes $a=b$ and $c=q_H\,a$. For $q_H<1$ the halo is oblate, that is, ``squashed'' 
in the same sense as the disk. The halo normalization relative to the thin disk
at ($R=R_\odot, Z=0$) is specified by $f_H$. From previous work, typical best-fit 
values are $n_H \sim$2.5-3.0, $f_H \sim10^{-3}$ and $q_H\sim0.5-1$. 

\subsection{       Dataset Preparation    }
\label{sec.dataset.preparation}

\begin{figure}
\scl{.6}
\plotone{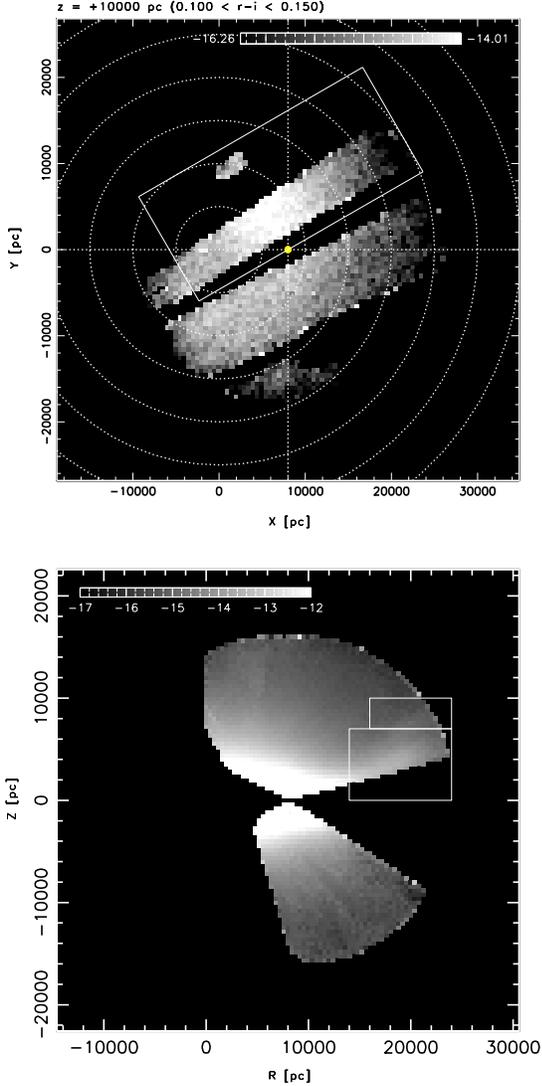}
\caption{The regions with large overdensities excluded from Galactic model fits. 
The pixels within the rectangle in the top panel are excluded to avoid contamination 
by the Virgo overdensity (Section~\ref{vlgv}). The pixels enclosed by the two 
rectangles in the bottom panel, centered at $R \sim$ 18 kpc, exclude the Monoceros stream.
\label{figexcl}}
\end{figure}

\begin{figure*}
\scl{.85}
\plotone{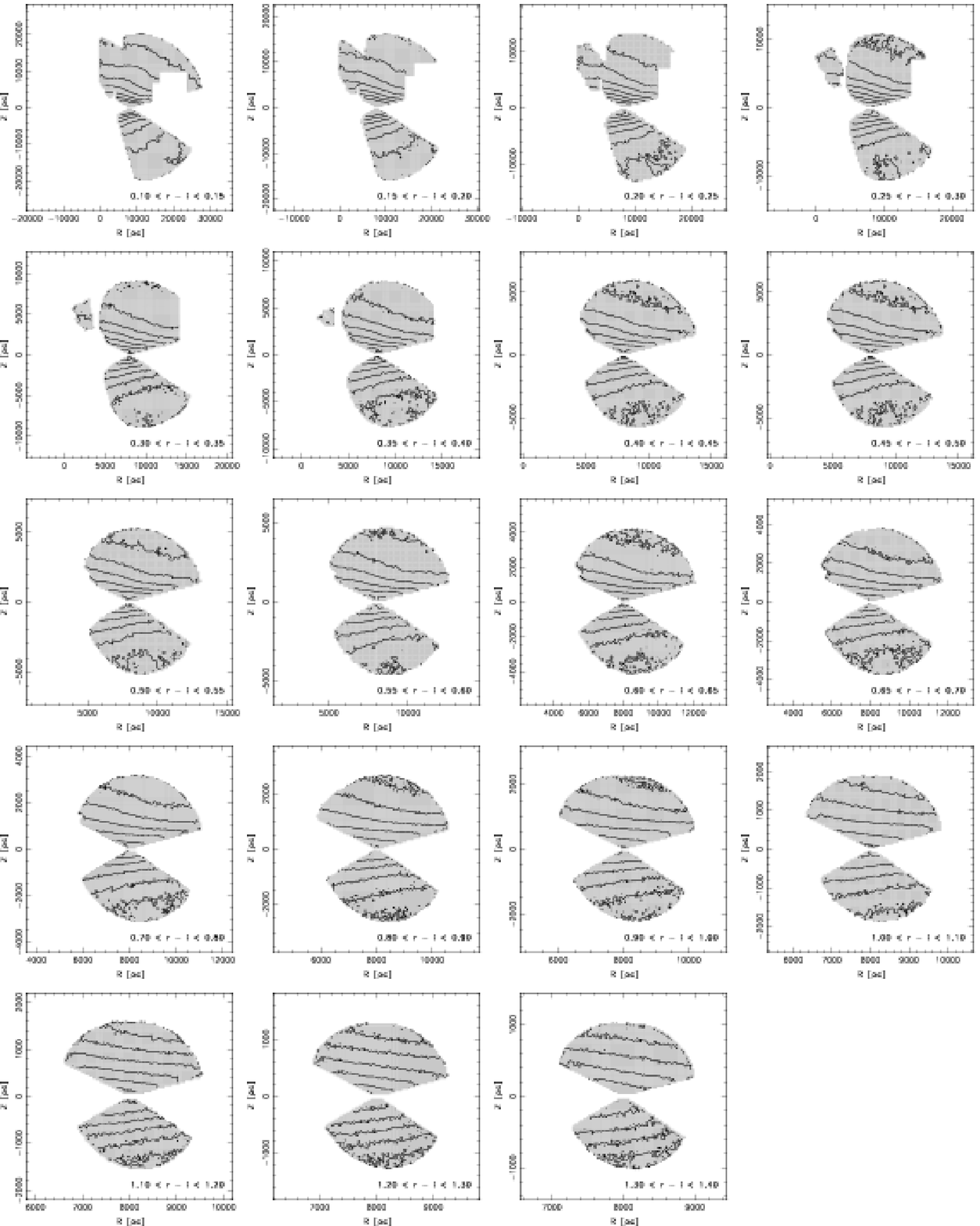}
\caption{``Cleaned up'' $(R, Z)$ maps of the Galaxy, analogous to figure \ref{RZmedians}, but
with pixels in obvious overdensities (fig.~\ref{figexcl}) excluded from azimuthal averaging. We
show the maps for all 19 color bins, with the bluest bin in the top left corner and the reddest bin
in the bottom right. The contours are the lines of constant density, spaced at constant logarithmic
intervals.
\label{RZcleaned}}
\end{figure*}

The fitting of models described by eqs.~\ref{galModel}--\ref{haloModel} will be 
affected by overdensities identified in Section~\ref{XYsection} and other, smaller
overdensities that may be harder to see at first. If unaccounted for, such 
overdensities will
almost certainly bias the best-fit model parameters. In general, as we discuss
later in Section~\ref{sec.clumpyness}, their effect is to artificially increase the scale 
heights of the disks, in order to compensate for the localized density excesses
away from the plane. We therefore exclude from the dataset the regions where there are 
obvious localized deviations from smooth background profile\footnote{Note that we are
excluding overdensities, but not underdensities, as there are physical reasons to expect
the Galaxy to have a smooth distribution with overdense regions 
(e.g., due to mergers, clusters, etc.).}. The excluded regions are shown in Figure~\ref{figexcl}.

We exclude the newly found large overdensity discernible in
Fig.~\ref{XYslices1} (the ``Virgo overdensity'') by masking the pixels that simultaneously satisfy:
\eqarray {
\label{exVirgo}
	-5 < X' / \mathrm{\,kpc} & < & 25 \nonumber \\
	Y' & > & -4 \mathrm{\,kpc} \nonumber \\
   	(X - 8 \mathrm{\,kpc})^2 + Y^2 + Z^2 & > & (2.5 \mathrm{\,kpc})^2 \nonumber 
}
where
\[ \left( \begin{array}{c}
	X' \\
	Y'
  \end{array} \right) 
  	=
  \left( \begin{array}{cc}
  \cos 30^\circ & - \sin 30^\circ \\
  \sin 30^\circ & \cos 30^\circ
  \end{array} \right)
  \left( \begin{array}{c}
	X \\
	Y
  \end{array} \right) 
\]
The third condition excludes from the cut pixels closer than $2.5$ kpc 
to the Sun, which are uncontaminated by the overdensity. The excluded region is 
is shown on Figure~\ref{figexcl} bounded by the rectangle in the top panel.

The Monoceros stream is  located at an approximately constant galactocentric
radius. We exclude it by masking out all pixels that satisfy either of 
the two conditions:
\eqarray {
	14 \mathrm{\,kpc} < R < 24 \mathrm{\,kpc} & \wedge & 0 < Z < 7 \mathrm{\,kpc} \nonumber \\ 
	16 \mathrm{\,kpc} < R < 24 \mathrm{\,kpc} & \wedge & 7 < Z < 10 \mathrm{\,kpc} \nonumber
}.
These correspond to the region bounded by two white rectangles in the bottom panel of
Fig.~\ref{figexcl}.

After the removal of Virgo and Monoceros regions, the initial fit for bins redder than $r-i = 1.0$
resulted in measured thin and thick scale heights of $H_1 \sim 280$ and $H_2 \sim 1200$. The
residuals of this fit showed clear signatures of at least two more major overdensities ($\sim 40$\%
above background), one near $(R,Z) \sim (6.5, 1.5)$~kpc and the other near $(R,Z) \sim (9, 1)$~kpc. 
We therefore went back and further excluded the pixels satisfying:
\eqarray {
	-90^\circ < \arctan(\frac{Z - 0.75\mathrm{kpc}}{R - 8.6\mathrm{kpc}}) < 18^\circ & \, \wedge \, & Z > 0 \nonumber \\ 
	R < 7.5\mathrm{kpc} & \, \wedge \, & Z > 0 \nonumber
}

The remaining pixels are averaged over the galactocentric polar angle $\phi$, to produce the
equivalent of $(R,Z)$ maps shown in fig.~\ref{RZmedians}. We additionally imposed a cut on
Galactic latitude, excluding all pixels with $b < 20^\circ$ to remove the stars observed close
to the Galactic disk. This excludes stars that may have been overcorrected for extinction 
(Section~\ref{extinction}), and stars detected in imaging runs crossing the Galactic plane
where the efficiency of SDSS photometric pipeline drops due to extremely crowded fields. Other,
less significant, $r-i$ bin-specific cuts have also been applied, for example
the exclusion of $|Z| > 2500$ pc stars in $r-i > 1.0$ bins to avoid contamination by halo stars.

We show all 19 ``cleaned up'' maps in figure~\ref{RZcleaned}. The contours denote the locations 
of constant density. The gray  areas show the regions with available SDSS data. 
Compared to Fig.~\ref{RZmedians}, the constant density contours are much more regular, 
and the effect of the Virgo overdensity is largely suppressed.
The regularity of the density distribution is particularly striking for redder bins (e.g., for
$r-i > 0.7$). In the bluest bin ($0.10 < r-i < 0.15$), there is a detectable departure from a smooth
profile in the top left part of the sampled region. This is the area of the $(R,Z)$ plane
where the pixels that are sampled far apart in $(X,Y,Z)$ space map onto adjacent pixels in 
$(R,Z)$ space. Either deviations from axial symmetry or small errors in photometric parallax 
relation (perhaps due to localized metallicity variations) can lead to
deviations of this kind. Unfortunately, which one of the two it is, is 
impossible to disentangle with the data at hand.

\subsection{       Model Fit       }
\label{sec.modelfit}

\subsubsection{       Fitting Algorithm    }

The model fitting algorithm is based on the standard Levenberg-Marquardt nonlinear $\chi^2$
minimization algorithm \citep{NumRecC}, with a multi-step iterative outlier rejection. 

The goal of the iterative outlier rejection procedure is to automatically and gradually remove 
pixels contaminated by unidentified overdensities, single pixels or small groups of pixels with large
deviations (such as those due to localized star clusters, or simply due to instrumental
errors, usually near the edges of the volume), and allow the fitter to ``settle'' towards the
true model even if the initial fit is extremely biased by a few high-$\sigma$ outliers.

The outlier rejection works as follows: after initial fit is made, the residuals are examined 
for outliers from the model higher than a given number of standard deviations, $\sigma_1$. Outlying
data points are excluded, the model is refitted, and all data points are retested with the new 
fit for deviations greater than $\sigma_2$, where $\sigma_2 < \sigma_1$. The procedure is repeated 
with $\sigma_3 < \sigma_2$, etc. The removal of outliers continues until the last step, where 
outliers higher than $\sigma_N$ are excluded, and the final model refitted. The parameters 
obtained in the last step are the best fit model parameters.

The $\sigma_i$ sequence used for outlier rejection is $\sigma_i = \{50, 40, 30, 20, 10, 5\}$. This
slowly decreasing sequence allows the fitter to start with rejecting the extreme outliers (which
themselves bias the initial fit), and then (with the model now refitted without these outliers,
and therefore closer to the true solution) gradually remove outliers of smaller and smaller 
significance and converge towards a solution which best describes the smooth background.

\subsubsection{       A  Measurement of Solar Offset        }

We begin the modeling by fitting a single exponential disk to the three reddest color bins to find
the value of the Solar offset $Z_\odot$. To avoid
contamination by the thick disk, we only use pixels with $|Z| < 300$ pc, and to avoid effects of
overestimated interstellar extinction correction for the nearest stars (Section~\ref{extinction}),
we further exclude pixels with $|Z| < 100$. We further exclude all pixels outside of $7600 < R <
8400$~pc range, to avoid contamination by clumpy substructure.

We obtain
\begin{eqnarray}
	Z_{\odot,\mathrm{bright}} & = & (25 \pm 5)\mathrm{\,pc} \\
	Z_{\odot,\mathrm{faint}} & = & (24 \pm 5)\mathrm{\,pc}
\end{eqnarray}
for the Solar offset, where the $Z_{\odot,\mathrm{bright}}$ is the offset obtained using the bright
photometric parallax relation, and $Z_{\odot,\mathrm{faint}}$ using the faint. The quoted
uncertainty is determined by simply assuming a 20\% systematic uncertainty in the adopted
distance scale, and does not imply a Gaussian error distribution (the formal random fitting error is
smaller than 1 pc).

Our value of the Solar offset agrees favorably with recent independent measurements
($Z_\odot = (27.5 \pm 6)\mathrm{\,pc}$, \citealt{Chen99}; $Z_\odot =
(27 \pm 4)\mathrm{\,pc}$, \citealt{Chen01}; $(24.2 \pm 1.7)\mathrm{\,pc}$ obtained from
trigonometric Hipparcos data by \citealt{Maiz-Apell01}). We keep the value of the Solar
offset fixed in all subsequent model fits.

\subsubsection{         Disk Fits         }
\label{sec.diskfits}

\begin{figure*}
\plotone{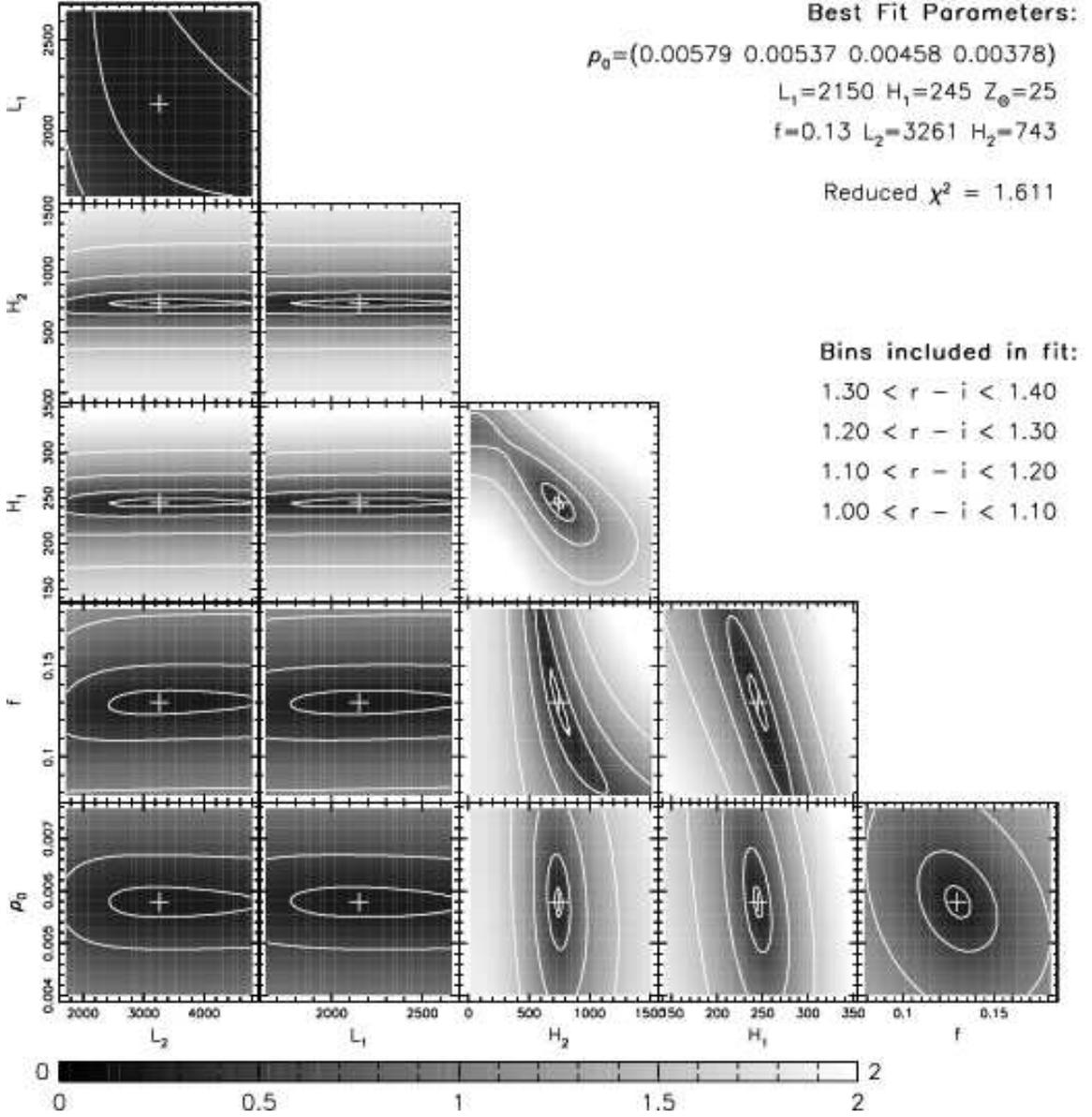}
\caption{
Two-dimensional cross sections of of reduced $\chi^2$ hyper-surface around best-fit 
values for $1.0 < r-i < 1.4$ data (Table~\ref{tbl.bright.joint}, first row). The fit was obtained assuming 
the ``bright'' photometric paralax relation (Equation~\ref{eq.Mr}). 
Analogous cross sections for fits obtained assuming Equation~\ref{eq.Mr.faint} 
(Table~\ref{tbl.faint.joint}, first row) show qualitatively same features.
The innermost contour is at $1.1 \times \chi^2_{min}$ level, while the rest are
logarithmically spaced in steps of 0.5 dex, starting at $\log \chi^2 = 0.5$.
\label{fig.bright.chi2.disk}}
\end{figure*}

\begin{deluxetable*}{rrrrrrrrr}
\tablecaption{Best Fit Values (Joint Fits, Bright Paralax Relation)\label{tbl.bright.joint}}
\tablehead{
	\colhead{$\chi^2$} & \colhead{Bin} & 
	\colhead{$\rho(R_{\odot},0)$} & \colhead{$L_1$} &
	\colhead{$H_1$} & \colhead{$f$} &
	\colhead{$L_2$} & \colhead{$H_2$} & \colhead{$f_H$}
}
\startdata
1.61 & $1.3 < r-i < 1.4$  & 0.0058 &  2150  &  245  &  0.13  &  3261  & 743 & \nodata \\ 
     & $1.2 < r-i < 1.3$  & 0.0054 & & & & & & \\ 
     & $1.1 < r-i < 1.2$  & 0.0046 & & & & & & \\ 
     & $1.0 < r-i < 1.1$  & 0.0038 & & & & & & \\ 
     & & & & & & & & \\ 
1.70 & $0.9 < r-i < 1.0$  & 0.0032 &  2862  &  251  &  0.12  &  3939  & 647 & 0.00507 \\ 
     & $0.8 < r-i < 0.9$  & 0.0027 & & & & & & \\ 
     & $0.7 < r-i < 0.8$  & 0.0024 & & & & & & \\ 
     & $0.65 < r-i < 0.7$ & 0.0011 & & & & & &  
\enddata
\tablecomments{Best fit values of Galactic Model parameters derived assuming the ``bright''
photometric paralax relation (Equation~\ref{eq.Mr}). The fit to $0.65 < r-i < 1.0$ 
bins (bottom row) includes the halo component. Its shape was kept fixed 
(Table~\ref{tbl.joint.haloonly}, top row) and only the normalization $f_H$ was 
allowed to vary.}
\end{deluxetable*}

\begin{deluxetable*}{rrrrrrrrr}
\tablecaption{Best Fit Values (Joint Fits, Faint Paralax Relation)\label{tbl.faint.joint}}
\tablehead{
	\colhead{$\chi^2$} & \colhead{Bin} & 
	\colhead{$\rho(R_{\odot},0)$} & \colhead{$L_1$} &
	\colhead{$H_1$} & \colhead{$f$} &
	\colhead{$L_2$} & \colhead{$H_2$} & \colhead{$f_H$}
}
\startdata
1.59 & $1.3 < r-i < 1.4$  & 0.0064 & 2037  &  229  &  0.14  &  3011  & 662 & \nodata \\ 
     & $1.2 < r-i < 1.3$  & 0.0063 & & & & & & \\ 
     & $1.1 < r-i < 1.2$  & 0.0056 & & & & & & \\ 
     & $1.0 < r-i < 1.1$  & 0.0047 & & & & & & \\ 
     & & & & & & & & \\ 
2.04 & $0.9 < r-i < 1.0$  & 0.0043 & 2620 & 225 & 0.12 & 3342 & 583 & 0.00474 \\ 
     & $0.8 < r-i < 0.9$  & 0.0036 & & & & & & \\ 
     & $0.7 < r-i < 0.8$  & 0.0032 & & & & & & \\ 
     & $0.65 < r-i < 0.7$  & 0.0015 & & & & & &  
\enddata
\tablecomments{Best fit values of Galactic Model parameters derived assuming the ``faint''
photometric paralax relation (Equation~\ref{eq.Mr.faint}). The fit to $0.65 < r-i < 1.0$ 
bins (bottom row) includes the halo component. Its shape was kept fixed 
(Table~\ref{tbl.joint.haloonly}, bottom row) and only the normalization $f_H$ was 
allowed to vary.}
\end{deluxetable*}

We utilize the $R-Z$ density maps of the four $r-i > 1.0$ bins to fit the double-exponential disk
model. These color bins sample the thin and thick disk, with a negligible halo
contribution (less than $\sim 1$\% for plausible halo models). Furthermore, the photometric relations
in this range of colors are calibrated to metallicities of disk dwarfs, thus making these bins
optimal for the measurement of disk model parameters.

We simultaneously fit all double-exponential disk model parameters ($\rho$, $H_1$, $L_1$, $f$,
$H_2$, $L_2$) to the data, for both bright and faint photometric parallax relations. To avoid
contamination by the halo, we only use the pixels with $|Z| < 2500$ pc. To avoid effects of
overestimated interstellar extinction correction for the nearest stars (Section~\ref{extinction}),
we further exclude pixels with $|Z| < 100$.

We jointly fit the data from all four color bins, and separately for each bin. In the former,
``joint fit'' case, only the densities $\rho(R_{\odot},0)$ are allowed to vary between the bins, while
the scale lengths, heights and thick-to-thin disk normalization $f$ are constrained to be the same
for stars in each bin. As the color bins under consideration sample stars of very similar mass, age 
and metallicity, we expect the same density profile in all bins\footnote{Note also that being 0.1mag 
wide, with a typical magnitude errors of $\sigma_r \gtrsim 0.02$mag the adjacent bins are
\emph{not} independent. The histograms in Figure~\ref{binspill} illustrate this well.}. 
The best fit parameters for the joint fit to $r-i > 1.0$ bins are given 
in top row of Tables~\ref{tbl.bright.joint}~and~\ref{tbl.faint.joint}, calculated assuming
the bright (Equation~\ref{eq.Mr}) and faint (Equation~\ref{eq.Mr.faint}) photometric parallax relation,
respectively. Two-dimensional cross sections of reduced $\chi^2$ hyper-surface around best-fit 
values are shown in Figure~\ref{fig.bright.chi2.disk} (for bright relation only -- analogous crossections
obtained with the faint relation look qualitatively the same).

In case of separate fits, all parameters are fitted independently for each color bin. Their variation
between color bins serves as a consistency check and a way to assess the degeneracies, significance
and uniqueness of the best fit values. The best-fit values are shown in the top four rows of 
Table~\ref{tbl.bright.individual.fits} (bright photometric parallax relation) and the top five\footnote{
The fit for $0.9 < r-i < 1.0$ bin when using the faint photometric relation and including 
a halo component (see Section~\ref{sec.diskhalofits}), failed to converge to physically 
reasonable value. We have therefore fitted this bin with disk components only.
} 
rows of Table~\ref{tbl.faint.individual.fits} (faint relation).

In all cases we are able to obtain good model fits, with reduced $\chi^2$ in
the range from $1.3$ to $1.7$. The best-fit solutions are mutually consistent. In particular, the thin disk scale height is well 
constrained to $H_1 = 250$~pc (bright) and $H_1 = 230-240$ (faint), as are the values of $\rho(R_{\odot},0)$
which give the same results in individual and joint fits at the $\sim 5\%$ level.

The thick-to-thin disk density normalization is $\sim 10\%$, with $f = 0.10-0.13$ (bright) 
and $f = 0.10-0.14$ (faint). The thick disk scale length solutions are in $H_2 = 750-900$~pc (bright) and
$H_2 = 660-900$~pc (faint) range. Thick disk normalization and scale heights appear less well constrained;
however, note that the two are fairly strongly correlated ($f$ vs $H_2$ panel in Figure~\ref{fig.bright.chi2.disk}).
As an increase in density normalization leads to a decrease in disk scale height and vice versa with no appreciable 
effect on $\chi^2$, any two models with so correlated differences of scale height and normalization
of up to $20\%$ to $30\%$ are practically indistiguishable. This interplay between $\rho$ and $H_2$ 
is seen in  Tables~\ref{tbl.bright.individual.fits}~and~\ref{tbl.faint.individual.fits}, most extremely 
for $1.1 < r-i < 1.2$ bin (Table~\ref{tbl.faint.individual.fits}, third row).
With this in mind, the fits are still consistent with a single thick disk scale 
height $H_2$ and density normalization $f$ describing the stellar number density distribution in 
all $r-i > 1.0$ color bins.

Constraints on disk scale lengths are weaker, with the goodness of fit and the values of other 
parameters being relatively insensitive on the exact values of $L_1$ and $L_2$ (Figure~\ref{fig.bright.chi2.disk},
first two columns). This is mostly due to a short observation baseline in the radial ($R$) direction. 
The best fit parameters lie in the range of $L_1=1600-2400$~pc, 
$L_2=3200-6000$~pc (bright) and $L_1=1600-3000$~pc, $L_2=3000-6000$~pc (faint parallax relation). 
Note that the two are anticorrelated (Figure~\ref{fig.bright.chi2.disk}, top left panel), and combinations
of low $L_1$ and high $L_2$, or vice versa can easily describe the same density field with similar 
values of reduced $\chi^2$ (the behavior seen in Tables~\ref{tbl.bright.individual.fits}~and~\ref{tbl.faint.individual.fits}).
The disk scale length fits in individual color bins are also consistent with there being
a single pair of scale lengths $L_1$ and $L_2$ applicable to all color bins.

\subsubsection{         Halo Fits         }

\begin{figure}
\plotone{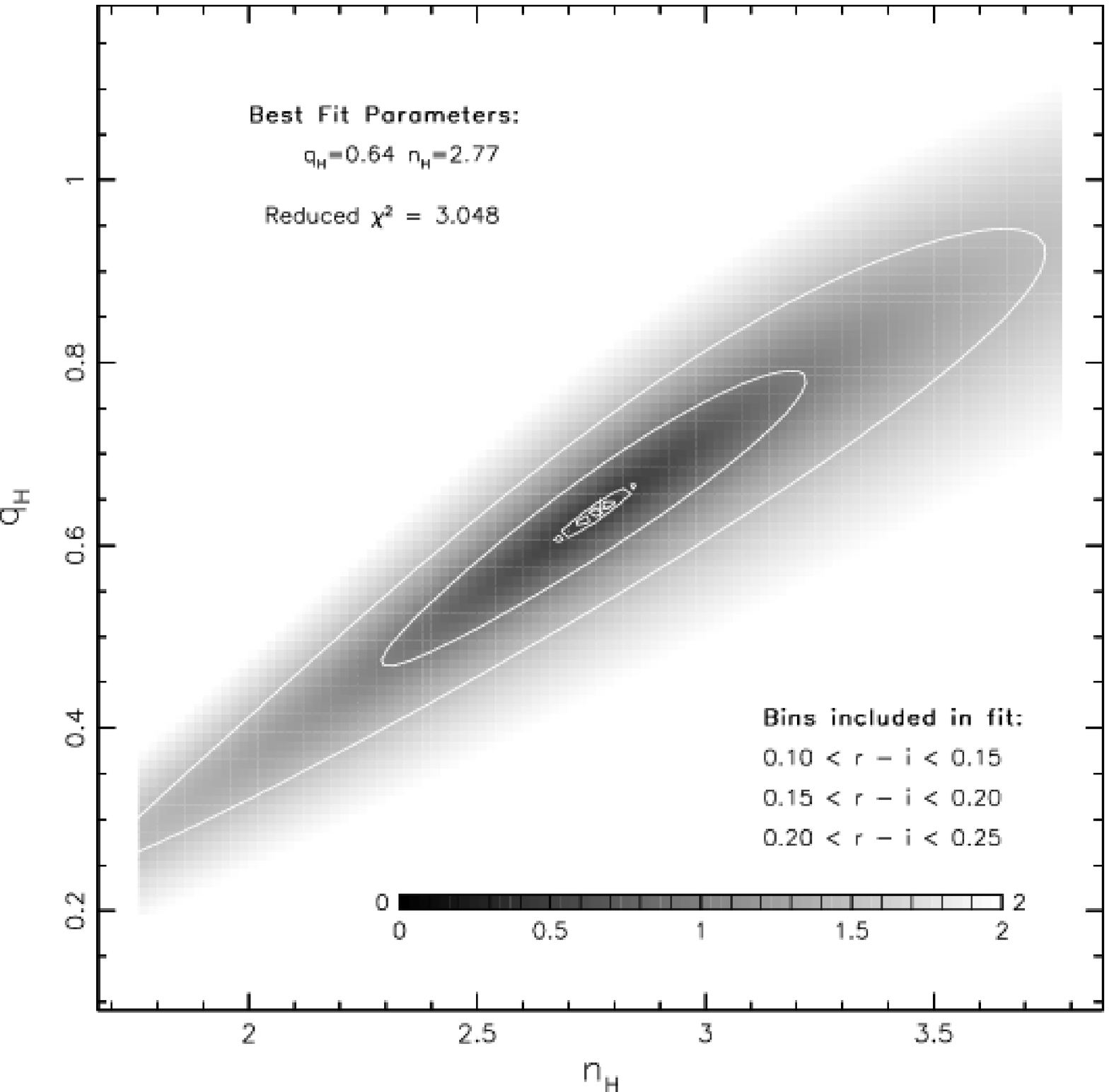}
\caption{
Reduced $\chi^2$ surface of halo parameters $n_H$ and $q_H$ around the best-fit values 
(Table~\ref{tbl.joint.haloonly}, first row). The innermost contour is at 
$1.1 \times \chi^2_{min}$ level, while the rest are logarithmically spaced in 
steps of 0.5 dex, starting at $\log \chi^2 = 0.5$.
\label{fig.bright.chi2.haloonly}
}
\end{figure}

\begin{figure*}
\scl{.7}
\plotone{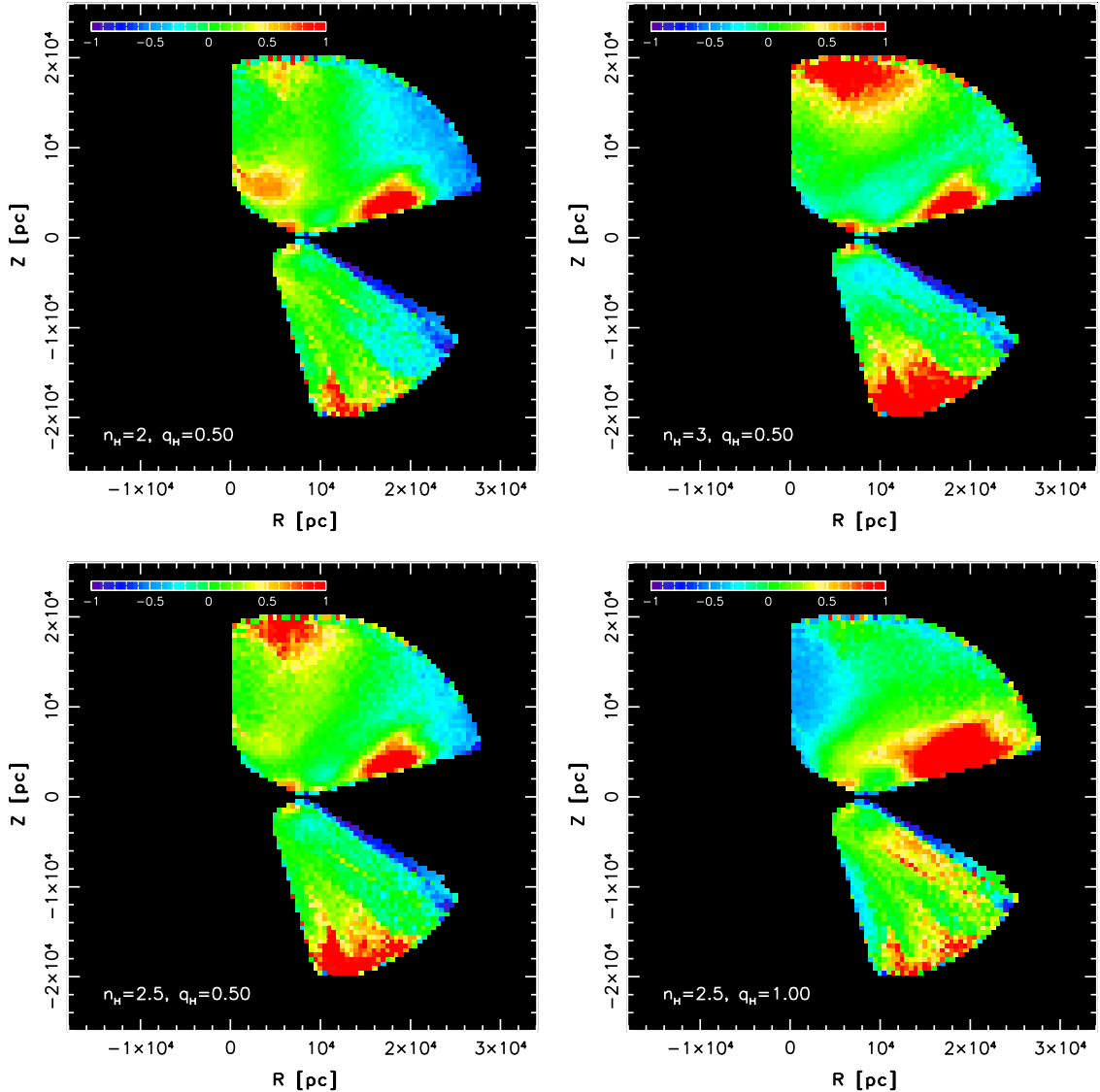}
\caption{
Data-model residuals, normalized to the model, for color bin $0.10 < r-i < 0.15$,
using for four different halo models. All four models have
identical thin and thick disk parameters, and only
the halo parameters are varied. Panels in the top row illustrate the changes 
in residuals when the halo power law index
$n_H$ is varied while keeping the axis ratio fixed. Panels of the bottom row
illustrate the effects of axis ratio $q_H$ change, while keeping the power
law index constant. While $n_H$ is not strongly constrained, the data strongly 
favor an oblate halo.
\label{haloPanels1}}
\end{figure*}

\begin{figure*}
\plotone{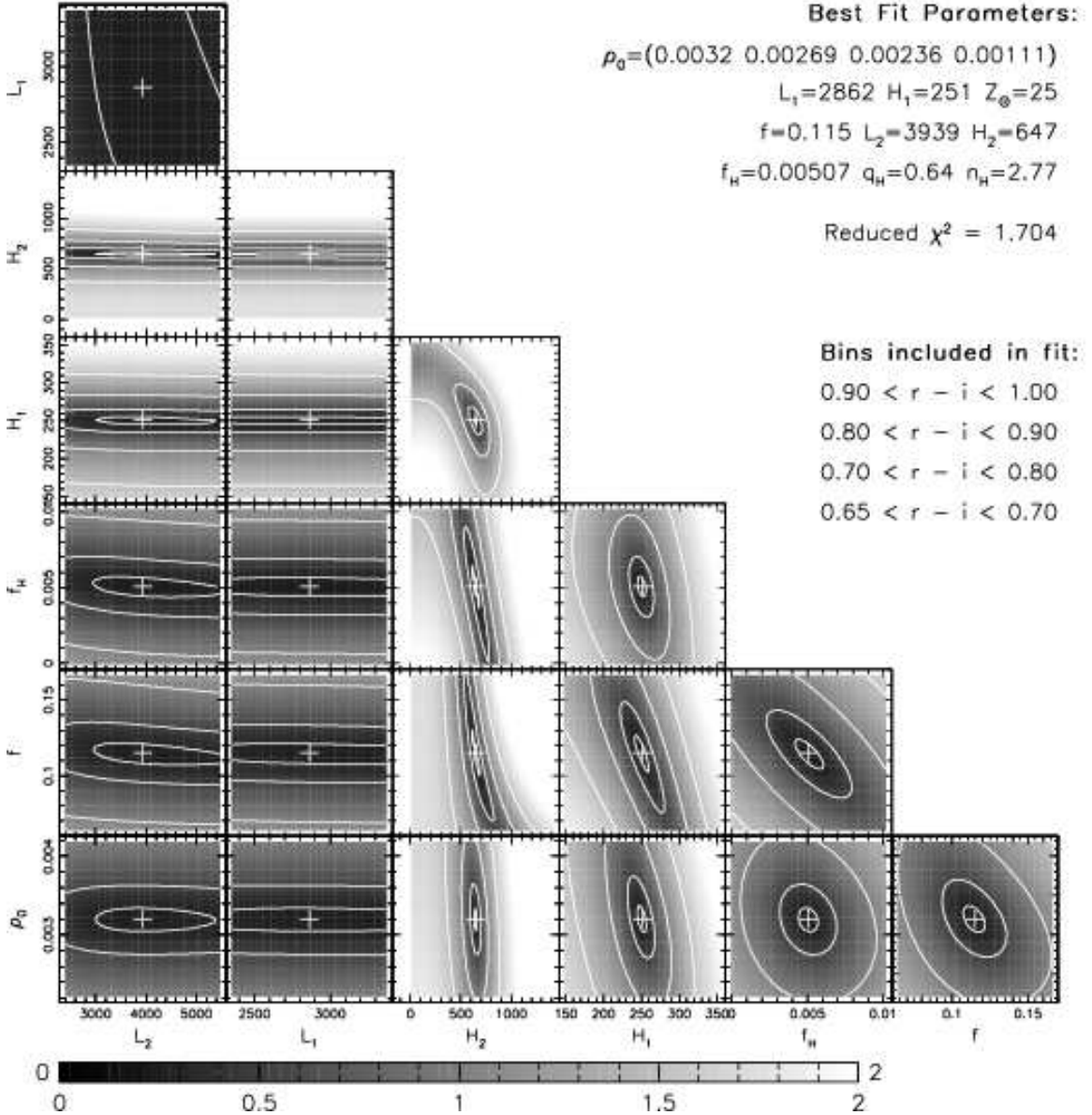}
\caption{Two-dimensional cross sections of of reduced $\chi^2$ hyper-surface around best-fit 
values for $0.65 < r-i < 1.0$ data (Table~\ref{tbl.bright.joint}, second row). The fit was obtained assuming 
the ``bright'' photometric paralax relation (Equation~\ref{eq.Mr}) and includes 
the contribution of the halo. Analogous cross sections for fits obtained assuming 
Equation~\ref{eq.Mr.faint} (Table~\ref{tbl.faint.joint}, second row) show qualitatively 
same features. The innermost contour is at $1.1 \times \chi^2_{min}$ level, while the rest are
logarithmically spaced in steps of 0.5 dex, starting at $\log \chi^2 = 0.5$.
\label{fig.bright.chi2.halo}}
\end{figure*}

\begin{deluxetable}{crrr}
\tablecaption{Halo Shape and Profile Fit\label{tbl.joint.haloonly}}
\tablehead{
	\colhead{Paralax Relation} & 
	\colhead{$\chi^2$} &
	\colhead{$q_H$} &
	\colhead{$n_H$}
}
\startdata
Bright &  3.05  &   $0.64 \pm 0.01$ &  $2.77 \pm 0.03$  \\ 
Faint  &  2.48  &   $0.62 \pm 0.01$ &  $2.78 \pm 0.03$ 
\enddata
\tablecomments{
Best fit values of halo power law index $n_H$ and axis ratio $q_H = c/a$, 
assuming the ``bright'' (top) and ``faint'' (bottom row) photometric 
paralax relation.}
\end{deluxetable}

For bluer color bins ($r-i < 1.0$) the probed distance range is larger, and the stellar 
halo component starts to appreciably contribute to the total density near the far edge 
of survey volume. As seen in the middle and bottom panel of Figure~\ref{rhoZ}, 
the disk-only solution becomes visibly unsatisfactory at $Z \gtrsim 4000$~kpc. Also,
the reduced $\chi^2$ values of disk-only models begin to climb to higher than a few
once we attempt to fit them to data in $r-i < 1.0$ bins.

Before we move on to adding and fitting the halo component, there are a few significant
caveats  that must be discussed, understood and taken into account. Firstly, 
the presence  of clumpiness and merger debris in the halo, if unaccounted for, 
will almost certainly bias (an make difficult, or even impossible to determine) 
the model parameters. An initial survey of the density field (Section~\ref{sec.maps}), 
the identification, and careful removal of identified overdensities 
(Section~\ref{sec.dataset.preparation}) are \emph{essential} for obtaining a reasonable fit.

Secondly, the photometric parallax relations (eqs.~\ref{eq.Mr.faint}~and~\ref{eq.Mr})
do not explicitly depend on stellar metallicity. Implicitly, as discussed in 
Section~\ref{sec.pp.metallicity}, they take metallicity into account by virtue of being 
calibrated to disk M-dwarfs on the red, and metal-poor halo stars at the 
blue end. This makes them correct for low metallicity stars ([Fe/H] $\lesssim$ -1.5) 
near $r-i \sim 0.15$, and high metallicity ([Fe/H] $\gtrsim -0.5$) at $r-i \gtrsim 0.7$.
They are therefore appropriate for the study of halo shape and parameters \emph{only at the blue end},
and disk shape and parameters \emph{only on the red end}. Conversely, they are inappropriate
for the study of disk shape and parameters at the blue, or halo shape and parameters at the 
red end. For the same reason, it is difficult to simultaneously fit the halo and the disk 
in the intermediate $r-i$ bins, as the application of photometric parallax relation inappropriate 
for the low metallicity halo induces distortions of halo shape in the density maps.

Therefore, to measure the shape of the halo, we only select the data points from the 
three bluest, $0.1 < r-i < 0.25$ bins, and only in regions of ($R, Z$) plane where a fiducial 
$q_H = 0.5$, $n_H=2.5$, $f_H=0.001$ halo model predicts the fraction of disk stars to be 
less than $5\%$. This allows us to fit for the power law index $n_H$, and the axis 
ratio $q_H$ of the halo. Because we explicitly excluded the disk, we cannot fit for the 
halo-to-thin-disk normalization $f_H$ (but see Section~\ref{sec.diskhalofits} later in
the text for a workaround).

The best fit parameters obtained for the halo are shown in Table~\ref{tbl.joint.haloonly} for
both the bright and faint photometric relation, and the reduced $\chi^2$ surface for the 
fit is shown in Figure~\ref{fig.bright.chi2.haloonly} (bright relation only -- the surface
looks qualitatively the same for the fit made assuming the faint relation).

The fits are significantly poorer than for the disks, with reduced $\chi^2 = 2-3$.
Formal best-fit halo parameters are $n_H=2.8$, $q_H=0.64$, but given the relatively
high and shallow minimum, and the shape of the $\chi^2$ surfaces in Figure~\ref{fig.bright.chi2.haloonly},
it is better to think of the fit results as constraining the parameters to a range of
values -- the power law index to $n_H = 2.5-3$, and the oblateness parameter $q_H = 0.5 - 0.8$.

Fig.~\ref{haloPanels1} shows residual maps for the bluest color bin and 
for four different halo models, with the thin and thick disk parameters
kept fixed at values determined using redder bins (Table~\ref{tbl.bright.joint}). 
Individual panels illustrate the changes in residuals when the halo power law index
is varied while keeping the axis ratio fixed (top row), and when the ellipticity
of the halo is changed, from oblate to spherical while keeping the power law
index $n_H$ fixed (bottom row). The Monoceros and Virgo overdensities, and the 
overdensity at $R\sim$6.5 kpc and $Z\sim$ 1.5 kpc, are clearly evident, but 
their detailed properties depend significantly on the particular halo model 
subtracted from the data.

We further find that a power-law halo model always over- or underestimates the stellar 
counts in the far outer halo (Figure~\ref{haloPanels1}), suggesting the use of a different 
profile may be more appropriate and consistent with ``dual-halo'' profiles favored by
(among others) \cite{Sommer-Larsen90,Allen91,Zinn93,Carney96,Chiba00} and more recently
discussed by \cite{Siegel02}.

However, no matter what the exact shape of the profile or 
the power law index is, only  significantly oblate halos provide good fits to the data 
(compare the bottom right to other panels in  Fig.~\ref{haloPanels1}). Specifically, 
given the reduced $\chi^2$ surface in Figure~\ref{fig.bright.chi2.haloonly}, spherical or 
prolate halo can be ruled out, and this remains to be the case 
irrespective of the details of the photometric parallax relation\footnote{Aspherical halos 
could be artificially favored by the $\chi^2$ analysis, 
as a way to parametrize away any existing halo inhomogeneity. However, given the analysis
of residuals in Section~\ref{sec.rhists}, we consider this to be a very unlikely explanation 
of the measured oblateness.}.

\subsubsection{         Simultaneous Disk and Halo Fits         }
\label{sec.diskhalofits}

\begin{deluxetable*}{rrrrrrrrr}
\tablecaption{Best Fit Values (Individual Fits, Bright Paralax Relation)\label{tbl.bright.individual.fits}}
\tablecolumns{9}
\tablehead{
	\colhead{Color bin} & \colhead{$\chi^2$} &
	\colhead{$\rho(R_{\odot},0)$} & \colhead{$L_1$} &
	\colhead{$H_1$} & \colhead{$f$} &
	\colhead{$L_2$} & \colhead{$H_2$} & \colhead{$f_H$}
}
\startdata
$1.3 < r-i < 1.4$  &  1.34  &   0.0062  &  1590  &  247  &  0.09  &  5989  & 909 & \nodata \\ 
$1.2 < r-i < 1.3$  &  1.31  &   0.0055  &  1941  &  252  &  0.11  &  5277  & 796 & \nodata \\ 
$1.1 < r-i < 1.2$  &  1.58  &   0.0049  &  2220  &  250  &  0.09  &  3571  & 910 & \nodata \\ 
$1 < r-i < 1.1$    &  1.64  &   0.0039  &  2376  &  250  &  0.10  &  3515  & 828 & \nodata \\ 
$0.9 < r-i < 1$    &  1.38  &   0.0030  &  3431  &  248  &  0.14  &  2753  & 602 & 0.0063 \\ 
$0.8 < r-i < 0.9$  &  1.48  &   0.0028  &  3100  &  252  &  0.10  &  3382  & 715 & 0.0039 \\ 
$0.7 < r-i < 0.8$  &  1.83  &   0.0024  &  3130  &  255  &  0.09  &  3649  & 747 & 0.0037 \\ 
$0.65 < r-i < 0.7$ &  1.69  &   0.0011  &  2566  &  273  &  0.05  &  8565  & 861 & 0.0043 
\enddata
\tablecomments{Best fit values of Galactic Model parameters, fitted separately for each $r-i$ bin
assuming the ``bright'' photometric paralax relation (Equation~\ref{eq.Mr}). In fits which
include the halo component, the shape of the halo was kept fixed (Table~\ref{tbl.joint.haloonly}, top
row), and only the normalization $f_H$ was allowed to vary.}
\end{deluxetable*}

\begin{deluxetable*}{rrrrrrrrr}
\tablecaption{Best Fit Values (Individual Fits, Faint Paralax Relation)\label{tbl.faint.individual.fits}}
\tablecolumns{9}
\tablehead{
	\colhead{Color bin} & \colhead{$\chi^2$} &
	\colhead{$\rho(R_{\odot},0)$} & \colhead{$L_1$} &
	\colhead{$H_1$} & \colhead{$f$} &
	\colhead{$L_2$} & \colhead{$H_2$} & \colhead{$f_H$}
}
\startdata
$1.3 < r-i < 1.4$  &  1.32  &   0.0064  &  1599  &  246  &  0.09  &  5800  & 893 & \nodata \\ 
$1.2 < r-i < 1.3$  &  1.40  &   0.0064  &  1925  &  242  &  0.10  &  4404  & 799 & \nodata \\ 
$1.1 < r-i < 1.2$  &  1.56  &   0.0056  &  2397  &  221  &  0.17  &  2707  & 606 & \nodata \\ 
$1 < r-i < 1.1$    &  1.71  &   0.0049  &  2931  &  236  &  0.10  &  2390  & 760 & \nodata \\ 
$0.9 < r-i < 1$    &  1.62  &   0.0043  &  3290  &  239  &  0.07  &  2385  & 895 & \nodata \\ 
$0.8 < r-i < 0.9$  &  1.69  &   0.0038  &  2899  &  231  &  0.08  &  2932  & 759 & 0.0021 \\ 
$0.7 < r-i < 0.8$  &  2.59  &   0.0034  &  2536  &  227  &  0.09  &  3345  & 671 & 0.0033 \\ 
$0.65 < r-i < 0.7$ &  1.92  &   0.0016  &  2486  &  241  &  0.05  &  6331  & 768 & 0.0039 
\enddata
\tablecomments{Best fit values of Galactic Model parameters, fitted separately for each $r-i$ bin
assuming the faint'' photometric paralax relation (Equation~\ref{eq.Mr.faint}). In fits which
include the halo component, the shape of the halo was kept fixed (Table~\ref{tbl.joint.haloonly},
bottom row), and only the normalization $f_H$ was allowed to vary.}
\end{deluxetable*}

Keeping the best fit values of halo shape parameters $q_H$ and $n_H$ constant,
we next attempt to simultaneously fit the thin and thick disk parameters and
the halo normalization, $f_H$, in four $0.65 < r-i < 1.0$ bins.

These bins encompass regions of $(R, Z)$ space where the stellar number
density due to the halo is not negligible and has to be taken into account.
Simultaneous fits of both the disk and all halo parameters are still unfeasible,
both because halo stars still make up only a small fraction of the total number density,
and due to poor applicability of the disk-calibrated photometric parallax relations in this 
$r-i$ range to low-metallicity halo stars. However, knowing the halo shape from the blue,
low-metallicity calibrated bins, we may keep $q_H$ and $n_H$ fixed and fit for the halo-to-thin-disk
normalization, $f_H$. Given the uncertainty in its current knowledge, thusly obtained 
value of $f_H$ is still of considerable interest despite the likely biases.

We follow the procedure outlined in Section~\ref{sec.diskfits}, and fit the data in $0.65 < r-i < 1.0$
bins jointly for all and separately in each color bin, with both bright and faint photometric
parallax relations.

The results of the joint fits are given in bottom rows of Tables~\ref{tbl.bright.joint}~and~\ref{tbl.faint.joint}.
Results for individual bins are given in bottom rows of Tables~\ref{tbl.bright.individual.fits}~and~\ref{tbl.faint.individual.fits}
for the bright and faint photometric relation, respectively.

We obtain satisfactory model fits, with reduced $\chi^2$ in $1.4$ to $2.0$ range. As was the case for
fits to $r-i > 1.0$ bins, the best fit disk parameter values are consistent between bins, 
and with the joint fit. The reduced $\chi^2$ surface cross-sections, shown in 
Figure~\ref{fig.bright.chi2.halo}, are qualitatively the same as those in 
Figure~\ref{fig.bright.chi2.disk} and the entire discussion of Section~\ref{sec.diskfits} 
about fit parameters and their interdependencies applies here as well.

Comparison of top and bottom rows in Tables~\ref{tbl.bright.joint}~and~\ref{tbl.faint.joint} shows 
consistent results between $r-i>1.0$ and $0.65 < r-i < 1.0$ bins. In particular, the scale heights of 
the thin disk are the same,
and the thick-to-thin disk normalization is the same to within $8-15$\%, still within fit uncertancies. The
scale lengths are still poorly constrained, and on average $10-30\%$ larger than in disk-only fits.
Given the poor constraint on scale lengths, it is difficult to asses whether this effect is physical,
or is it a fitting artifact due to the addition of stellar halo component. The scale height of the thick 
disk, $H_2$ is $\sim 14$\% smaller than in disk-only fits. This is likely due to the reassignment to 
the halo of a fraction of stellar number density previously assigned to the thick disk.

For $f_H$, the halo-to-thin-disk normalization at ($R=8$~kpc, $Z=0$), the best fit values
are in $0.3-0.6$\% range, with the best fit value for joint fits being $f_H = 0.5$\% both for the bright
and faint parallax relation. In particular, note how insensitive $f_H$ is on the choice
of photometric parallax relation. In this region of $r-i$ colors, the average difference between 
the bright and faint parallax relations is $\Delta M_r = 0.25$~mag; therefore even in case of 
uncertainties of $\sim$ half a magnitude, the change in $f_H$ will be no greater than $\sim10-20$\%.

\subsection{       Analysis        }

The Galactic model parameters as fitted in the preceding Section are biased\footnote{Or ``apparent'',
in the terminology of \citealt{Kroupa93}} by unrecognized stellar multiplicity, finite 
dispersion of the photometric parallax relation and photometric errors. They are further made 
uncertain by possible systematics in calibration of the photometric parallax relations,
and a simplified treatment of stellar metallicities.

In this section, we analyze all of these (and a number of other) effects on a series of
Monte Carlo generated mock catalogs, and derive the corrections for each of them.
We also look at the resolved local overdensities found in the data, discuss the question
of possible statistical signatures of further unresolved overdensities, and questions
of uniqueness and degeneracy of our best-fit model.

After deriving the bias correction factors, we close the Section by summarizing and writing out 
the final debiased set of best fit SDSS Galactic model parameters, together with their 
assumed uncertainties.

\subsubsection{         Monte-Carlo Generated Mock Catalogs         }

\begin{deluxetable*}{lcrrrrr}
\tabletypesize{\scriptsize}
\tablecaption{Monte Carlo Generated Catalog Fits\label{tbl.simulations}}
\tablecolumns{7}
\tablehead{
	\colhead{Simulation} & \colhead{$\chi^2$} &
	\colhead{$L_1$} &
	\colhead{$H_1$} & \colhead{$f$} &
	\colhead{$L_2$} & \colhead{$H_2$}
}
\startdata
True Model  			&  \nodata &   2500          &  240          &  0.10              &  3500          & 770 \\ 
Perfect Catalog			&  1.03    &   $2581 \pm 44$ &  $244 \pm 2$  &  $0.094 \pm 0.009$ &  $3543 \pm 69$ & $791 \pm 18$ \\ 
Photometric and Paralax Errors  &  0.95    &   $2403 \pm 40$ &  $230 \pm 2$  &  $0.111 \pm 0.010$ &  $3441 \pm 57$ & $725 \pm 13$ \\ 
25\% binary fraction		&  0.97    &   $2164 \pm 39$ &  $206 \pm 1$  &  $0.119 \pm 0.011$ &  $3199 \pm 47$ & $643 \pm 9$ \\ 
50\% binary fraction		&  0.97    &   $1986 \pm 34$ &  $193 \pm 1$  &  $0.115 \pm 0.011$ &  $2991 \pm 41$ & $611 \pm 7$ \\ 
100\% binary fraction		&  1.02    &   $1889 \pm 31$ &  $178 \pm 1$  &  $0.104 \pm 0.010$ &  $2641 \pm 31$ & $570 \pm 6$  
\enddata
\tablecomments{The true model parameters (top row), and the best fit values of model parameters
recovered from a series of Monte-Carlo generated catalogs. These test the correctness of
data processing pipeline and the effects of cosmic variance (``Perfect Catalog''), the effects
of photometric paralax dispersion and photometric errors (``Photometric and Paralax Errors''),
and the effects of varying fraction of unresolved binary stars in the data (last three rows).}
\end{deluxetable*}

To test the correctness of the data processing and fitting procedure and derive the correction
factors for Malmquist bias, stellar multiplicity and uncertainties due to photometric
parallax systematics, we developed a software package for generating realistic mock star
catalogs. These catalogs are
fed to the same data-processing pipeline and fit in the same manner as the real data.

The mock catalog generator, given an arbitrary Galactic model (which in our case is defined
by eqs.~\ref{galModel}--\ref{haloModel}, a local position-independent luminosity function, 
and binary fraction), generates a star catalog within an arbitrarily complex footprint 
on the sky. The code can also include realistic magnitude-dependent photometric 
errors (Figure~\ref{magerr2}, bottom panel) and the errors due to Gaussian dispersion $\sigma_{M_r}$
around the photometric parallax mean, $M_r(r-i)$.

Using this code, we generate a series of mock catalogs within the footprint of the SDSS data used in this study 
(Figure~\ref{fig.skymap}) using a fiducial model with parameters listed in the top row of Table~\ref{tbl.simulations}.
For the luminosity function, we use the \citet{Kroupa93} luminosity function, transformed
from $\phi(M_V)$ to $\phi(M_r)$ and renormalized to $\rho(R=8000,Z=0) = 0.04$~stars~pc$^{-3}$~mag$^{-1}$
in the $1.0 < r-i < 1.1$ bin. As we will be making comparisons between the simulation 
and $r-i>1.0$ bins, we do not include the halo component ($f_H = 0$).

For all tests described in the text to follow, we generate the stars in $0.7 < r-i < 1.6$ color 
and $10 < r < 25$ magnitude range, which is sufficient
to include all stars that may possibly scatter into the survey flux ($15 < r < 21.5$) and disk color
bins ($1.0 < r-i < 1.4$) limits, either due to photometric errors, uncertainty in the photometric
parallax relation, or an added binary companion. To transform from distance to magnitude, we use the
bright photometric parallax relation (eq.~\ref{eq.Mr}).

\subsubsection{         Correctness of Data Processing and Fitting Pipeline         }

We first test for the correctness of the data processing and fitting pipeline, by generating 
a ``perfect'' catalog. Stars in this catalog have no photometric errors added, and their
magnitudes and colors are generated using eqs.~\ref{eq.Mr}~and~\ref{eq.locus}.

We fit this sample in the same manner as the real data in Section~\ref{sec.diskfits}.
The results are given in the second row of Table~\ref{tbl.simulations}. The fit recovers the original 
model parameters, with the primary source of error being the ``cosmic variance'' due 
to the finite number of stars in the catalog.

This test confirms that fitting and data processing pipelines introduce no
additional uncertainty to best-fit model parameters. It also illustrates
the limits to which one can, in principle, determine the model parameters from our 
sample assuming  a) that stars are distributed in a double-exponential disk and
b) the three-dimensional location of each star is perfectly known. These limits are 
about $1-2$\%, significantly smaller than all other sources of error.

\subsubsection{         Effects of Malmquist Bias         }
\label{sec.malmquist.effects}

We next test for the effects of photometric errors, and the errors due to the finite width of the 
photometric parallax relation. We model the photometric errors as Gaussian, with a 
magnitude-dependent dispersion $\sigma_r$ measured from the data (Figure~\ref{magerr2}, bottom panel). 
Median photometric errors range from $\sigma_r = 0.02$ on the bright to $\sigma_r = 0.12$ on the
faint end. We assume the same dependence holds for $g$ and $i$ band as well. 
We model the finite width of the photometric parallax relation as a Gaussian $\sigma_{M_r} = 0.3$ 
dispersion around the mean of $M_r(r-i)$. The two sources of effective photometric error add up 
in quadrature and act as a source of a Malmquist bias, with the photometric parallax relation 
dispersion giving the dominant effect (eq.~\ref{eq.disterr}).

The best fit parameters obtained from this sample are given in the third row of Table~\ref{tbl.simulations}. 
The thin and thick disk scale heights are underestimated by $\sim 5\%$. The density normalization
$f$ is overestimated by $\sim 10$\% (note however that this is still within the statistical uncertainty).
The scale lengths are also slightly underestimated, with the effect  less pronounced for the thick disk.

We conclude that the Malmquist bias due to photometric errors and the dispersion around the photometric
parallax relation has a relatively small effect on the determination of Galactic model parameters, 
at the level of $\sim 5$\%.

\subsubsection{         Effects of Unrecognized Multiplicity         }
\label{sec.binarity}

Unrecognized multiplicity biases the density maps and the determination of Galactic model parameters
by systematically making unresolved binary stars, when misidentified as a single star, appear 
closer then  they truly are. It's effect is most strongly dependent on the fraction of 
observed ``stars'', $f_m$, that are in fact unresolved multiple systems.

We model this effect by simulating a simplified case where all multiple systems are binaries. Because
the fraction of binary systems is poorly known, we generate three mock catalogs with varying fractions
$f_m$ of binary systems misidentified as single stars, and observe the effects of $f_m$ on the
determination of model parameters. Photometric errors and photometric parallax dispersion 
(as discussed in Section~\ref{sec.malmquist.effects}) are also mixed in.

The results are given in the last three rows of Table~\ref{tbl.simulations}. The effect of
unresolved binary systems is a systematic reduction of all spatial scales of the model. Measured 
disk scale heights are underestimated by as much as 25\% ($f_m = 1$), 20\% ($f_m = 0.5$) and 
15\% ($f_m = 0.25$). Measured scale lengths are similarily biased, with the thin disk
scale length being underestimated by 25, 20, and 13\% and the thick disk scale length by 25, 15, and 9\%
for $f_m = 1, 0.5, 0.25$, respectively. The thick disk density normalization is mildly 
overestimated ($\sim 10$\%) but not as strongly as the disk scales, and still within statistical
uncertainty.

\subsubsection{         Effects of Systematic Distance Determination Error        }
\label{sec.systematics}

\begin{deluxetable*}{lcrrrrr}
\tablecaption{Effects of $M_r(r-i)$ Calibration Errors\label{tbl.fits.plusminus}}
\tablewidth{6in}
\tablecolumns{7}
\tablehead{
	\colhead{Simulation} & \colhead{$\chi^2$} &
	\colhead{$L_1$} &
	\colhead{$H_1$} & \colhead{$f$} &
	\colhead{$L_2$} & \colhead{$H_2$}
}
\startdata
$M_r(r-i) - 0.5$ &  1.18 &   $3080 \pm 55$ &  $305 \pm 4$  &  $0.123 \pm 0.011$ &  $4881 \pm 78$ & $976 \pm 19$ \\ 
$M_r(r-i)$       &  0.95 &   $2403 \pm 40$ &  $230 \pm 2$  &  $0.111 \pm 0.010$ &  $3441 \pm 57$ & $725 \pm 13$ \\ 
$M_r(r-i) + 0.5$ &  1.23 &   $1981 \pm 32$ &  $198 \pm 2$  &  $0.138 \pm 0.013$ &  $3091 \pm 52$ & $586 \pm 11$ 
\enddata
\tablecomments{
Effects of systematic error in the calibration of photometric paralax
relation. Middle row lists the parameters recovered assuming the correct paralax relation
(eq.~\ref{eq.Mr}), from a Monte Carlo generated catalog with realistic photometric
errors and dispersion $\sigma_{M_r} = 0.3$mag around the mean of $M_r(r-i)$. This is the
same catalog as in row 3 of Table~\ref{tbl.simulations}. The first and last row show parameters
recovered when a paralax relation which systematically under-/overestimates the absolute magnitudes
by 0.5 magnitudes (over-/underestimates the distances by $\sim 23$\%) is assumed.}
\end{deluxetable*}

We next measure the effect of systematically over- or underestimating the distances to stars
due to absolute calibration errors of the photometric parallax relation. This can already 
be judged by comparing the values of model parameters determined from fits using the bright and faint 
photometric parallax relations (Tables~\ref{tbl.bright.joint}~and~\ref{tbl.faint.joint}), but 
here we test it on a clean simulated sample with a known underlying model.

We generate a mock catalog by using the bright photometric parallax relation (eq.~\ref{eq.Mr}) 
to convert from distances to magnitudes, and mix in SDSS photometric and parallax dispersion 
errors (Section~\ref{sec.malmquist.effects}). We process this catalog by assuming a parallax relation 0.5 
magnitudes brighter, and 0.5 magnitudes fainter than Equation~\ref{eq.Mr}, effectively 
changing the distance scale by $\pm 23$\%.

Fit results are shown in Table~\ref{tbl.fits.plusminus}, including for comparison in the middle row the 
parameters recovered using the correct $M_r(r-i)$ relation. The effect of systematic
distance errors is to comparably increase or decrease measured geometric scales. The thin and thick
disk scale heights increase by 33\% and 34\%, and the scale lengths by 28\% and 42\%, respectively,
if the distances are overestimated by 23\%. If they are underestimated by the same factor, the parameters
are reduced by 14\% and 19\% (thin and thick disc scale height), 18\% and 10\% (thin and thick disk 
scale lengths). Interestingly, both increasing and decreasing the distance scale results in an 
\emph{increase} of measured normalization, by a factor of $\sim 10-25$\%.

\subsubsection{         Test of Cylindrical Symmetry         }
\label{sec.cylsym}

\begin{figure*}
\scl{.85}
\plotone{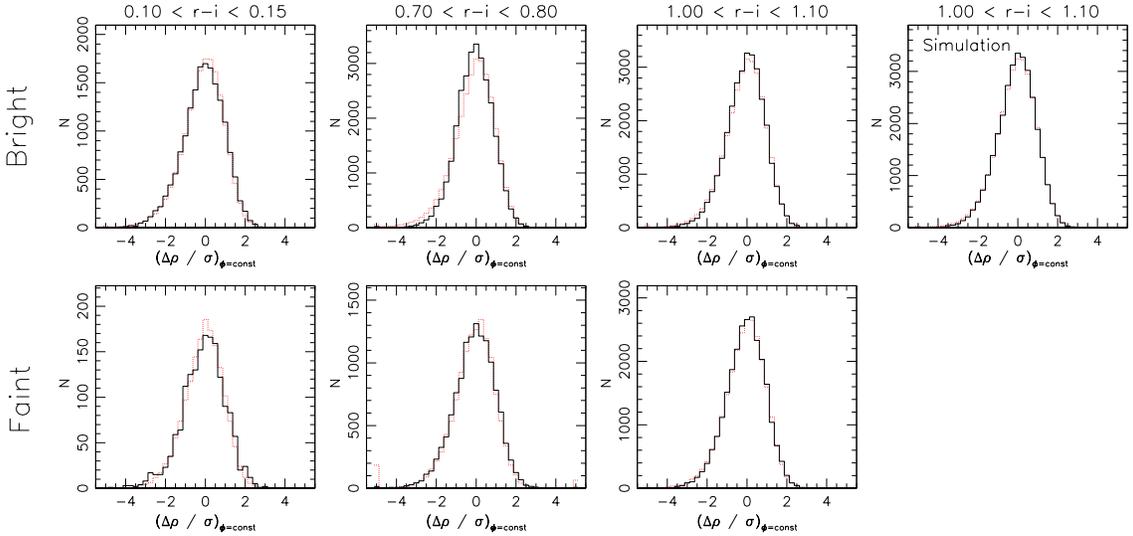}
\caption{Distribution of noise-normalized deviations of density in pixels $(X,Y,Z)$ from 
the mean density measured along their corresponding annuli $(R = \sqrt{X^2 + Y^2}, Z)$. Black 
solid histogram shows the data. Red dotted histogram shows a Poisson noise model. 
Histograms in the top row and
bottom rows have been calculated assuming the bright (Equation~\ref{eq.Mr}) and faint
(Equation~\ref{eq.Mr.faint}) photometric paralax relation, respectively. The rightmost
panel in the top row shows the same distributions derived from a Monte Carlo simulated catalog,
with 25\% unresolved binary fraction, $\sigma_{M_r} = 0.3$ paralax dispersion and SDSS
photometric errors.
\label{fig.phimeans}}
\end{figure*}

In Section~\ref{sec.rzmaps} we argued based on the shapes of isodensity contours in 
Figures~\ref{XYslices1}--\ref{XYslices2b} and in particular in Figure~\ref{figcyl} that once 
the large overdensities are taken out, the Galactic density distribution 
is cylindrically symmetric. Therefore it was justifiable for the purpose of determining the overall 
Galactic stellar number density distribution, to measure the density along the same 
galactocentric annuli $R$ and only consider and model the distribution of stars in two-dimensional 
$R-Z$ plane.

Using Figure~\ref{fig.phimeans} we quantitatively verify this assumption. In the
panels of the top row as solid black histograms we plot the distribution of
\eq{
	\frac{\Delta\rho}{\sigma} = \frac{\rho(R,\phi,Z) - \overline{\rho}(R, Z)}{\sigma_P(R,\phi,Z)} \nonumber
}
for four $r-i$ color bins\footnote{Analogous histograms of other $r-i$ bins share the same 
features.}. This is the difference of the density measured in a pixel at $(R, \phi, Z)$ and the mean density
$\overline{\rho}(R, Z)$ at annulus $(R, Z)$, normalized by the expected Poisson fluctuation
$\sigma_P(R,Z) = \sqrt{N(R,\phi,Z)} / V(R,\phi,Z)$.

The dotted red histogram in the panels shows a Poisson model of noise-normalized deviations 
expected to occur due to shot noise only. If all pixels were well sampled ($N \gtrsim 50$ stars, 
which is not the case here), this distribution would be a $\mu=0$, $\sigma=1$ Gaussian.

The data and the Poisson model show a high degree of agreement. However, in a strict statistical 
sense, for all but the $1.3 < r-i < 1.4$ bin the data and the model are inconsistent with being drawn
from the same distribution at a 5\% confidence level. This is not very surprising, as the 
effects of unresolved multiplicity and other observational errors may modify the residual distribution.
We verify this by examining the same statistic calculated from a Monte Carlo generated catalog 
with 25\% unresolved binary fraction and mixed in photometric error (Table~\ref{tbl.simulations}, fourth row).
The resulting ``observed'' and Poisson model histograms for the simulated catalog are shown 
in the top right-most panel of Figure~\ref{tbl.simulations}. They show the same behavior as seen 
in the data.

One may ask if these distributions could be used to further test (and/or constrain) the photometric 
parallax relations, as under- or overestimating the distances will break the cylindrical 
symmetry of Galaxy and distort the isodensity contours.
The answer is, unfortunately, no. In the bottom row of Figure~\ref{fig.phimeans} we show the 
distributions analogous to those in the top row, but calculated from maps obtained using the 
faint parallax relation (Equation~\ref{eq.Mr.faint}). They are very similar to those in the top row, 
with no deviations that we can conclusively attribute to the change in photometric parallax 
relation, although there is an intriguing slightly better Data-Model agreement in
$0.7 < r-i < 0.8$ color bin for the faint, than for the bright relation. This is initially 
surprising, as one would intuitively expect the erroneous distance 
estimate to map the density from different real Galactocentric annuli ($R, Z$) to the 
same observed ($R_o, Z_o$), and therefore widen the residual distribution. However, this 
effect (as we verified using the Monte Carlo generated catalogs) is indiscernible for the 
combination of density distribution seen in the Milky Way, and the portion in $(R,Z)$ space where 
we have clean data. The region near $R = 0$ where the assumption of cylindrical symmetry is 
most sensitive to errors in distance determination is contaminated by debris from the Virgo 
overdensity, making it unusable for this particular test.

\subsubsection{         Resolved Substructure in the Disk         }
\label{sec.clumpyness}

\begin{figure*}
\scl{.75}
\plotone{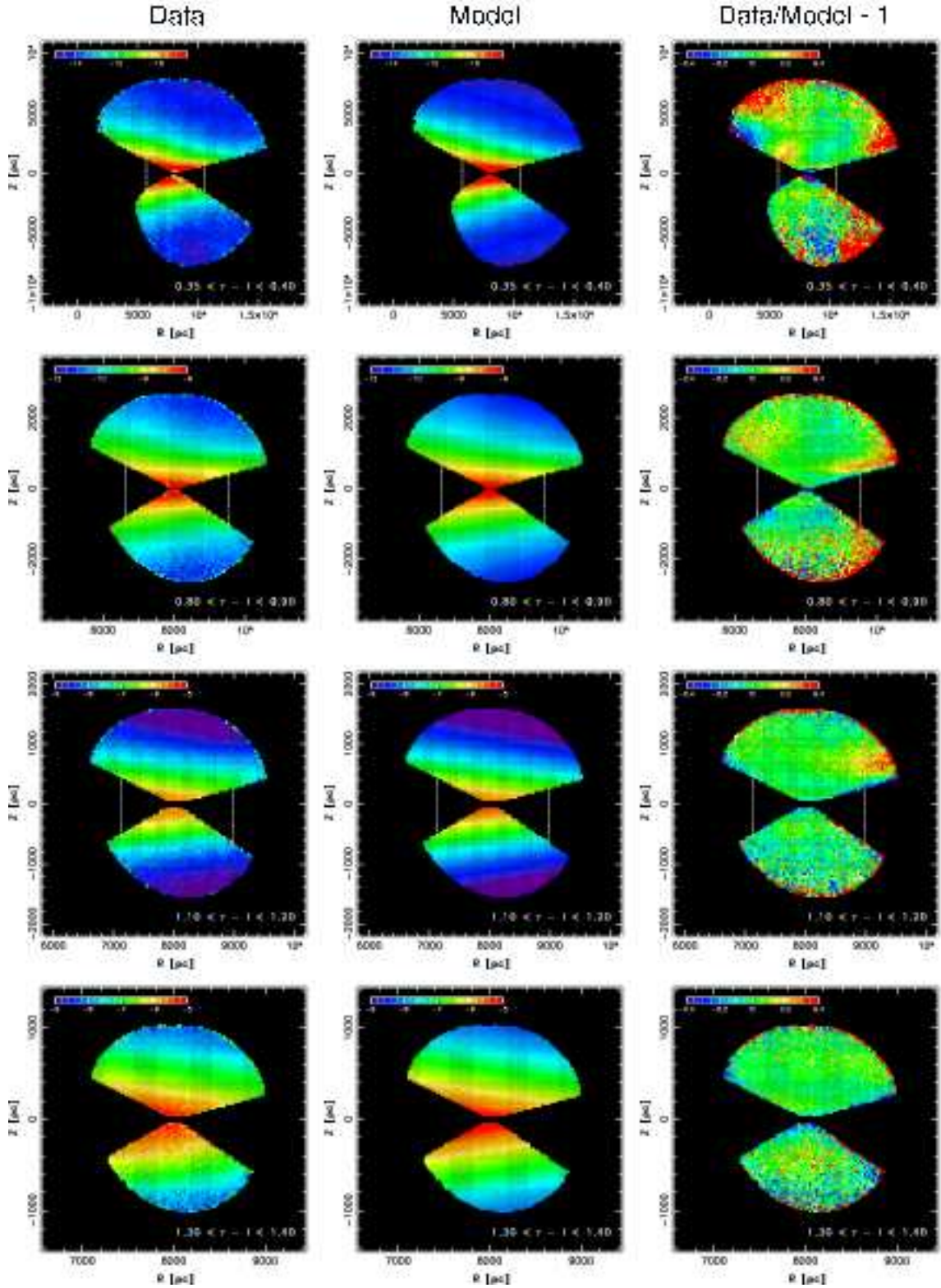}
\caption{Examples of model fits for four color bins, one per each row. Note the different scales.
The left panel of each row shows the data, the middle panel the best-fit model and the right 
panel shows (data-model)
residuals, normalized to the model. The residuals are shown on a linear stretch, from -40\% to
+40\%. Note the excellent agreement of the data and the model for reddest color bins (bottom row),
and an increasing number of overdensities as we move towards bluer bins. In the residuals map
for the $0.35 < r-i < 0.40$ bin (top row) the edges of the Virgo overdensity (top right) and 
the Monoceros stream (left), the overdensity at $(R \sim 6.5, Z \sim 1.5)$ kpc and a small 
overdensity at $(R \sim 9.5, Z \sim 0.8)$ kpc (a few red pixels) are easily discernible. The 
apparently large red overdensity in the south at $(R \sim 12, Z \sim -7)$ kpc
is an instrumental effect and not a real feature.
\label{haloPanels3}}
\end{figure*}

\begin{figure*}
\scl{.75}
\plotone{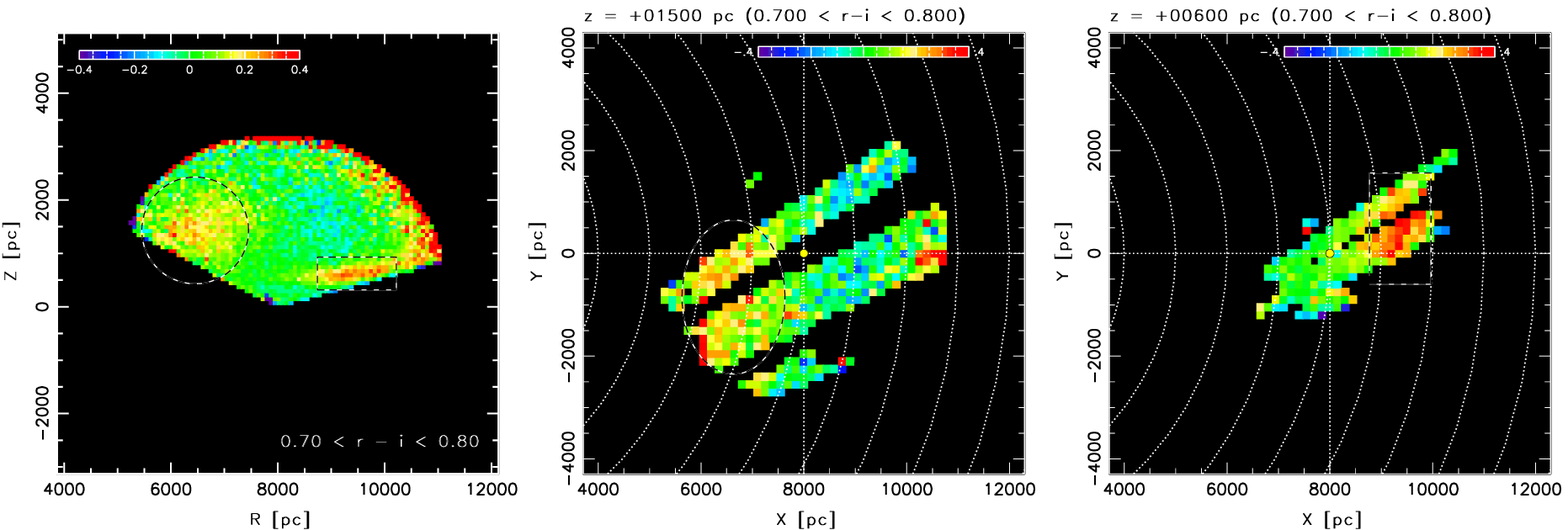}
\caption{Ring-like deviations from the underlying density distribution detected after the
best fit model subtraction. Left panel shows the data-model residuals in the R-Z plane, 
normalized to the model, for the $0.7 < r-i < 0.8$ bin. Two overdensities detected on 
Figure~\ref{haloPanels3} are clearly visible and marked by a dashed circle and rectangle. 
In the $X-Y$ plane, shown in the middle panel, the $R\sim 6.5$~kpc feature reveals itself 
as a ring-like $\sim 20$\% density enhancement over the smooth background at $Z \sim 1.5$~kpc. 
Similarly on the right panel, the $R \sim 9.5$ feature is detectable as a 
strong $\sim 50$\% enhancement in the $Z = 600$~pc slice.
\label{fig.clumps}}
\end{figure*}

The panels in Fig.~\ref{haloPanels3} illustrate the fitting results and the
revealed clumpy substructure. The columns, from left to right, show
the data, the model and the model-normalized residuals. The bottom three rows are results 
of fitting a disk-only model, while the top row also includes a fit for the halo.

While the best-fit models are in good agreement with a large fraction of
the data, the residual maps show some localized features. The most prominent
feature is found at practically the same position ($R\sim$6.5 kpc and
$Z\sim$ 1.5 kpc) in all color bins, and in figure \ref{haloPanels3} is the
most prominent in the top right panel
The feature itself is not symmetric with respect to the Galactic plane, though a
weaker counterpart seems to exist at $Z<0$. It may be connected to the feature observed
by \citet{Larsen96} and \citet{Parker03} in the POSS I survey at 
$20^\circ < l < 45^\circ$, $b \sim 30^\circ$ and which they interpreted at the time as a 
signature of thick disk asymmetry.
We also show it in an $X-Y$ slice on the center panel of Figure~\ref{fig.clumps}, where
it is revealed to have a ring-like structure, much like the Monoceros stream in 
Figure~\ref{XYslices2a}.
Another smaller overdensity is noticeable in all but the reddest Data-Model panel of figure
\ref{haloPanels3} at $R \sim 9.5$ kpc and $Z \sim 0.8$ kpc, apparently extending for
$\sim$1 kpc in the radial direction. When viewed in an $X-Y$ slice, it is also consistent 
with a ring (Figure~\ref{fig.clumps}, right panel); however, due to the smaller 
area covered in $X-Y$ plane, an option of it being a localized clumpy overdensity 
is not possible to exclude.

If this substructure is not removed from the disk as we have done in 
Section~\ref{sec.dataset.preparation}, it becomes a major source of bias in
determination of model parameters. The effect depends on the exact location, size
and magnitude of each overdensity, and whether the overdensity is inside the survey's
flux limit for a particular color bin. For example, the effect of ($R\sim$6.5,
$Z\sim1.5$) overdensity was to increase the scale of the thick disk, while reducing 
the normalization, to compensate for the excess number density at higher $Z$ values.
The effect of $R \sim 9.5$ overdensity was similar, with an additional increase in
scale length of both disks to compensate for larger than expected density at $R > 9$~kpc.
Furthermore, these effects occur only in bins where the overdensities are visible, leading
to inconsistent and varying best-fit values across bins. Removal of the clumps 
from the dataset prior to fitting the models (Section~\ref{sec.dataset.preparation}) 
restored the consistency.

\subsubsection{         Statistics of Best-fit Residuals         }
\label{sec.rhists}

\begin{figure*}
\scl{.85}
\plotone{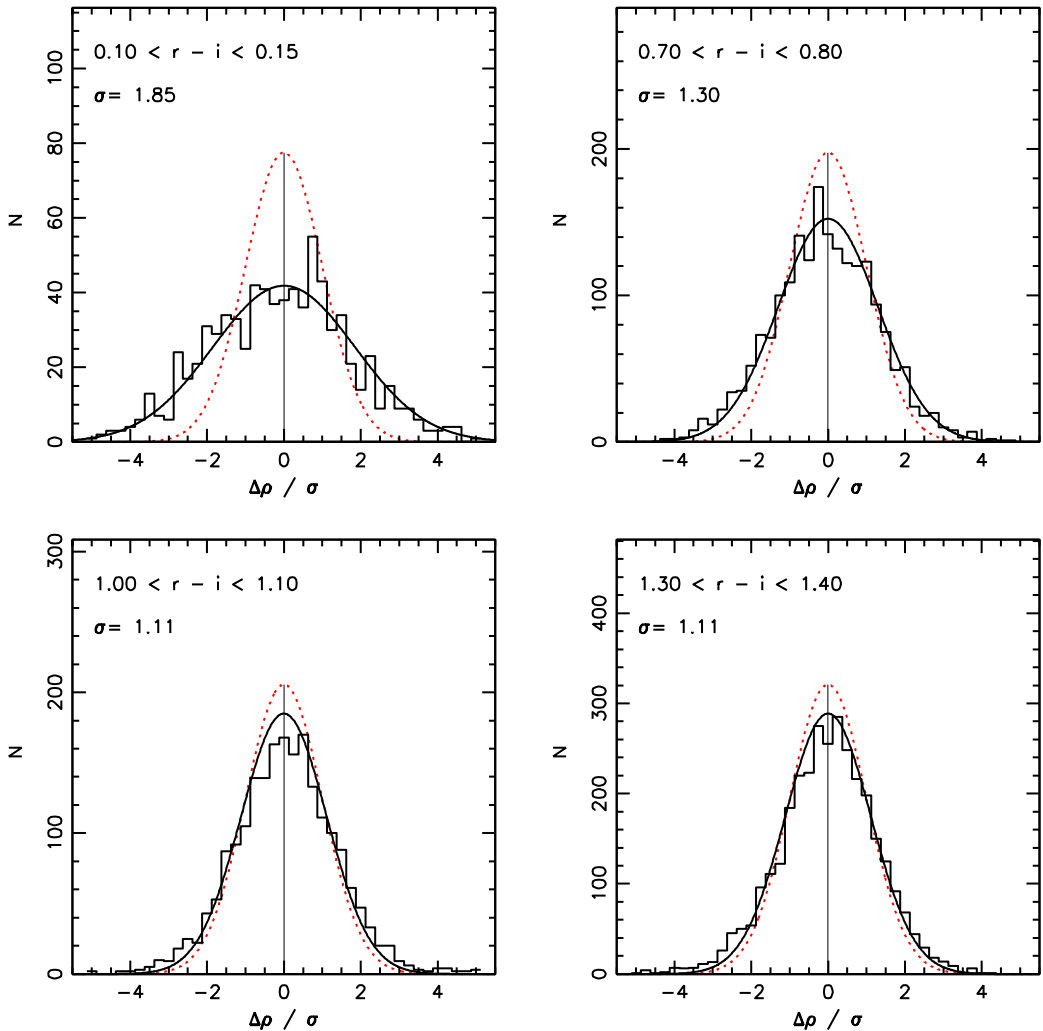}
\caption{Distribution of residuals in $(R, Z)$ plane pixels. The solid black histogram 
shows the data, and the overplotted solid black curve is a Gaussian distribution with
dispersion $\sigma$ determined from the interquartile range of the data. For comparison,
the dotted red line shows a $\sigma=1$ Gaussian, the expected distribution if the
residuals were due to shot noise only.
\label{fig.rhists.rz}}
\end{figure*}

\begin{figure}
\scl{.55}
\plotone{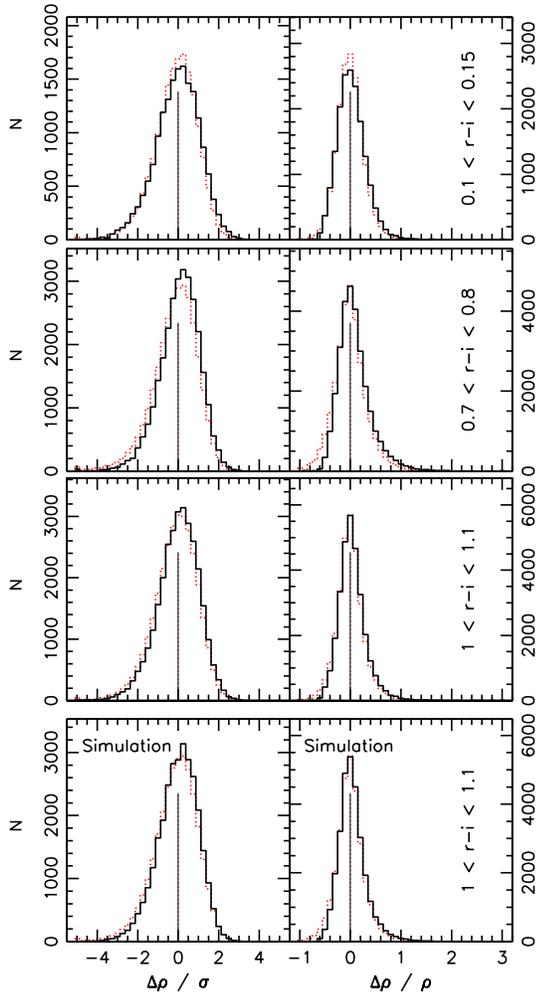}
\caption{Left column: the Poisson-noise normalized distribution of residuals in 
three-dimensional $(X,Y,Z)$ pixels for three representative color bins. Right column: 
model-normalized distribution of residuals in each pixel, (Data - Model) / Model. The 
solid black histograms show the data, while the dotted red histograms show the expectation 
from residuals due to Poisson noise only.
The bottom row shows the same distributions derived from a Monte Carlo simulated catalog,
with 25\% unresolved binary fraction, $\sigma_{M_r} = 0.3$ paralax dispersion and SDSS
photometric errors.
\label{fig.rhists}}
\end{figure}

\begin{figure}
\scl{.85}
\plotone{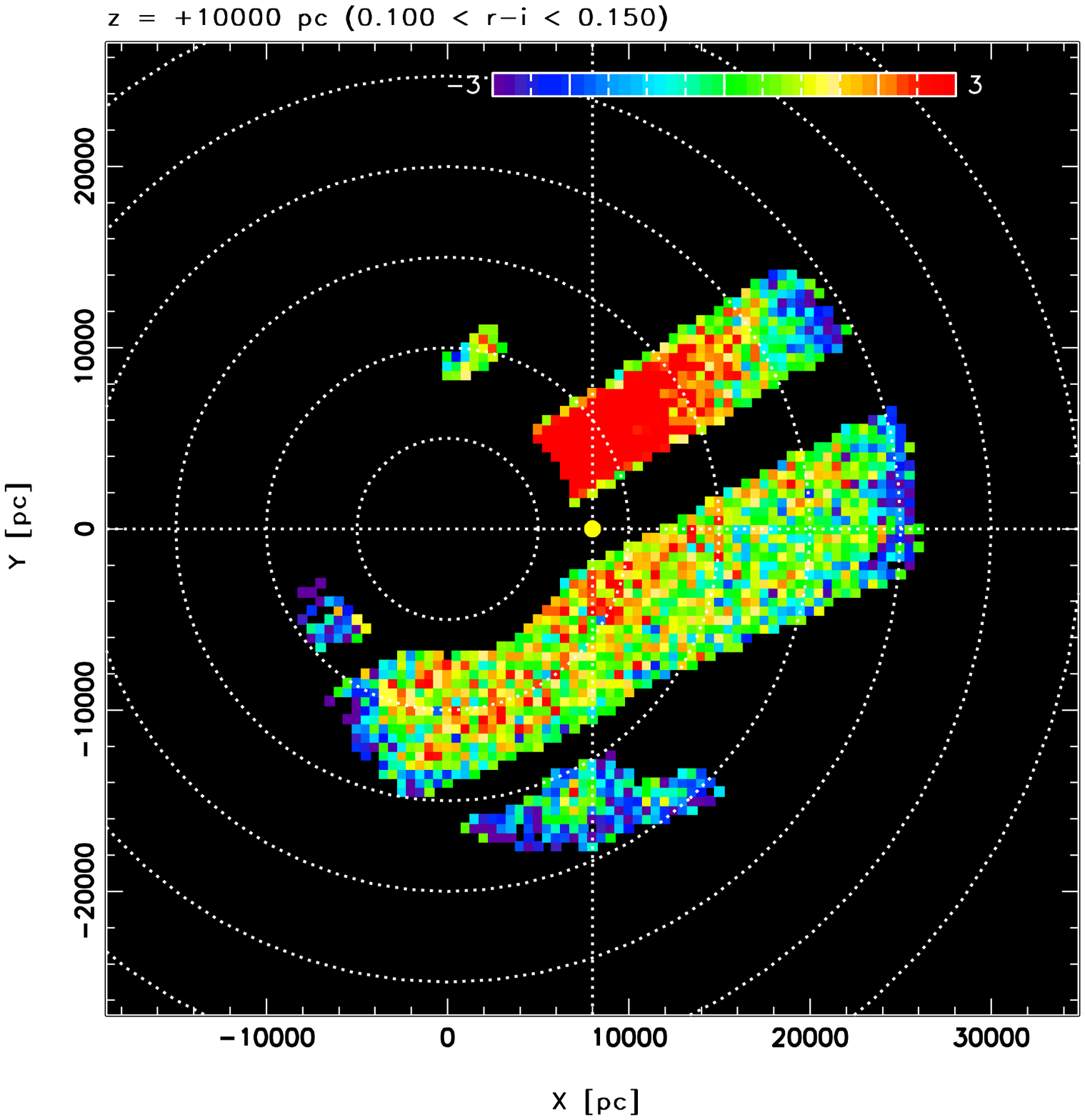}
\caption{A $Z = 10$~kpc $X-Y$ slice of data from the $0.10 < r-i < 0.15$ color bin. Only
pixels with less than 5\% of the density coming from the Galactic disk component are shown. The 
colors encode noise-normalized residuals in each pixel, saturating at $-3\sigma$ (purple) 
and $+3\sigma$ (red). The large red spot in the top part of the figure is due to the Virgo
overdensity (this region was excluded during model fitting; it is shown here for completeness only).
\label{fig.resid.xyslice}}
\end{figure}

The best-fit residuals of data where no apparent substructure was detected
may hold statistical evidence of unrecognized 
low-contrast overdensities and clumpiness. If there are no such features in the data, 
the distribution of residuals will be consistent with a Poisson noise model. Conversely,
if a substantial degree of unresolved clumpiness exists, the distribution of residuals
will be wider and may be skewed compared to the distribution expected from Poisson noise only.

We begin by inspecting the statistics of residuals in $R,Z$ plane, shown in Figure~\ref{fig.rhists.rz}
for four color bins representative of the general behavior. The solid black histogram
shows the distribution of Data-Model deviation normalized by the shot noise. As pixels in
$(R,Z)$ plane are well sampled (typically, $N_{stars} > 50$ in more than 95\% of pixels),
the shot noise induced errors are Gaussian, and the residuals are expected to be normally 
distributed ($N(\mu=0,\sigma=1)$, dotted red curve) for a perfect model fit. The distribution 
seen in the data is wider than the expected Gaussian, by 10\% at the red end 
and 85\% at the blue end.

Two-dimensional $(R, Z)$ plane pixels contain stars from all (potentially distant)
$X-Y$ positions in the observed volume which map to the same $R$, thus making the 
the residuals difficult to interpret. We therefore construct analogous distributions of residuals 
for pixels in 3D $(X, Y, Z)$
space. They are shown in the panels of left column of Figure~\ref{fig.rhists} (solid black histograms).
As this time not all $(X, Y, Z)$ pixels are well sampled, a full Poissonian noise model is
necessary to accurately model the distribution of residuals. We overplot it
on panels of Figure~\ref{fig.rhists} as a dotted red histograms. In the left column of 
the same figure, we also plot the measured distribution of model normalized residuals
(solid black histogram), and the Poisson model prediction for residuals due to shot-noise only 
(dotted red histogram). To judge the effects of observational errors and unresolved
multiplicity, the bottom two panels show the distributions measured from a Monte Carlo 
generated catalog with 25\% unresolved binary fraction and photometric error 
(Table~\ref{tbl.simulations}, fourth row). Comparison of data and Poisson models, and the
observed and simulated distributions, leads us to conclude that across all examined color bins, 
the distribution of deviations is consistent with being caused by shot noise only.

This is in apparent conflict with the analysis of residuals in 2D $(R, Z)$ plane.
The key in reconciling the two is to notice that different spatial scales are
sampled in 3D and 2D case. The 3D analysis samples scales comparable to the pixel size. 
The effective sampling scale is made variable and somewhat larger by the smearing in 
line-of-sight direction due to unrecognized stellar multiplicity, but is still on order 
of not more than a few pixel sizes. On the other hand, the effective scale in 2D is the 
length of the arc over which 3D pixels were averaged to obtain the $R-Z$ maps. This is on order of few 
tens of percent of the faint volume-limiting distance (Table~\ref{tbl.bins}, column $D_{1}$) for each bin.
The deviations seen in 2D maps are therefore indicative of data-model mismatch on large 
scales, such as those due to large scale overdensities or simply due to the mismatch 
of the overall shape of the analytic model and the observed density distribution.

In support of this explanation in Figure~\ref{fig.resid.xyslice} we plot a rainbow-coded 
shot noise normalized map of residuals in pixels at $Z=10$~kpc slice, $0.1 < r-i < 0.15$ color bin.
On large scales a small but noticeable radial trend in the residuals is visible, going from slightly underestimating 
the data (larger median of residuals, more red pixels) at smaller $R$ towards overestimating the data
near the edge of the volume at higher $R$ (smaller median of residuals, more blue pixels). This
trend manifests itself as widening of residual distribution (and increase in $\chi^2$) in
Figure~\ref{fig.rhists.rz}.

The small scale fluctuations are visible as the ``noisiness'' of the data. 
They are locally unaffected by the large-scale trend, and consistent with just Poisson 
noise superimposed on the local density background. If examined in bins along the $R$ direction, 
the large scale trend does leave a trace: the median of residuals is slightly higher than expected 
from the Poisson model at low $R$ and lower than expected at high $R$. But when the residuals of 
all pixels are examined together, this signal disappears as the opposite shifts from lower and 
higher radii compensate for each other. This leaves the residual distribution in 
Figure~\ref{fig.rhists} consistent with being entirely caused by shot-noise.

We conclude from this admittedly crude but nevertheless informative analysis that i) it 
rules out significant clumpiness on scales comparable to the pixel size of each color bin 
ii) demonstrates there are deviations on scales comparable to radial averaging size, indicating 
the functional forms of the model do not perfectly capture the large-scale 
distribution, and iii) shows that these deviations are negligible for the disk and pronounced 
for the halo, pointing towards a need for halo profiles more complicated than a single power
law.

\subsubsection{    Wide Survey Area and Model Fit Degeneracies     }
\label{sec.degeneracies}

\begin{figure}
\scl{.45}
\epsscale{.9}
\plotone{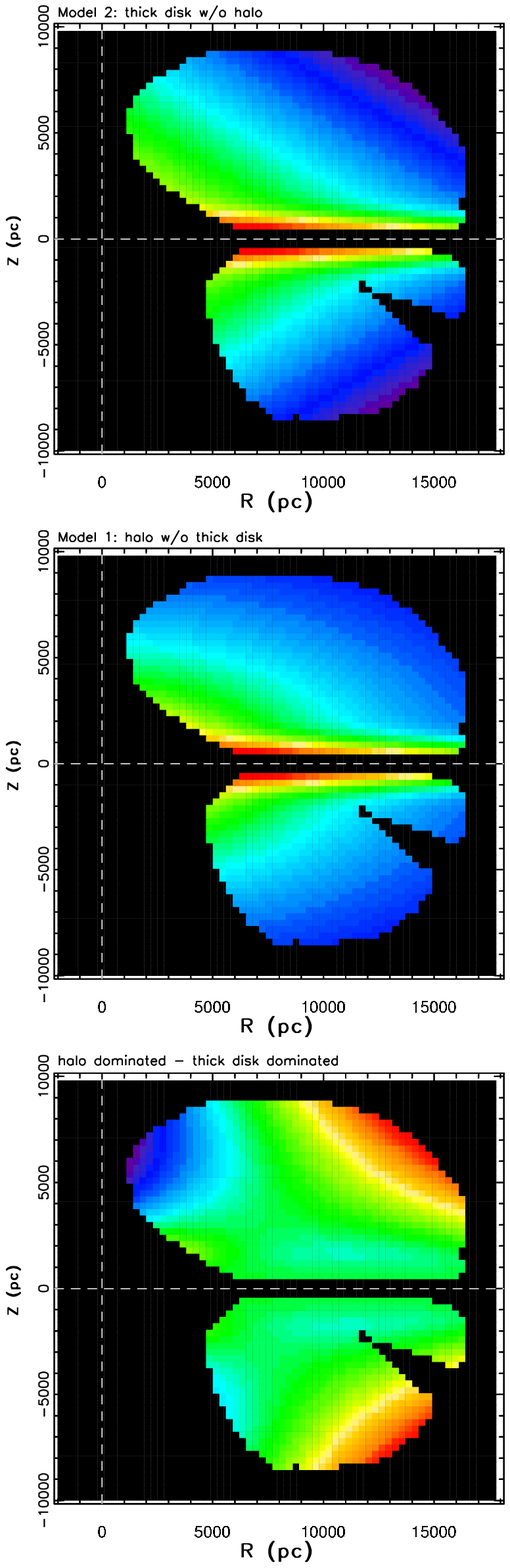}
\caption{An illustration of the degeneracies in fitting models for 
stellar distribution. The top panel shows a thin disk plus thick disk 
model, without any contribution from the halo (volume density on a 
logarithmic stretch, from blue to red, shown only for the regions
with SDSS data), and the middle panel shows a single disk plus 
an oblate halo model. Both models are fine-tuned to produce nearly 
identical counts for $R=8$ kpc and $|Z|<8$ kpc. The bottom panel shows 
the difference between the two models (logarithmic stretch for $\pm$ 
a factor of 3, from blue to red, the zero level corresponds to green color). 
The models are distinguishable only at $|Z|>3$ kpc and $R$ significantly 
different from 8 kpc.
\label{models2D}}
\end{figure}

\begin{figure*}
\scl{1}
\plotone{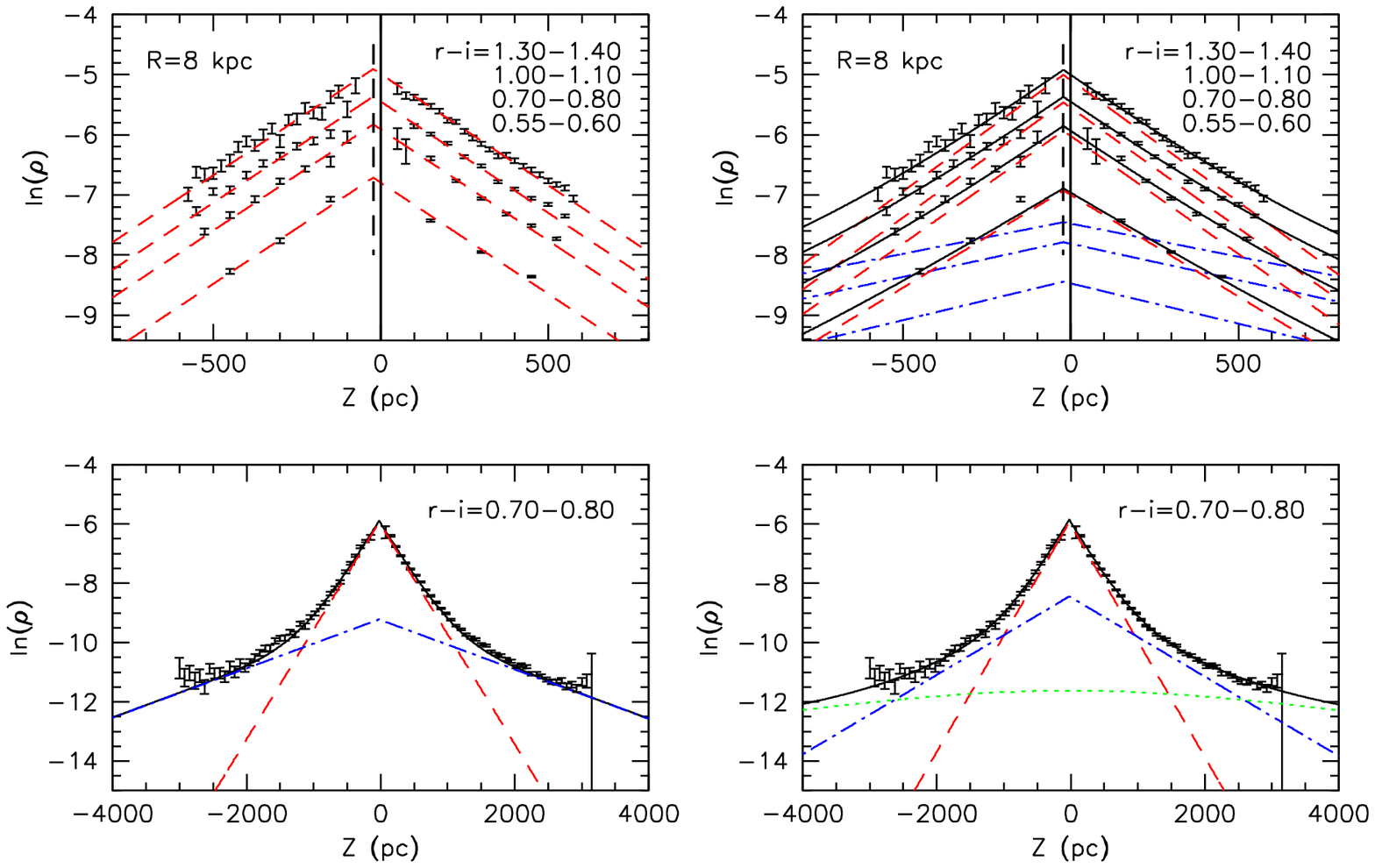}
\caption{
An illustration of degeneracies present in fitting of Galactic models. The two panels in
the left column are the same top two panels of Fig.~\ref{rhoZ}. The panels to the right
show the same data, but are overplotted with best fit models from 
Table~\ref{tbl.bright.individual.fits}. In spite of substantially different best fit
values, the two models are virtually indistinguishable when fitting the $R=8{\rm kpc}, \pm Z$
direction of the data.
\label{rhoZcomp}}
\end{figure*}

In a model with as many as ten free parameters, it is not easy to assess the 
uniqueness of a best-fit solution, nor to fully understand interplay between 
the fitted parameters. We show two illuminating examples of fitting degeneracies.

In Fig.~\ref{models2D} we plot the density distributions for two significantly 
different models: a thin plus thick disk model without a halo, and a single 
disk plus halo model. Despite this fundamental intrinsic difference, it is 
possible to fine-tune the model parameters to produce nearly identical $Z$ 
dependence of the density profiles at $R=8$~kpc. As shown
in the bottom panel, significant differences between these two models
are only discernible at $|Z|>3$~kpc and $R$ significantly different from 
$8$~kpc.

Secondly, in the left column of Fig.~\ref{rhoZcomp} we reproduce the top two panels of 
Fig.~\ref{rhoZ}. The density profile is well described by two exponential disks of scale 
heights $H_1 = 260$ and $H_2 = 1000$ and normalization of 4\%. In the right 
column of the figure we plot the same data, but overplotted with best fit models 
from Table~\ref{tbl.bright.individual.fits}. The scales in this model are $H_1 = 245$ 
and $H_2 = 750$, with thick-to-thin normalization of 13\%, and the bottom right
panel also includes a contribution of the halo. Although significantly different,
the two models are here virtually indistinguishable.

This is a general problem of pencil beam surveys with a limited sky coverage. A
single pencil beam and even a few pencil beams (depending on the quality of the 
data and positioning of the beams) cannot break such model degeneracies. We speculate 
that this in fact is likely the origin of some of the dispersion in disk parameter values
found in the literature (e.g., \citealt{Siegel02}, Table 1; \citealt{Bilir06}).

In our case, while we have not done a systematic search for degenerate models leading
to similar $\chi^2$ given our survey area, we have explored the possibility
by attempting a 100 refits of the data starting with random initial parameter 
values. In case of fits to individual bins, we find local $\chi^2$ minima,
higher by $\sim 20-30$\% than the global minimum, with parameter values
noticeably different from the best fit solution. However, when jointly fitting
all $r - i > 1.0$ color bins, in all cases the fit either fails to converge,
converges to a local $\chi^2$ minimum that is a factor of few higher than
the true minimum (and produces obviously spurious features in maps of residuals), 
or converges to the same best-fit values given in Tables~\ref{tbl.bright.joint}.

SDSS therefore seems largely successful in breaking the degeneracies caused 
by the limited survey area and photometric precision, leaving local departures
from exponential profiles as the main remaining source of uncertainty in
best-fit model parameters.

\subsubsection{     Physical Basis for the Density profile Decomposition into Disks and the Halo    }

\begin{figure}
\scl{.7}
\plotone{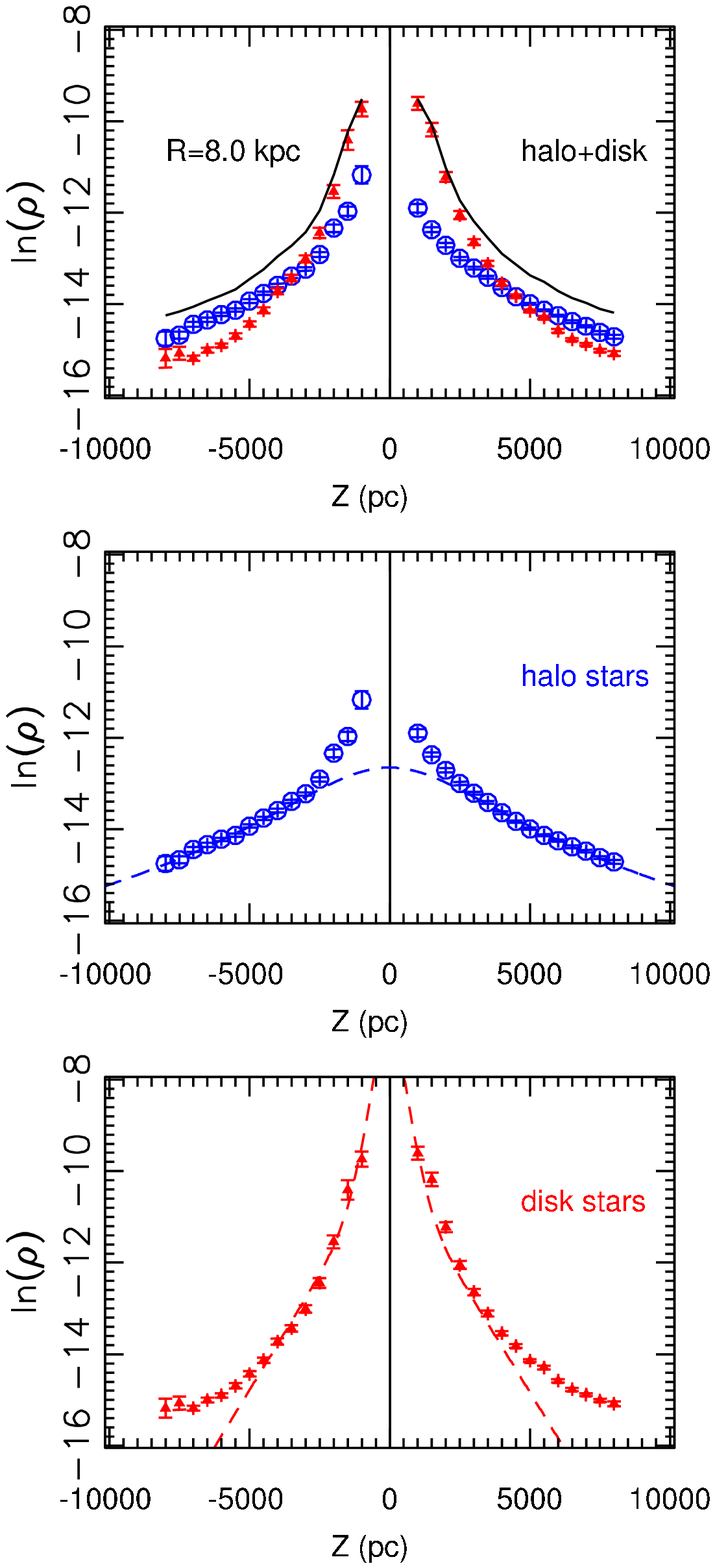}
\caption{The vertical ($Z$) distribution of SDSS stellar
counts for $R=8$ kpc, and $0.10<r-i<0.15$ color bin. Stars
are separated by their $u-g$ color, which is a proxy for
metallicity, into a sample representative of the halo 
stars (low metallicity, $0.60<u-g<0.95$, circles) and a sample  
representative of the disk stars (high metallicity, $0.95<u-g<1.15$, 
triangles). The line in the top panel shows the sum of the counts 
for both subsamples. The counts for each subsample are shown separately 
in  the middle and bottom panels, and compared to the best
fit models, shown as lines. Note that the disk stars are
more concentrated towards the Galactic plane. Due to a simple
$u-g$ cut, both samples are expected to suffer from contamination: 
close to the Galactic plane ($|Z|<$ 2 kpc) the halo sample is 
contaminated by the disk stars, while further away from the plane 
($|Z|>$ 5 kpc) the disk sample is contaminated by halo stars.
\label{rhoZmetal}}
\end{figure}

Although the density profiles shown in bottom right panel of Fig.~\ref{rhoZcomp}
and the bottom panel of Fig.~\ref{rhoZ} appear with high signal-to-noise ratios, 
it may be somewhat troubling that as our range of observed distances expands, 
we need to keep introducing additional components to explain the data. Are these
components truly physically distinct systems, or largely phenomenological 
descriptions with little physical basis?

The question is impossible to answer from number density data alone, and two
companions papers use metallicity estimates (Paper II) and kinematic information (Paper III) 
to address it. Here we only look at a
subset of this question, namely the differentiation between
the disk and halo components. Disk stars (Population I and intermediate Population II) 
have metallicities on average higher by about 1-2 dex than that of the 
halo. Such a large difference in metallicity affects
the $u-g$ color of turn-off stars (e.g., \citealt{Chen01}). 
An analysis of SDSS colors for Kurucz model atmospheres suggests that stars 
at the tip of the stellar locus with $0.7 < u-g \la 1$ necessarily have
metallicities lower than about $-1.0$. These stars also have markedly
different kinematics further supporting the claim that they are halo stars 
(Paper II and III).

We select two subsamples of stars from the $0.10<r-i<0.15$ color bin:
low metallicity halo stars with $0.60<u-g<0.95$, and high metallicity
disk stars with $0.95<u-g<1.15$. This separation is of course only
approximate and significant mixing is expected both at the faint 
end (disk stars contaminated by the more numerous halo stars) and 
at the bright end (halo stars contaminated by the more numerous disk stars).
Nevertheless, the density profiles for these two subsamples,
shown in Fig.~\ref{rhoZmetal}, are clearly different. In particular,
the disk profile is much steeper, and dominates for $Z\la3$ kpc, while
the halo profile takes over at larger distances from the Galactic plane.
This behavior suggests that the multiple components visible in the 
bottom panel in Fig.~\ref{rhoZ} are not an over-interpretation of
the data.

In addition to supporting a separate low-metallicity halo component, this 
test shows that a single exponential disk model is insufficient to explain 
the density profile of high-metallicity stars. This is ordinarily remedied
by introducing the thick disk. However, with only the data presented here, 
we cannot deduce if the division into thin and thick disk has a physical 
basis or is a consequence of our insistence on exponential functions to 
describe the density profile.

\subsubsection{         The Corrected Best Fit Parameters          }
\label{sec.bestfit}

\begin{figure*}
\plotone{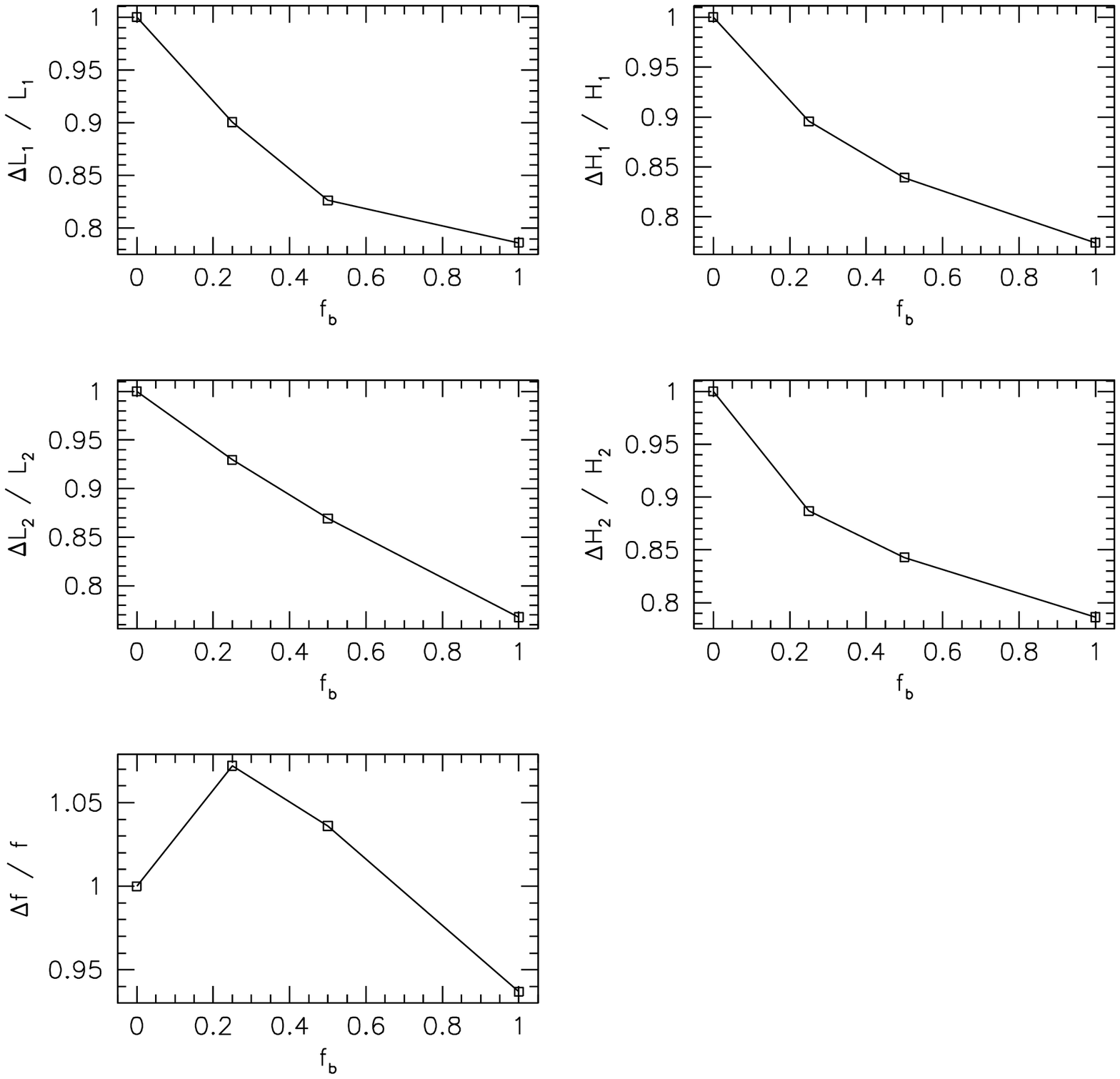}
\caption{Effect of unrecognized binarity on fits of model parameters, derived from 
simulations listed in Table~\ref{tbl.simulations}. Each of the panels shows the 
change of a particular model parameter when a fraction $f_b$ of observed ``stars'' are 
unrecognized binary systems.
\label{fig.binaryfitsplot}}
\end{figure*}

\begin{deluxetable*}{cccc}
\tablewidth{5in}
\tablecolumns{4}
\tablecaption{The Galactic Model\label{tbl.finalparams}}
\tablehead{
	\colhead{Parameter} & \colhead{Measured} & 
	\colhead{Bias-corrected Value} & \colhead{Error estimate}
}
\startdata 
$Z_0$  &  25		& \nodata	& $20$\%	\\
$L_1$  &  2150		& 2600		& $20$\%	\\
$H_1$  &  245		& 300		& $20$\%	\\
$f$    &  0.13		& 0.12		& $10$\%	\\
$L_2$  &  3261		& 3600		& $20$\%	\\
$H_2$  &  743		& 900		& $20$\%	\\
$f_h$  &  0.0051	& \nodata	& $25$\%	\\
$q$    &  0.64		& \nodata	& $\la0.1$ 	\\
$n$    &  2.77		& \nodata	& $\la0.2$      \\ 	
\enddata 
\tablecomments{
Best-fit Galactic model parameters (see eqs.~\ref{galModel}--\ref{haloModel}), as
directly measured from the apparent number density distribution maps (2$^{\rm nd}$ column)
and after correcting for a 35\% assumed binary fraction and Malmquist bias due to
photometric errors and dispersion around the mean of the photometric paralax relation 
(3$^{\rm rd}$ column).}
\end{deluxetable*}

In Section~\ref{sec.modelfit}, we have used two samples of stars to fit the 
parameters of the disk:
the $1.0 < r-i < 1.4$ sample of M dwarfs, and $0.65 < r-i < 1.0$ sample
of late K / early M dwarfs. Best fit results obtained from the two samples 
are very similar, and consistent with the stars being distributed in
two exponential disks with constant scales across the spectral types
under consideration.

The fit to $0.65 < r-i < 1.0$ sample required an
addition of a third component, the Galactic halo. This, combined with
the photometric parallax relations that are inappropriate for low metallicity 
stars in this color range, may bias the determination of thick disk parameters. For
example, while the measured scale height of the thick disk in $0.65 < r-i < 1.0$ range
is $\sim 10$\% lower than in $1.0 < r-i < 1.4$ range, it is difficult
to say whether this is a real effect, or interplay of the disk and
the halo.

Furthermore, we detected two localized overdensities in the thick disk region
(Section~\ref{sec.clumpyness}). While every effort was made to remove them from
the data before fitting the model, any residual overdensity that was not 
removed may still affect the fits. If this is the case, the $0.65 < r-i < 1.0$ bins 
are likely to be more affected than their redder counterparts, being that they cover 
a larger volume of space (including the regions where the overdensities were found).

For these reasons, we prefer the values of disk parameters as determined from
$1.0 < r-i < 1.4$ sample, as these are a) unaffected by the halo and b) least
affected by local overdensities.

Other dominant sources of errors are (in order of decreasing importance) i)
uncertainties in absolute calibration of the photometric parallax relation, 
ii) the misidentification of 
unresolved multiple systems as single stars, and iii) Malmquist bias
introduced by the finite width of $M_r(r-i)$ relation. Given the currently 
limited knowledge of the true photometric parallax relation (Figure~\ref{fig.Mr}), 
there is little one can do but try to pick the best one consistent with the existing data, and
understand how its uncertainties limit the accuracy of derived parameters.
Out of the two relations we use (bright, eq.~\ref{eq.Mr}, and faint, 
eq.~\ref{eq.Mr.faint}), we prefer the bright normalization as it is
consistent with the kinematic data (Paper III) and the analysis done with
wide binary candidates (Section~\ref{sec.widebinaries}) shows its shape to be correct to 
better than 0.1mag for $r-i > 0.5$. If we are mistaken, as discussed in 
Section~\ref{sec.systematics}, errors in $M_r$ of $\Delta M_r = \pm 0.5$ 
will lead to errors of $20-30$\% in parameter estimation. Given 
Figure~\ref{fig.Mr} and the analysis of wide binary candidates in Section~\ref{sec.widebinaries}
we believe this to be the worst case scenario, and 
estimate that the error of each scale parameter is unlikely to be 
larger than $\pm 20$\%.

The dependence of best-fit parameters derived from mock catalogs on multiplicity (binarity)
is shown in Figure~\ref{fig.binaryfitsplot}. The challenge in correcting for multiplicity 
is knowing the exact fraction of observed ``stars'' which are unresolved multiple systems. 
While it is understood that a substantial  fraction of Galactic field stars are in 
binary or multiple systems, its exact value, dependence on spectral type, population, and other
factors is still poorly known. Measurements range from 20\% for late type 
(L, M, K dwarfs -- \citealt{Reid06}; \citealt{Reid97}; \citealt{Fischer92})
to upward of 60\% for early types (G dwarfs; \citealt{Duquennoy91}). The
composition (mass ratios) of binaries also poorly constrained, but appears to
show a preference towards more equal-mass companions in late spectral
types (\citealt{Reid06}, Fig.~8). Given our least biased disk parameters were derived from 
from the M-dwarf sample ($r-i > 1.0$), we choose to follow \citet{Reid97}
 and adopt a binary fraction of 35\%.

We accordingly revise $L_1, H_1$ and $H_2$ upwards by 15\% and $L_2$ by 10\% 
to correct for multiplicity (see Figure~\ref{fig.binaryfitsplot}). We further include
additional 5\% correction due to Malmquist bias (Section~\ref{sec.malmquist.effects}), 
and for the same reason correct the density normalization by -10\%. The final values of 
measured and corrected parameters are listed in Table~\ref{tbl.finalparams}.

\section{                              The Virgo Overdensity               }
\label{vlgv}

The $X-Y$ projections of the number density
maps at the heights above 6 kpc from the Galactic plane show a strong deviation from expected 
cylindrical symmetry. In this Section we explore this remarkable feature in more detail. We refer 
to this feature as \emph{``the Virgo overdensity''} because the highest detected overdensity 
is in the direction of constellation Virgo, but note that the feature is
detectable over a thousand square degrees of sky.

\subsection{             The Extent and Profile of the Virgo overdensity     }

\begin{figure*}
\plotone{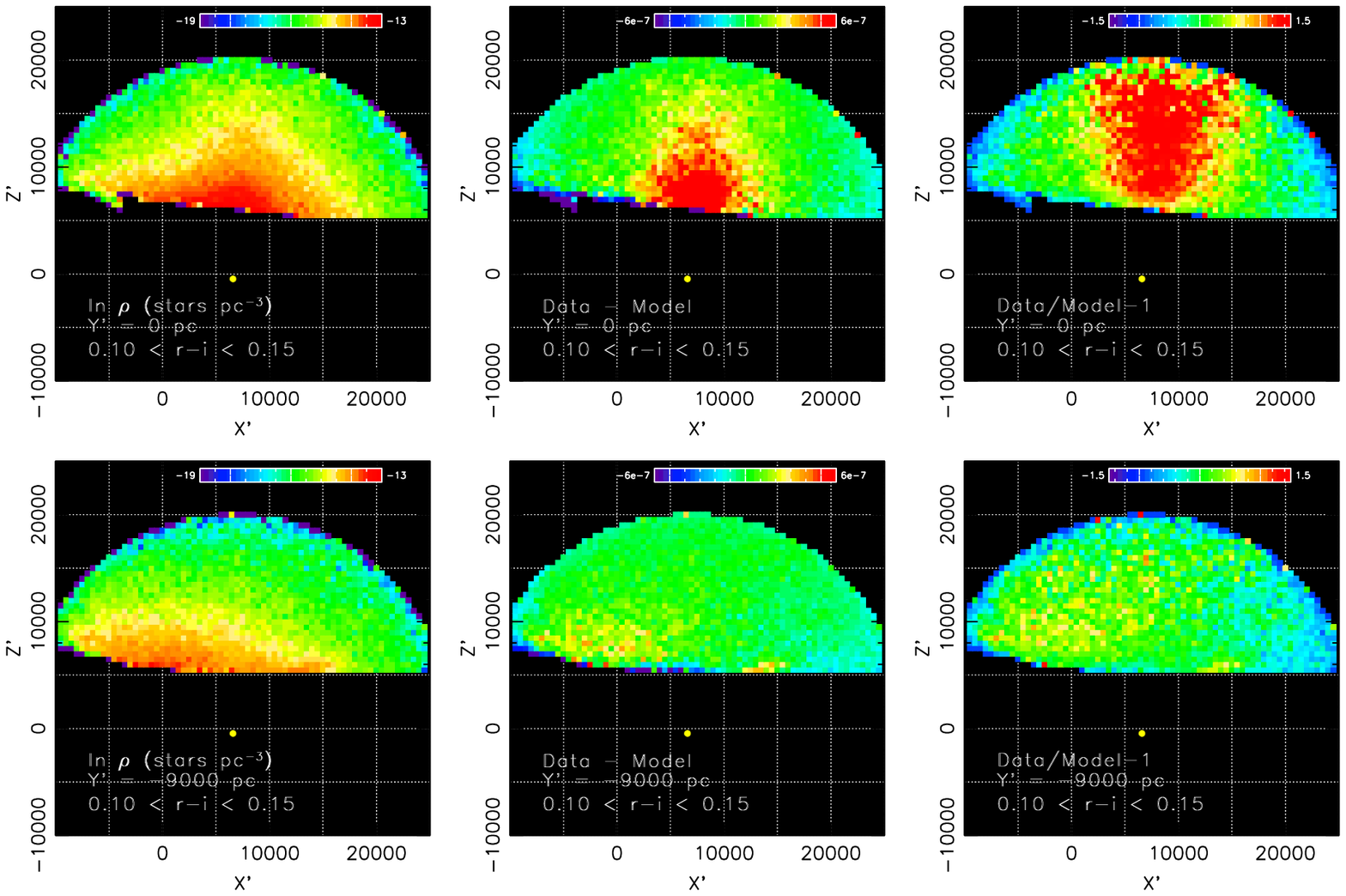}
\caption{The top left panel shows the distribution of stellar number density similar to that in 
Fig.~\ref{RZmedians}, except that here we only show the data from a 
narrow $Y'=0$ slice in a $X', Y', Z'$ coordinate system defined by rotating 
the $X, Y, Z$ galactocentric system counterclockwise by $\phi = 30^\circ$ around the $Z$ axis. In these 
coordinates, the $Y'=0$ plane cuts vertically through the center of the Virgo overdensity.
The top middle panel shows {\it the difference} of the observed density and a best-fit model 
constrained using the data from the $Y<0$ region. The right panel show the same difference 
but {\it normalized to the model}. The bottom panels display analogous slices taken 
at $Y' = -9$~kpc. Compared to the top row, they show a lack of any discernable substructure.
\label{vlgvPanels2}}
\end{figure*}

\begin{figure*}
\scl{.85}
\plotone{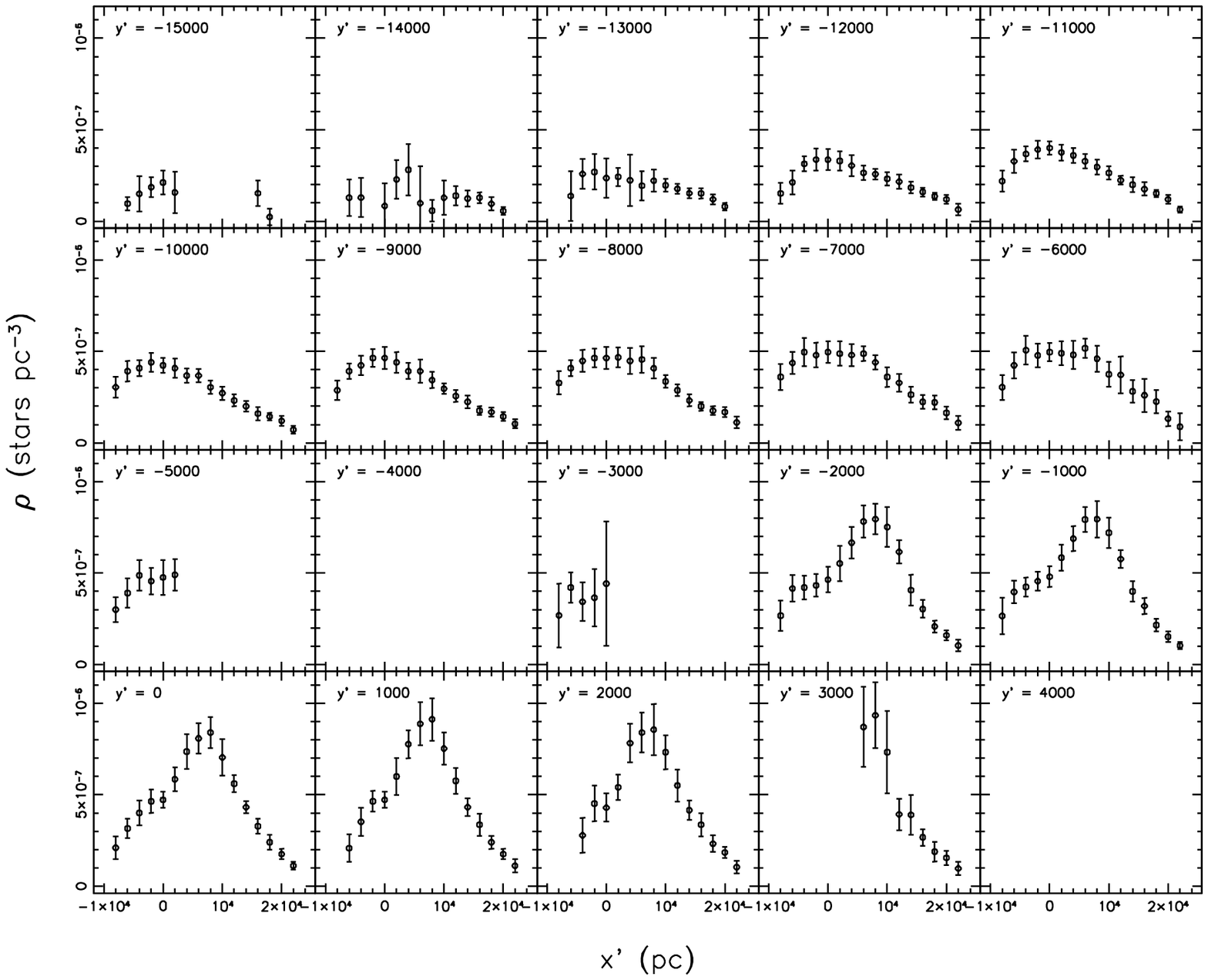}
\caption{The distribution of stellar number density
for $0.10 < r-i < 0.15$ color bin at $Z=10$~kpc above the Galactic plane. Each
panel shows a narrow $Y'$ crossection in coordinate system defined by $\phi=30^\circ$ 
(see the caption of Figure~\ref{vlgvPanels2}). Note a clear and strong density excess
around $X' \sim 8$~kpc in $Y' > 0$ panels, coming from the Virgo overdensity.
\label{rhoRS}}
\end{figure*}

To quantify the extent and profile of the Virgo overdensity, we consider the data in an $X'-Z$ plane,
perpendicular to the Galactic symmetry axis, and is rotated from the $X-Z$ plane by $\phi=30^\circ$ clockwise 
around the $\hat{Z}$ axis. In Figures~\ref{XYslices1}, this plane would be seen edge on, as a
straight line at a $30^\circ$ angle from the $X$ axis, passing through the Galactic center and
the approximate center of the Virgo overdensity. Note that in this plane the distance measured along 
the $X'$ axis is just the cylindrical galactocentric radius $R$.

In the top left panel of Figure~\ref{vlgvPanels2} we show the corresponding 
number density map for the bluest color bin. Isodensity contours show a significant deviation 
from the expected monotonic decrease with $X' (=R)$. 
Instead, they reveal the existence of an overdense region around $X' \sim$ 7-8 kpc 
and $Z \sim 10$~kpc. This overdensity is also visible in the density profiles at $Z = 10$~kpc 
above the plane, shown in $Y' > -3$~kpc panels of Fig.~\ref{rhoRS}. As discernible from these 
figures, the Virgo overdensity is responsible for at least a factor of 2 number density 
excess at $Z=10$~kpc.

To analyze this feature in more detail, we subtract a best-fit Galactic model from the data shown
in the top right panel of Figure~\ref{vlgvPanels2}. We first fit a model described by 
Equations~\ref{galModel}--\ref{haloModel} to the observations having $Y<0$ (or equivalently,
$180^\circ < l < 360^\circ$). As evidenced by Fig.~\ref{XYslices1}, this region does not 
seem significantly affected by the overdensity. We show the difference of the data from
top right panel of Figure~\ref{vlgvPanels2} and the so obtained model in the top 
middle panel of the same figure. The top right panel shows the same difference but 
normalized to the model.

The model-normalized map reveals much more clearly the extent and location of the 
overdensity. A significant density excess (up to a factor of 2) exists over the entire sampled 
range of $Z$ ($6 < Z/{\rm kpc} < 20$). Importance of the overdensity, relative to the 
smooth Milky Way halo background, increases as we move away from the Galactic plane.
This increase is however mainly due to a fast power-law decrease of the number 
density of the halo, which causes the increase in Virgo-to-MW ratio. The number density of stars 
belonging to the overdensity actually increases \emph{towards} the Galactic plane, 
as seen in the top middle panel.

For comparison, analogous density and residual plots from a parallel plane at $Y'=-9$ kpc 
is shown in the bottom row of Figure~\ref{vlgvPanels2}. 
These show no large scale deviations from the model. The density contours rise smoothly 
and peak near $X' = 0$, the point closest to the Galactic center. The same is seen
in $Y' < -5$~kpc slices of Figure~\ref{rhoRS}.

Because no local maximum of the overdensity is detected as $Z$ approaches the observation 
boundary at $Z=6$ kpc, with the data currently available we are unable to quantify its 
true vertical ($Z$) extent. It is possible that it extends all the way into the Galactic 
plane and, if it is a merging galaxy or a stream, perhaps even to the southern Galactic hemisphere.
In the direction of Galactic radius, the Virgo overdensity is detected in the 
$2.5 < X'/{\rm kpc} < 12.5$ region. The $X'$ position\footnote{Note that in this $Y'=0$ plane 
$X' \equiv R$, the galactocentric cylindrical radius.} of maximum density appears to shifts slightly 
from $X' \sim$6 kpc at $Z=6$ kpc to $X' \sim$7 kpc at $Z=15$ kpc. 
The width (``full-width at half density'') decreases by a factor of $\sim 2$ as $Z$ increases 
from 6 to 20 kpc. While not a definitive proof, these properties are consistent with a 
merging galaxy or stream. 

The thickness of the overdensity in the direction perpendicular to the plane of the image in 
Figure~\ref{vlgvPanels2} (the $Y'$ direction) is no less than $\sim 5$~kpc. As in the case of the 
$Z$ direction, the true extent remains unknown because of the current data availability. Note that
the size of the overdensity seen in the maps in the direction of the line of sight towards the
Sun is a combination of true size and the smearing induced by the photometric measurement and
parallax errors (Fig.~\ref{magerr2}) and (most significantly) the effects of unrecognized 
stellar multiplicity. The true line of sight extent is therefore likely smaller, by at least 30-35\%.

\subsection{              Direct Verification of the Virgo Overdensity  }

\begin{figure*}
\plotone{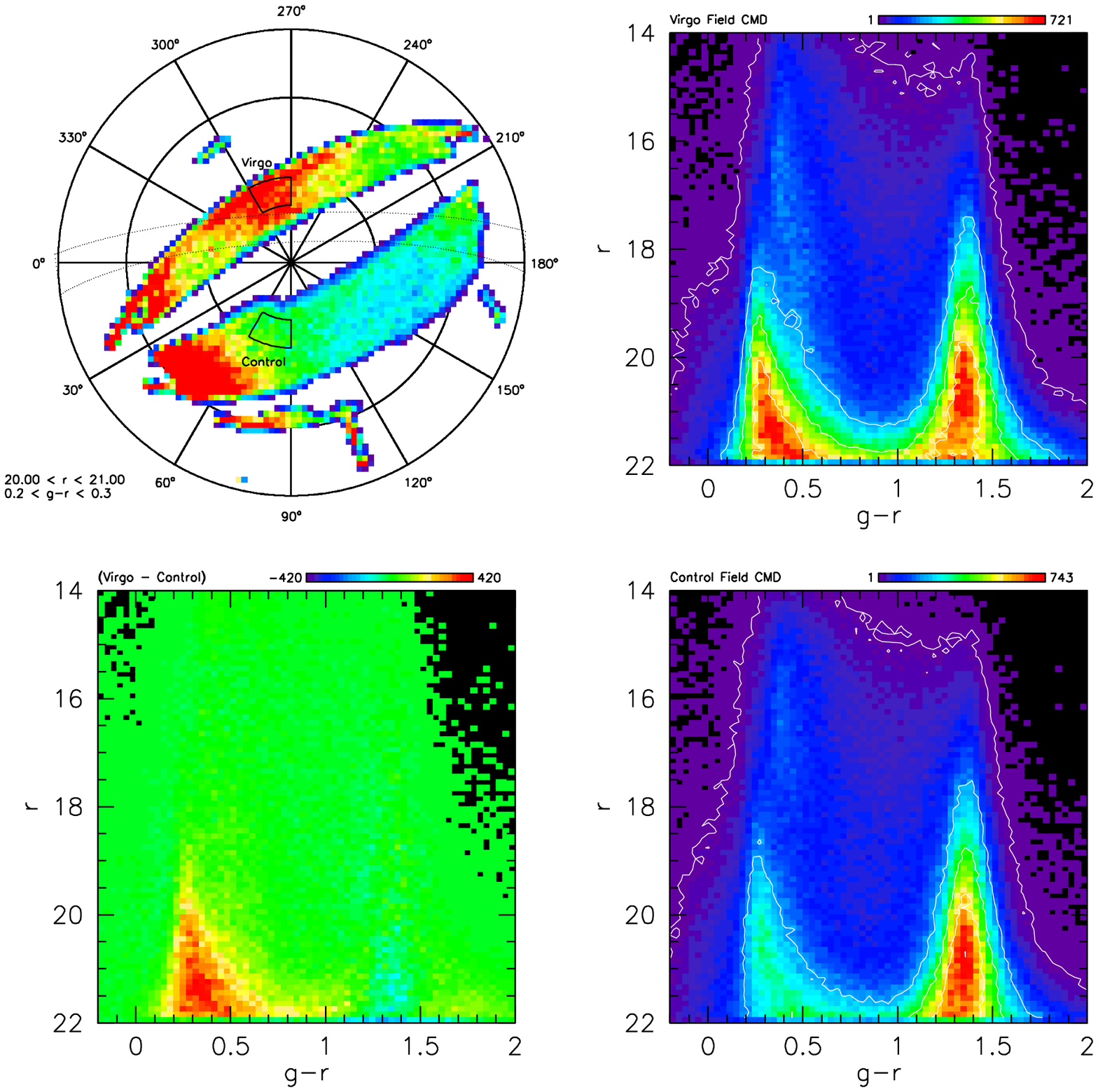}
\caption{The top left panel shows the sky density of stars with $b>0^\circ$, 
$0.2 < g-r < 0.3$ and $20 < r < 21$ in the Lambert projection (concentric 
circles correspond to constant Galactic latitude; equal area corresponds 
to equal solid angle on the sky) of Galactic coordinates (the north Galactic 
pole is in the center, $l$=0 is towards the left, and the outermost circle is $b=0^\circ$).
The number density is encoded with a rainbow color map and increases from blue to red. 
Note that the sky density
distribution is {\it not} symmetric with respect to the horizontal $l=0,180$ line. 
When the stellar color range is sufficiently red (e.g. $0.9 < g-r < 1.0$), this 
asymmetry disappears (not shown). The two right panels show the Hess diagrams for 
two 540~deg$^2$ large regions towards $(l=300^\circ,b=60^\circ, top)$ and $(l=60^\circ,
b=60^\circ, bottom)$, marked as polygons in the top left panel. The bottom left 
panel shows the difference of these Hess diagrams -- note the strong statistically 
significant overdensity at $g-r\sim0.3$ and $r\ga20$. The pixel size in
each of the three Hess diagrams is $(d(g-r), dr) = (0.033, 0.1)$.
\label{mosaic}}
\end{figure*}

\begin{figure}
\plotone{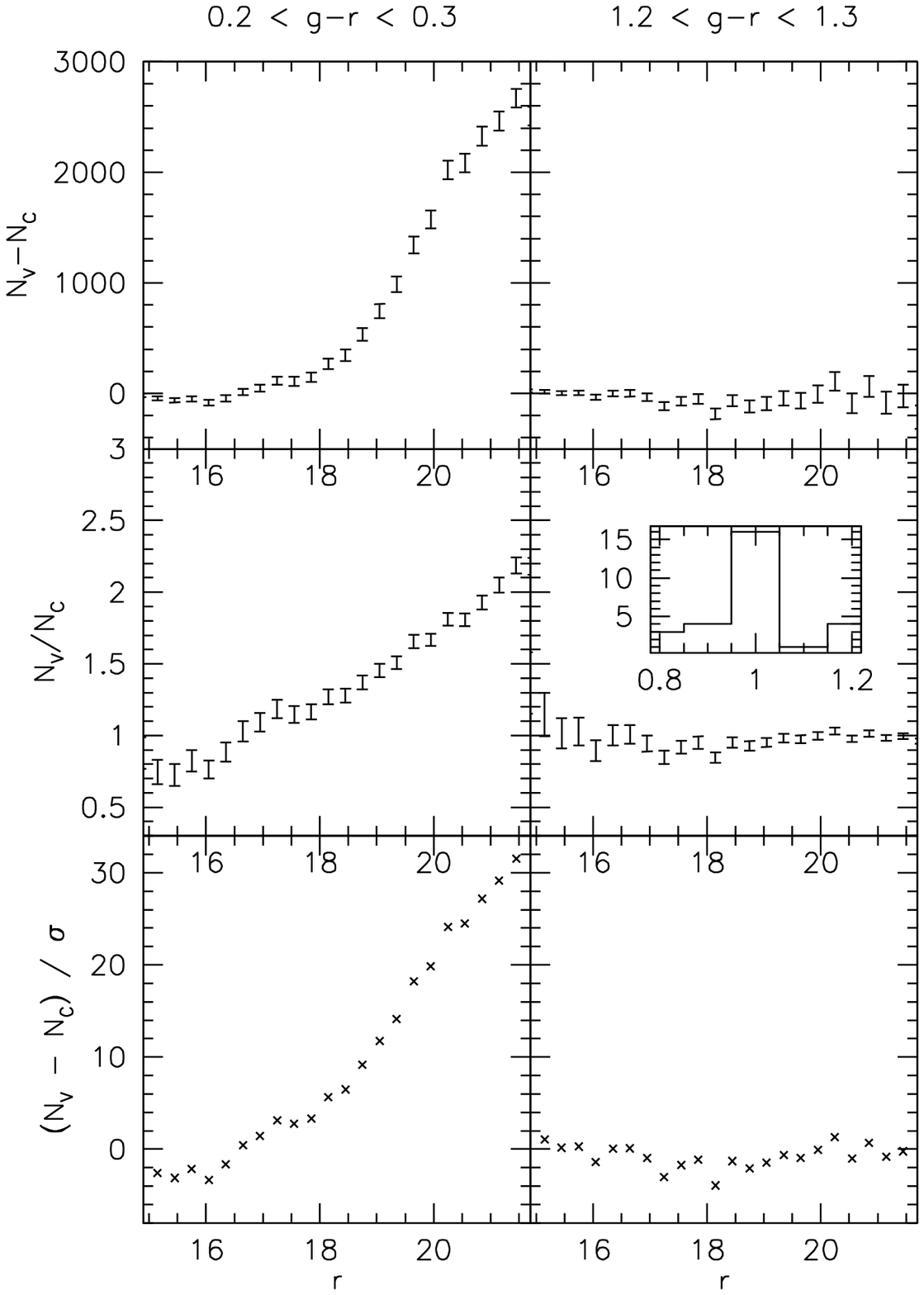}
\caption{Quantitative analysis of the Hess diagram difference shown in the bottom left
panel in Fig.~\ref{mosaic}. The left column corresponds to the color bin $0.2 < g-r < 0.3$
that is representative of the Virgo overdensity, and the right column is a control sample 
with stars satisfying $1.2 < g-r < 1.3$. The top panels show the counts difference as a 
function of apparent magnitude, and the middle panels shows the counts ratio. The inset 
in the middle right panel shows a histogram of the counts ratio for $r<21.5$. The bottom 
panels show the counts difference normalized by the expected Poisson fluctuations. 
Note that for red stars the counts are indistinguishable, while for blue stars the 
difference is highly statistically significant. 
\label{cmdcuts}}
\end{figure}

Significant data processing was required to produce maps such as the one revealing 
the Virgo overdensity (e.g. the top panels in Fig.~\ref{vlgvPanels2}). In order
to test its existence in a more direct, and presumably more robust, way, we examine
the Hess diagrams constructed for the region of the sky that includes the maximum overdensity,
and for a control region that appears unaffected by the Virgo feature. The boundaries
of these two regions, which are symmetric with respect to the $l=0$ line, the 
corresponding Hess diagrams, and their difference, are shown in Fig.~\ref{mosaic}.

The top left panel of Fig.~\ref{mosaic} shows the northern (in Galactic coordinates) 
sky density of stars with $0.2 < g-r < 0.3$ and $20 < r < 21$ in the Lambert equal area
projection of Galactic coordinates (the north Galactic pole is in the center, 
$l$=0 is towards the left, and the outermost circle is $b=0^\circ$). This map projection does not
preserve shapes (it is not conformal, e.g. \citealt{Gott05}), but it preserves areas - the area of each
pixel in the map is proportional to solid angle on the sky, which makes it particularly suitable
for study and comparison of counts and densities on the celestial sphere.
The color and magnitude constraints select stars in a $D \sim 18$ kpc 
heliocentric shell, and can be easily reproduced using the publicly available SDSS database. 
The Virgo overdensity is clearly visible even with these most basic color and magnitude cuts, 
and extends over a large patch of the sky, roughly in the $l=300^\circ, b=65^\circ$ direction.
The overall number density distribution is clearly {\it not} symmetric with respect to
the horizontal $l=0,180$ line. For example, in a thin shell at $r \sim 21$mag there are 
$1.85\pm 0.03$ times
more stars in the $l=300^\circ, b=65^\circ$ direction,
than in the corresponding symmetric ($l=60^\circ, b=65^\circ$) direction, a $\sim28\sigma$ deviation
from a cylindrically symmetric density profile. When the color range is sufficiently red (e.g. $0.9
< g-r < 1.0$), and in the same magnitude range, the asymmetry disappears (not shown). These
stars have a smaller absolute magnitude, are therefore much closer, and do not go far enough
to detect the overdensity.

The two right panels in Fig.~\ref{mosaic} show the Hess diagrams for two 540~deg$^2$ large
regions towards $(l=300^\circ,b=60^\circ)$ and $(l=60^\circ,
b=60^\circ)$, and the bottom left 
panel shows their difference. The difference map reveals a strong
overdensity at $g-r\sim0.3$ and $r\ga20$. A more quantitative
analysis of the Hess diagram difference is shown in Fig.~\ref{cmdcuts}. 
For red stars the counts in two regions are indistinguishable, while for blue stars 
the counts difference is highly statistically significant. There is no indication 
for a turnover in blue star number count difference, which strongly suggests 
that the Virgo overdensity extends beyond the SDSS faint limit.
We conclude that the Hess diagram analysis robustly proves the existence
of a significant star count overdensity towards $l=300^\circ, b=65^\circ$, from
approximately $r \sim 18$ mag to $r \sim 21.5$ mag.

From the diagram in bottom left panel of Fig.~\ref{mosaic}, a crude
estimate of the surface brightness of the overdensity can be made by summing up the fluxes of all
stars in the CMD and dividing the total flux by the area observed. To isolate the overdensity,
we only count the fluxes of stars satisfying $0.2 < g-r < 0.8$ and $18 < r < 21.5$. {\it This will
effectively be a lower limit,} because we will miss stars dimmer than the limiting magnitude ($r
= 21.5$), and bright giants ($r < 18$). We obtain a value of:
\eq {
	\Sigma_r = 32.5\, \mathrm{mag\, arcsec}^{-2}
}
This is about a magnitude and a half fainter than the surface brightness of Sagittarius dwarf
northern stream ($\Sigma_V \sim 31~\mathrm{mag\,arcsec}^{-2}$; \citealt{Martinez-Delgado01},
\citealt{Martinez-Delgado04}).

Assuming the entire overdensity covers $\sim{}1000$~deg$^2$ of the sky (consistent with what is seen
in the top left panel of Fig.~\ref{mosaic}), and is centered at a distance of $D \sim 10$ kpc,
from the surface brightness we obtain an estimate of the integrated absolute $r$ band magnitude,
$M_r = -7.7$~mag. This corresponds to a total luminosity of $L_r = 0.09 \cdot 10^6 L_\odot$, where we
calculate the absolute $r$ band magnitude of the Sun to be $M_{r\odot} = 4.6$, using eqs. 2 and 3
from \cite{Juric02}, and adopting $(B-V)_\odot = 0.65$ and $V_\odot = 4.83$ from \citet{GalacticAstronomy}.
This luminosity estimate is likely uncertain by at least a factor of a few. Most of the
uncertainty comes from the unknown exact distance and area covered by the overdensity. Uncertainty
due to the flux limit depends on the exact shape of the luminosity function of stars making up the
overdensity, but is likely less severe. For example, assuming that the luminosity function of the
overdensity is similar to that of the Solar neighborhood (\citealt{Reid02}, Table~4), 
and that our sample of overdensity stars is incomplete for $g-r > 0.5$ (see bottom left panel of
Fig.~\ref{mosaic}), the corrected luminosity and absolute magnitude are $L_r = 0.10 \cdot 10^6
L_\odot$ and $M_r = -7.8$ (note that only the red end of the luminosity function is relevant here). 
Taking a more conservative incompleteness bound of $g-r > 0.3$, the
luminosity grows to $L_r = 0.11 \cdot 10^6 L_\odot$ (22\% difference), or $M_r = -8$, in terms of
absolute magnitude. Again, {\it these are all lower limits.}

\subsection{ Metallicity of the Virgo Overdensity  }

\begin{figure*}
\plotone{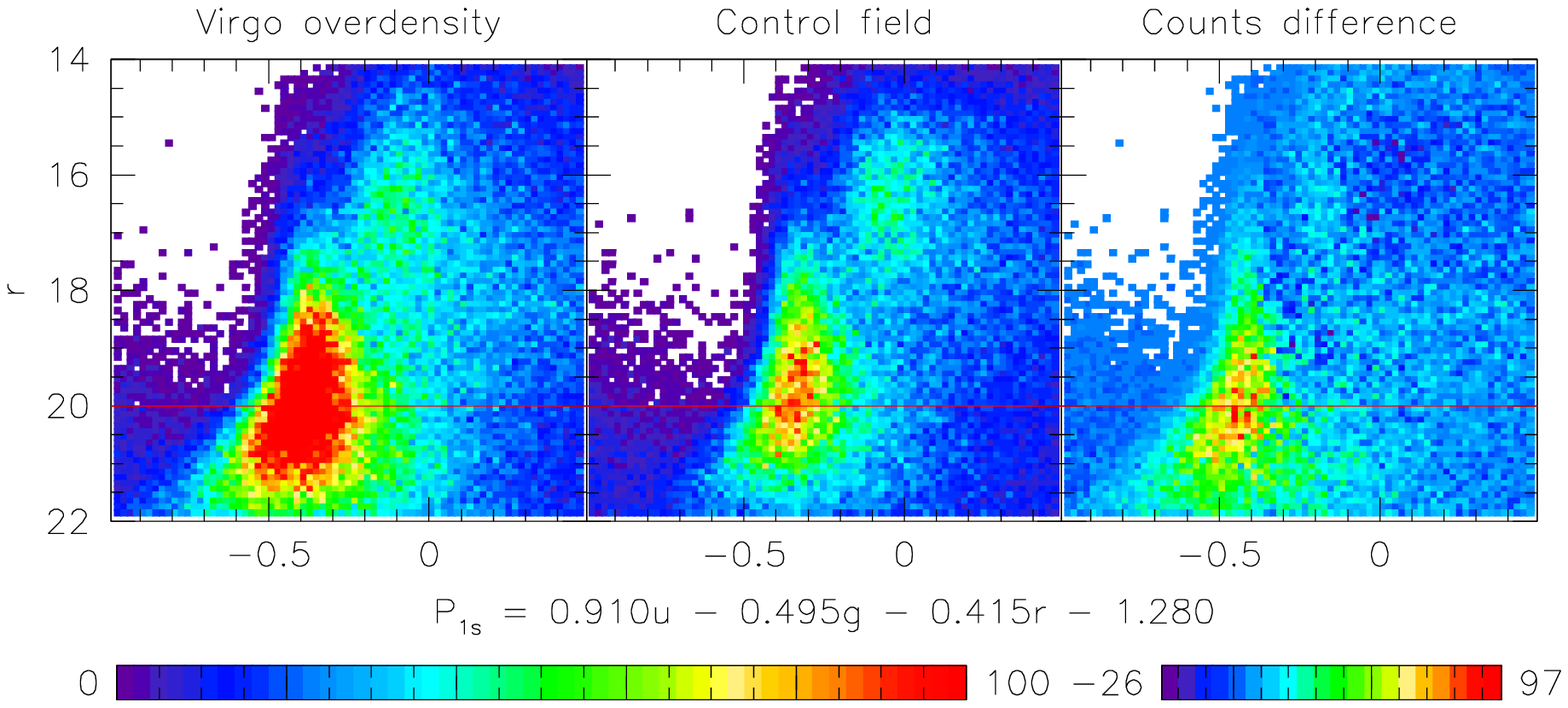}
\caption{Hess diagrams of $P_{1s}$ color vs. $r$ magnitude for the Virgo overdensity field (left),
the control field as defined on figure \ref{mosaic} (middle), and their difference (right). The
colors encode star counts within the fields. A significant excess of stars with $P_{1s}
< -0.2$ is present in the Virgo overdensity field. There is no statistically significant difference in star
counts for stars having $P_{1s} > -0.2$, implying that the stars that constitute of the Virgo
overdensity have metallicities lower than disk stars, and closer to metallicities characteristic
of the halo.
\label{p1svsr}}
\end{figure*}

The SDSS u band measurements can be used to gauge metallicity of the Virgo overdensity. 
As already discussed in Section~\ref{rhoZsec}, stars at the tip of the stellar
locus ($0.7 < u-g \la 1$) typically have metallicities lower than about $-1.5$.
This $u-g$ separation can be improved by using instead the principal axes in 
the $g-r \, \mathrm{vs.} \, u-g$ color-color diagram \citep{Ivezic04}
\eqarray{
	P_{1s} = 0.415 (g-r) + 0.910 (u-g) - 1.28 \\
	P_{2s} = 0.545 (g-r) - 0.249 (u-g) + 0.234
}
The main sequence stars can be isolated by requiring 
\eq{
   	-0.06 < P_{2s} < 0.06
}
and the low-metallicity turn-off stars using 
\eq{
   	-0.6 < P_{1s} < -0.3,
}
with $P_{1s} = -0.3$ approximately corresponding to $[Fe/H]=-1.0$. 

In Fig.~\ref{p1svsr} we show Hess diagrams of $P_{1s}$ color vs. $r$ magnitude for the Virgo
overdensity field and the control field, and their difference. A significant excess of
stars with $P_{1s} < -0.3$ exists in the Virgo overdensity field, while there is no statistically
significant difference in star counts for stars having $P_{1s} > -0.3$. The
observed $P_{1s}$ distribution implies metallicities lower than those of thick
disk stars, and similar to those of the halo stars (see also Paper II).

\subsection{          Detections of Related Clumps and Overdensities          }

\begin{figure*}
\plotone{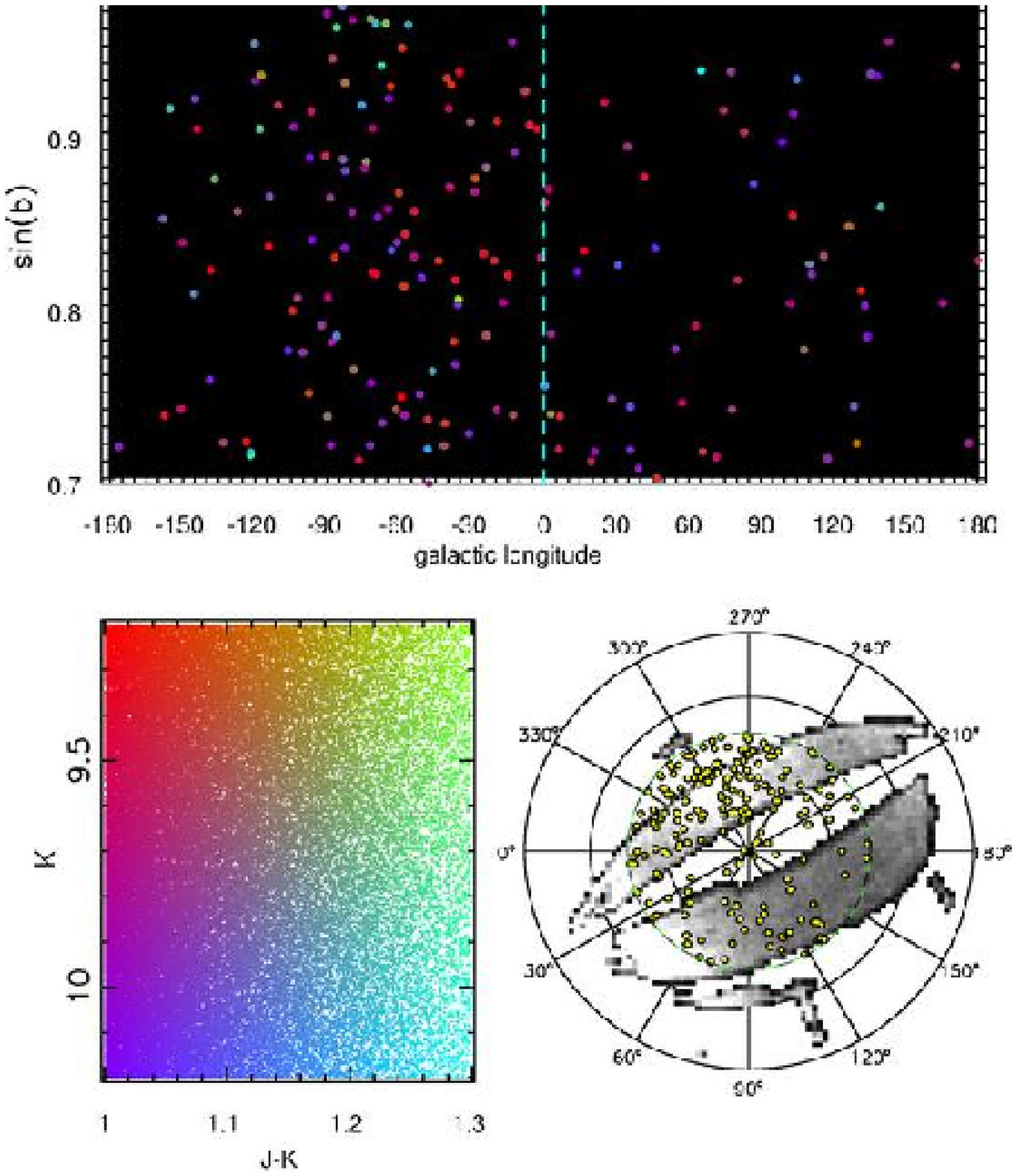}
\caption{The sky distribution of 189 2MASS M giant candidates with $b>45^\circ$, selected by
9.2$<K<$10.2 and 1.0$<J-K<$1.3. The symbols in the top panel are color-coded using their $K$ band
magnitude and $J-K$ color, according to the scheme shown in the bottom panel (which shows all 75,735
candidates from the whole sky). The symbols in bottom right panel show the same sample as in the
top panel in Lambert projection, with the SDSS density map from Fig.X shown as the gray
scale background. At $sin(b)>0.8$, there are 2.5 times as many stars with $l<0$ than with $l>0$.
This asymmetry provides an independent confirmation of the Virgo overdensity revealed by the SDSS
data.
\label{2mass}}
\end{figure*}

There are a number of stellar overdensities reported in the literature that 
are probably related to the Virgo overdensity. \citet{Newberg02} searched for halo 
substructure in SDSS equatorial strips ($\delta \sim 0$) and reported a density peak at 
$(l, b) \sim (297, 63)$.
They tentatively concluded that this feature is {\em ``a stream or other diffuse concentration 
of stars in the halo''} and pointed out that follow-up radial velocity measurements are 
required to ascertain that the grouping is not a product of chance and statistics of small numbers.

Detections of RR Lyrae stars are particularly useful because they are excellent
standard candles. Using RR Lyrae detected by the QUEST survey, \citet{Vivas01}, see
also \citealt{Zinn03}) discovered an overdensity at $\sim 20$ kpc from the Galactic 
center at $(l, b) \sim (314, 62$) (and named it the ``$12\fh4$ clump''). The same clump 
is discernible in the SDSS candidate RR Lyrae
sample \citep{Ivezic00,Ivezic03,Ivezic03a,Ivezic03c}. More recently, the NSVS 
RR Lyrae survey \citep{Wozniak04} detected an overdensity in the same direction, and at distances 
extending to the sample faint limit, corresponding to about $\sim$6 kpc (P. Wozniak, 
private communication).

2MASS survey offers an important advantage of an all-sky view of the Milky
Way. We have followed a procedure developed by \citet{Majewski03} to
select M giant candidates from the public 2MASS database. We use M giant 
candidates that belong to the Sgr dwarf stream to fine-tune selection
criteria. We also estimate the mean K band absolute magnitude by tying 
it to the stream distance implied by RR Lyrae stars \citep{Ivezic03c,Ivezic03}
We adopt 1.0$<J-K<$1.3 and 9.2$<K<$10.2 as the color-magnitude selection 
of M giant candidates at a mean distance of 10 kpc. 

Using a sample of 75,735 candidates selected over the whole sky (dominated
by stars in the Galactic plane), we study their spatial distribution in 
the high galactic latitude regions (see  Fig.~\ref{2mass}). We find a
significant excess of candidate giants in the Virgo overdensity area, 
compared to a region symmetric with respect to the $l=0$ line, with the 
number ratio consistent with the properties of the Virgo overdensity inferred 
from SDSS data. For example, in a subsample restricted to
55$^\circ<b<$80$^\circ$, there are 66 stars with 240$<l<$360, and only 21
stars with 0$<l<$120, with the former clustered around $l\sim$300. There is 
no analogous counts asymmetry in the southern Galactic hemisphere.

\subsection{         A Merger, Tri-axial Halo, Polar Ring, or? }

The Virgo overdensity is a major new feature in the Galactic halo: even within the limited sky coverage
of the available SDSS data, it extends over a thousand square degrees of sky. Given the
well defined overdensity outline, low surface brightness and luminosity, its most plausible
interpretation is a tidally disrupted remnant of a merger event involving the Milky Way and a
smaller, lower-metallicity dwarf galaxy. However, there are other possibilities.

An attempt may be made to explain the detected asymmetry by postulating a non-axisymmetric
component such as a triaxial halo. This alternative is particularly interesting because 
\citet{Newberg05}, who essentially used the same data as analyzed here, have suggested
that evidence for such a halo exists in SDSS starcounts. A different data analysis method
employed here -- the three-dimensional 
number density maps -- suggests that the excess of stars associated with the Virgo overdensity 
is {\it not} due to a triaxial halo. The main argument against such a halo is that,
despite its triaxiality, it still predicts that the density decreases with the 
distance from the Galactic center. But, as shown in Figs.~\ref{XYslices1} and \ref{rhoRS},
the observed density profile has a local maximum that is {\it not} aligned with 
the Galactic center. This can still be explained by requiring the axis of the halo 
not to be coincident with the Galactic axis of rotation. However, even this model requires 
the halo density to attain maximal value in the Galactic center, and as
seen from figure~\ref{vlgvPanels2} a modest linear extrapolation of Virgo 
overdensity to $Z=0$ still keeps it at $R \sim 6$~kpc away from the Galactic center. Unless
one is willing to resort to models where the center of the stellar halo and the center of the
Milky Way disk do not coincide, tri-axial halo cannot explain the geometry of Virgo overdensity.

Although this makes the explanation of Virgo as a signature of triaxial halo unlikely, 
it does not preclude the existence of such a halo.
Unfortunately, it would be very difficult to obtain a reliable measurement of the halo
triaxiality with the currently available SDSS data because of contamination
by the Virgo overdensity and uncertainties about its true extent. As more SDSS and other data become
available in other parts of the sky, it may become possible to mask out the overdensity 
and attempt a detailed halo fit to reveal the exact details of its shape and structure.

Another possible explanation of the overdensity is a ``polar ring'' around the 
Galaxy. This possibility seems much less likely than the merger scenario because there is
no visible curvature towards the Galactic center at high $Z$ in Fig.~\ref{vlgvPanels2}.
Indeed, there seems to be a curvature in the opposite sense, where the bottom ($Z \sim 6$ kpc) 
part of  the overdense region appears to be about $0.5-1$ kpc closer to the Galactic center 
than its high-$Z$ part. In addition, there is no excess of 2MASS M giant candidates in 
the southern sky that could be easily associated with the northern Virgo overdensity\footnote{
Note that the polar rings explanation is also unlikely for theoretical reasons as these are 
thought to originate in large galactic collisions which would leave its imprint on other components
of the Milky Way as well. We discuss it as an option here from purely observational standpoint,
and mainly for completeness.}.

Finally, the coincidence of this overdensity and the Virgo galaxy
supercluster \citep{Binggeli99} could raise a question whether the overdensity
could be due to faint galaxies that are misclassified as stars. While
plausible in principle, this is most likely not the case because the
star/galaxy classifier is known to be robust at the 5\% level to at least
$r=21.5$ \citep{Ivezic02}, the overdensity is detected
over a much larger sky area (1000~deg$^2$ vs. $\sim 90$~deg$^2$), and the
overdensity is confirmed by apparently much brighter RR Lyrae stars
and M giants. 

\section{                            Discussion                        } 
\label{Disc}

\subsection{  A Paradigm Shift }

Photometric parallax methods have a long history of use in studies of the Milky Way structure
(e.g., \citealt{Gilmore83}, \citealt{Kuijken89b}, \citealt{Chen01}, \citealt{Siegel02}). 
An excellent  recent example of the application of this method to pre-SDSS data is the
study by \citet{Siegel02}. While their and SDSS data
are of similar photometric quality, the sky area analyzed here is over 400 times 
larger than that analyzed by Siegel et al. This large increase in sample
size enables a shift in emphasis from modelling to direct model-free {\it mapping} of the complex
and clumpy Galactic density distribution. Such mapping and analysis of the maps
allows for identification and removal of clumpy substructure, which is a {\it necessary
precondition} for a reliable determination of the functional form and best-fit
parameters of the Galactic model.

This qualitative paradigm shift was made possible by the availability of SDSS data. SDSS is superior
to previous optical sky surveys because of its high catalog completeness and precise multi-band
CCD photometry to faint flux limits over a large sky area. In particular, the results presented here
were enabled by several distinctive SDSS characteristics:
\begin{itemize}
\item
A large majority of stars detected by the SDSS are main-sequence stars, which have a fairly 
well-defined color-luminosity relation. Thus, accurate SDSS colors can be used to 
estimate luminosity, and hence, distance, for each individual star. Accurate photometry 
($\sim0.02$ mag) allows us to reach the intrinsic accuracy of photometric parallax relation, 
and estimate distances to single stars within 15-20\%, and estimate the relative distances 
of stars in clumps of similar age and metallicity to better than 5\%.
\item
Thanks to faint flux limits ($r \sim 22$), distances as large as 15--20~kpc are probed using 
numerous main sequence stars ($\sim 48$~million). At the same time, 
the large photometric dynamic range and the strong dependence of stellar luminosities 
on color allow constraints ranging from the Sun's offset from the Galactic
plane ($\sim 25$ pc) to a detection of overdensities at distances beyond 10 kpc.
\item 
Large sky area observed by the SDSS (as opposed to pencil beam surveys), 
spanning a range of galactic longitudes and latitudes, enables not only a good coverage 
of the ($R,Z$) plane, but also of a large fraction of the Galaxy's volume. The full
three-dimensional analysis, such as slices of the maps in $X-Y$ planes, reveals
a great level of detail.
\item 
The SDSS $u$ band photometric observations can be used to identify stars with 
sub-solar metallicities, and to study the differences between their distribution and 
that of more metal-rich stars. 
\end{itemize}

\subsection{ The Best-Fit Galactic Model }

\begin{figure}
\plotone{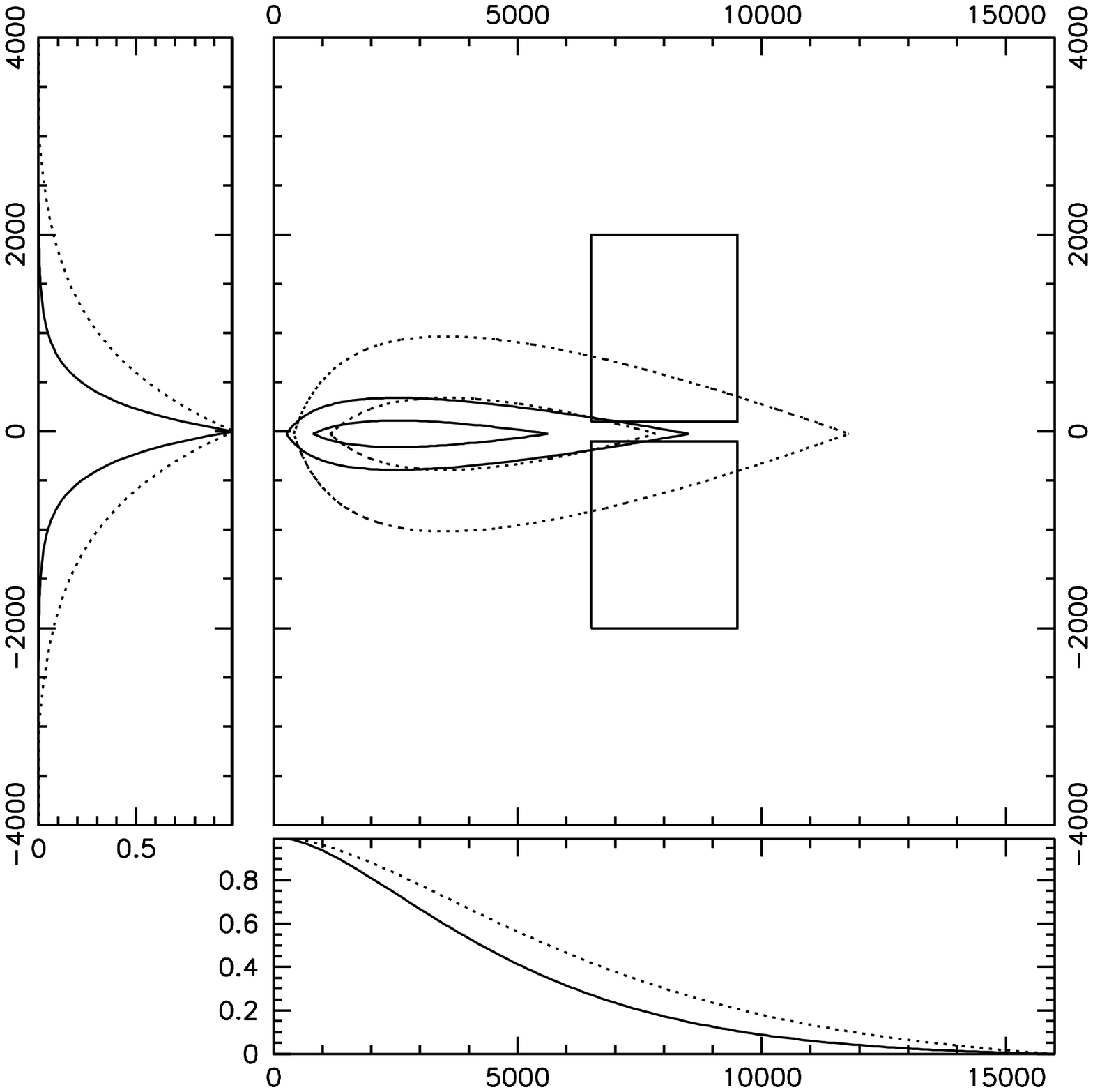}
\caption{The mass contribution to thin (solid) and thick (dotted) disks from different radii and 
heights in the Galaxy. Center panel shows the isodensity contours of thin (solid) and thick (dotted) disk
having the bias-corrected parameters of Table~\ref{tbl.finalparams}. Two rectangles centered around
$R = 8$~kpc enclose the range in radii and vertical distances $Z$ from the disk from which the model 
parameters were measured.
The bottom panel shows the cumulative fraction of disk mass enclosed \emph{outside} a given
radius $R$. Similarly, the side panel shows the fraction of disk mass enclosed at heights 
$|Z| > |Z_{\rm given}|$. Note that while our coverage in $Z$ direction is adequate, in the $R$ direction
we sample a region that contains less than 20\% of the total disk mass, and extrapolate 
the obtained density laws towards the Galactic center where most of the mass lies.
\label{fig.galmass}}
\end{figure}

When the exponential disk models are used to describe the gross behavior of the stellar number 
density distribution, we derive the best-fit parameter values summarized in Table~\ref{tbl.finalparams}. 

Before proceeding to compare these results to the literature, we note that a proper 
comparison with previous work is sometimes difficult due to the lack 
of clarity of some authors regarding which effects were (or were not)
taken into account when deriving the model parameters. Of particular concern is
the problem of unrecognized multiplicity: uncorrected for, or if using a significantly
different binary fraction, it will affect the disk scales by up to $30\%$
(Section~\ref{sec.binarity}). In the discussion to follow we assumed, if
not explicitly mentioned otherwise, that all appropriate corrections were taken into 
account by the authors of the studies against which compare our results.

The derived $300$~pc vertical scale of the thin disk (corrected for an assumed
35\% binary fraction) is about $10\%$ lower than the cannonical $325$pc value, and
near the middle of the range of values found in the recent literature ($240 - 350$pc,
\citet{Robin96,Larsen96,Buser99,Chen01,Siegel02}). Similarly, the scale height of the 
thick disk is in the range found by \citet{Siegel02,Buser99} and \cite{Larsen96},
and about $20\%$ higher than the $580-790$pc range spanned by measurements of \cite{Robin96,Ojha99}
and \cite{Chen01}. We note that uncorrected for unrecognized multiplicity, our 
thin and the thick disk scale estimates ($245$ and $740$, respectively) would
be closer to the lower end of the range found in the literature. 

We find the local thick disk normalization of $\sim 12\%$, larger than most previous estimates
but similar to recent determinations by \citet{Chen01} and \citet{Siegel02} ($\gtrsim 10$\%).
Models with normalizations lower than $10\%$ show increasingly large $\chi^2$ and, in particular, 
the combinations of parameters characteristic of early ``low normalization/high thick disk scale height''
models (e.g., \citealt{Gilmore83,Robin86,Yoshii87,Yamagata92,Reid93}) are strongly
disfavored by the SDSS data. The root cause of the apparent discrepancy
may be sought for in the fact that all of these studies were pencil-beam surveys
operating on a single or at most a few lines of sight, usually towards the NGP.
However, a single or even a few pencil beams are insufficient to break the degeneracies
inherent in the multiparameter Galactic model (Section~\ref{sec.degeneracies}).
While adequately describing the observed lines of sight, these pencil-beam
best-fit parameters are local minima, unrepresentative of the entire Galaxy.
Only by using a wider and deeper sample, such as the one presented here, were we able 
to break the degeneracy and derive a globally representative model.

The value of thin disk scale length is in agreement with the recent estimates
by \citet{Ojha99}, \citet{Chen01} and \citet{Siegel02}, and lower than the traditionally
assumed $3-4$kpc. The scale length of the thick disk is longer than that of 
the thin disk. The qualitative nature of this result is robust: variations of the assumed 
photometric parallax relation, binary fraction or the exact choice of and size 
of the color bins, leave it unchanged. Quantitatively, the ratio of best-fit 
length scales is close to $1.4$, similar (within uncertainties) to typical scale length
ratios of $\sim 1.25$ seen in edge-on late-type disk galaxies \citep{Yoachim06}.

Assuming that exponential density laws correctly describe the density distribution 
all the way to the Galactic center, our model implies that $\sim 23$\%
of the total luminosity (and stellar mass) in K and M-dwarfs is contained in the thick disk. Despite being
an extrapolation from a small region containing only a few tens of percent
of the total mass of the disk (Figure~\ref{fig.galmass}), this is in good agreement with
observations of thick disks in external edge-on galaxies (\citealt{Yoachim06}, Figure~24).

\subsection{ Detection of Resolved Substructure in the Disk } 

Although important for understanding the current structure of the Milky Way, the disk mass 
ratios, details of density laws, or the exact values of Galactic model parameters are \emph{insufficient
by themselves} to attack the question of mechanism of Galactic disk formation (both thin and thick).
It is the departures from these laws that actually hold more information about the formation than 
the laws themselves.

Thick disk formation scenarios can be broadly be divided into three classes: (1) slow kinematic
heating of stars from the thin disk \citep{Spitzer53,Barbanis67,Sellwood84}, (2) pressure-supported 
slow collapse immediately after an ELS-like
monolithic collapse (e.g., \citealt{Larson76}) or (3) merger-induced stirring of thin disk
material and/or direct accretion of stellar content of the progenitor 
\citep{Quinn93,Abadi03,Brook04}. Scenarios (1) and (2) are usually disfavored due to the inability
of either the giant molecular clouds or the spiral structure to excite the stars to orbits observed
in the thin disk (e.g. \citealt{Quillen00}), and the apparent lack of vertical metallicity
gradient in the thick disk (\citealt{Gilmore95}; see however Paper II
for evidence to the contrary).

The third scenario recently garnered increased attention, with detailed 
theoretical simulations of the formation of realistic galaxies in $\Lambda$CDM hierarchical 
merger picture context \citep{Abadi03,Brook04}, and the observation of properties of thick disks 
\citep{Dalcanton02,Yoachim06} and even a counter-rotating thick disk in FGC 227 \citep{Yoachim05}.
A simulation reported by \citet{Abadi03}, while not \emph{directly} comparable to the Milky 
Way (their galaxy is spheroid-dominated) is especially illuminating regarding the qualitative 
mechanisms that may build up the thick disk. Three of their
conclusions are of particular consequence to our work: i) the thick disk is formed by direct 
accretion of stellar content from satellites on low inclination orbits, ii) the stars from 
a single disrupted satellite are not uniformly radially mixed, but rather form a torus-like structure
at radii where the final disruption occurs, and iii) if formed through the same process,
the disk of the Milky Way disk may still hold signatures of such early accretion events.

Our finding that the thin and thick disk structure, similarly to that of the halo, is complicated 
by localized overdensities and 
permeated by ring-like departures from exponential profiles may lend credence
to the mechanism described by
\citet{Abadi03}. In addition to already known Monoceros stream, we found evidence 
for two more overdensities in the thick disk region (Figure~\ref{fig.clumps}),
both consistent with rings or streams in the density maps. While unlikely to
be the relics from the age of thick disk formation
(they would need to survive for $\sim 8-10$~Gyr) it is plausible they, like the Monoceros stream,
are remnants of smaller and more recent accretion events analogous to those that formed the
thick disk.

In case of the Monoceros stream, the three-dimensional maps offer an effective method to study 
its properties. The maps demonstrate this feature
is well localized in the radial direction, which rules out the hypothesis that 
this overdensity is due to disk flaring. The maps also show that the Monoceros 
stream is not a homogeneously dense ring that surrounds the Galaxy, providing
support for the claim by \citet{Rocha-Pinto03} that this structure is a 
merging dwarf galaxy (see also \citealt{Penarrubia05} for a comprehensive
theoretical model). In Paper II, we demonstrate that stars in the Monoceros
stream have metallicity distribution more metal-poor than thick disk stars,
but more metal-rich than halo stars.

Discoveries of this type of substructure point to a picture of the thick disk filled
with streams and remnants much in the same way like the halo. A crude extrapolation of three 
disk overdensities seen in our survey volume ($|Z| < 3$ kpc, $R < 15$ kpc) to
the full Galactic disk leads to a 
conclusion that there may be up to $\sim$15 - 30 clumpy substructures of this type 
in the Galaxy. These ``disk streams'' are still likely to carry,
both in their physical (metallicity) and kinematic properties, 
some information on their progenitors and history.

\subsection{ Stellar Halo } 

We find it possible to describe the halo of the Milky Way by an oblate $r^{-n_H}$ power-law ellipsoid, 
with the axis ratio $c/a \equiv q_H \sim 0.5 - 0.8$ and the power-law index of $n_H = 2.5-3$ (with
the formal best fit parameters $q_H=0.64$ and $n_H=2.8$ for galactocentric radii
$\la$20 kpc). These values are consistent with previous studies: specifically, they are in excellent 
agreement with Besancon program values ($q_H = 0.6 - 0.85$, $n_H = 2.44-2.75$; \citealt{Robin00}), with 
a more recent measurement of $q_H = 0.6$ and $n_H = 2.75$ by \cite{Siegel02}, and
with the previous SDSS estimate of $q_H \sim 0.55$ \citep{Chen01}. The convergence of best-fit values
is encouraging, especially considering the differences in methodologies
(direct fitting vs. population synthesis modelling) and the data (photometric systems,
limiting magnitudes, types of tracers, and lines of sight) used in each of these studies.

The goodness of halo model fit is poorer than that of the disk fits (reduced $\chi^2 \sim 2-3$).
Similar problems with halo fits were previously noticed in studies of smaller samples of
kinematically and metallicity-selected field stars \citep{Hartwick87,Sommer-Larsen90,Allen91,Preston91,Kinman94,Carney96,Chiba00},
globular clusters \citep{Zinn93,Dinescu99} and main sequence stars \citep{Gilmore85,Siegel02},
and are unlikely to be explained away by instrumental or methodological reasons alone \citep{Siegel02}.
Our own analysis of why this is so (Section~\ref{sec.rhists} and Figure~\ref{fig.resid.xyslice}) 
points towards a need for a more complex density distribution profile.
For example, instead of a single power law, a two-component ``dual-halo'', in which the 
stars are divided into a spherical and a flattened sub-component, may be invoked to 
explain the observations (e.g., \citealt{Sommer-Larsen90}).

Such models, when applied to starcounts, do show improvements
over a single power law \citep{Siegel02}. Furthermore, this division may be theoretically motivated
by an attempt to unify the ELS and \cite{SearleZinn} pictures of Galaxy formation: the
flattened subcomponent being a result of the initial monolithic collapse, and the spherical 
component originating from subsequent accretion of satellites \citep{Sandage90,Majewski93,Norris94}.
While this explanation is \emph{circumstantially} supported by the detection of ongoing accretion 
in the halo today (e.g. \citealt{Yanny00,Ivezic00,Vivas01,Majewski03,Belokurov06} and references therein), 
we would prefer a more \emph{direct} line of evidence for it, derived from observations of halo
stars themselves.

For example, one may hope the component coming from accretion is visibly irregular, 
streamlike, and/or clumpy, thus lending credence to the hypothesis of its origin. 
However, our examination of the distribution of residuals in Section~\ref{sec.rhists} 
revealed no signal of unresolved clumpy substructure in the halo on $1 - 2\sim$ kpc scales.
Instead, we found the large reduced $\chi^2$ is best explained by a poor choice of
density law profile (a single power law). A double power-law, or a more complicated
profile such as the one used by \cite{Preston91}, would likely better fit the data.

The clumpiness may still be prevalent, but on a different spatial scale, or smalled in amplitude 
and harder to detect with a simple analysis employed here. We leave a detailed study of scale-dependent 
clumpiness in the halo and its possible two-component nature for a subsequent
study.

\subsection{ The Virgo Overdensity } 

We report the discovery of Virgo overdensity. Despite its large angular size
and proximity, its low surface brightness kept it from being recognized by smaller surveys. Given the low
surface brightness, its well defined outline, and low metallicity, the most plausible explanation of
Virgo overdensity is that it is a result of a merger event involving the Milky Way and a smaller, 
lower-metallicity dwarf galaxy. For now, based on existing maps, we are unable to differentiate whether 
the observed overdensity is a tidal stream, a merger remnant, or both. However, it is evident that 
the Virgo overdensity is surprisingly large, extending in vertical ($Z$) direction to the boundaries 
of our survey ($6 < Z < 15$~kpc), and $\sim 10$~kpc in $R$ direction. It is also exceedingly faint, 
with a lower limit on surface brightness of $\Sigma_r = 32.5\, \mathrm{mag\, arcsec}^{-2}$.

A potential connection of Virgo overdensity and the Sagittarius stream is discussed in a 
followup paper by \cite{Martinez-Delgado07}. Their N-body simulations of the 
Sagittarius stream show that the Virgo overdensity resides in the region of space 
where the leading arm of the Sagittarius stream is predicted to cross the Milky Way 
plane in the Solar neighborhood. This tentative Virgo-Sagittarius association
needs to be confirmed by measurement of highly negative radial velocities for the 
stars of the Virgo overdensity.

A similar diffuse structure, the Triangulum-Andromeda feature (hereafter, TriAnd), was recently identified by
\citet{Rocha-Pinto04} and \citet{Majewski04} in the southern Galactic hemisphere, as an overdensity 
of M giants observed with 2MASS. They find an excess in M giant number density over a large area of the sky
($100^\circ < l < 150^\circ$, $-40^\circ < b < -20^\circ$). TriAnd, just as the Virgo structure
presented here, is very diffuse and shows no clear core. \citet{Rocha-Pinto04} estimate the
distance to TriAnd of at least $\sim 10$ kpc. Recently, additional tenuous structures were
discovered in the same region of the sky \citep{Majewski04,Martin07},
pointing to the possibility that diffuse clouds such as Virgo and TriAnd are quite common in the
Galactic halo.

Assuming that the Virgo overdensity is a part of a larger previously unidentified stream, it would 
be of interest to
look for a possible continuation in the southern Galactic hemisphere. Our preliminary analysis of
2MASS M-giants data did not reveal a similarly large density enhancement 
in the south. It would also be interesting to follow the stream towards the Galactic north, 
beyond the $Z \sim 20$~kpc limit of our survey, where a signature of
overdensity has been revealed by RR Lyrae stars \citep{Duffau06}. Above all,
the understanding of the Virgo overdensity would greatly benefit from
measurements of proper motion and radial velocity of its constituent stars.

\subsection{ Mapping the Future }

This study is only a first step towards a better understanding of the Milky Way 
enabled by modern large-scale surveys.
Star counting, whether interpreted with traditional modeling methods, or
using number density maps, is limited by the number of observed stars, the 
flux limit and sky coverage of a survey, and the ability to differentiate 
stellar populations. All these data aspects will soon be significantly improved.

First, the SDSS has entered its second phase, with a significant fraction of
observing time allocated for the Milky Way studies (SEGUE, the Sloan Extension for 
Galaxy Understanding and Exploration, \citealt{Newberg03}). In particular, the 
imaging of low galactic latitudes and a large number of stellar spectra  optimized 
for Galactic structure studies will add valuable new data to complement this work. 
In addition, the SDSS kinematic data, both from radial velocities and from proper 
motions (determined from astrometric comparison of the SDSS and the Palomar 
Observatory Sky Survey catalog, \citealt{Munn04}) is already yielding significant 
advances in our understanding of the thin and thick disk, and halo kinematic 
structure (Paper III, in prep.). 

Another improvement to the analysis presented here will come from the GAIA
satellite mission (e.g. \citealt{Wilkinson05}). GAIA will provide geometric 
distance estimates and spectrophotometric measurements for a large number of stars brighter 
than $V\sim20$. Despite the relatively bright flux limit, these data will be invaluable 
for calibrating photometric parallax relation, and for studying the effects of metallicity, 
binarity and contamination by giants. At the moment, the uncertainties of the photometric parallax
relation are the single largest contributor to uncertainties in the derived parameters of Galactic
models, and improvements in its calibration are of great interest to all practitioners in this
field.

A further major leap forward will be enabled by upcoming deep synoptic sky surveys,
such as Pan-STARRS \citep{Kaiser02} and LSST \citep{Tyson02}. Pan-STARRS has already
achieved first light
with its first telescope, and the four-telescope version may become
operational around 2010. If approved for construction in 2009, the LSST may obtain
its first light in 2014. These surveys will provide
multi-band optical photometry of better quality than SDSS over practically 
the entire sky (LSST will be sited in Chile, and Pan-STARRS is at Hawaii; note
that Pan-STARRS will not use the $u$ band filter). One of their advantages will 
be significantly deeper data -- for example, the LSST will enable studies such as 
this one to a 5 magnitudes fainter limit, corresponding to a distance limit of 150 kpc 
for the turn-off stars. LSST proper motion measurements will constrain tangential
velocity to within 10 km/s at distances as large as that of the Virgo overdensity 
reported here ($\sim$10 kpc). These next-generation maps will be based on
samples including several billion stars and will facilitate not only the accurate
tomography of the Milky Way, but of the whole Local Group.

\vskip 0.4in \leftline{Acknowledgments}

We thank Princeton University, the University of Washington and the Institute for
Advanced Study for generous financial support of this research.
M. Juri\'{c} gratefully acknowledges support from the Taplin Fellowship and from NSF
grant PHY-0503584. \v{Z}. Ivezi\'{c} 
and B. Sesar acknowledge support by NSF grant AST-0551161 to LSST for design
and development activity. Donald P. Schneider acknowledges support by
NSF grant AST-0607634. We especially thank the anonymous referee for numerous 
helpful comments and suggestions which have significantly improved this manuscript.

    Funding for the creation and distribution of the SDSS Archive has been provided by the Alfred P.
Sloan Foundation, the Participating Institutions, the National Aeronautics and Space Administration,
the National Science Foundation, the U.S. Department of Energy, the Japanese Monbukagakusho, and the
Max Planck Society. The SDSS Web site is http://www.sdss.org/.

    The SDSS is managed by the Astrophysical Research Consortium (ARC) for the Participating
Institutions. The Participating Institutions are The University of Chicago, Fermilab, the Institute
for Advanced Study, the Japan Participation Group, The Johns Hopkins University, the Korean
Scientist Group, Los Alamos National Laboratory, the Max-Planck-Institute for Astronomy (MPIA), the
Max-Planck-Institute for Astrophysics (MPA), New Mexico State University, University of Pittsburgh,
University of Portsmouth, Princeton University, the United States Naval Observatory, and the
University of Washington.

This publication makes use of data products from the Two Micron All Sky Survey, which is a joint
project of the University of Massachusetts and the Infrared Processing and Analysis
Center/California Institute of Technology, funded by the National Aeronautics and Space
Administration and the National Science Foundation.

\bibliographystyle{apj}
\bibliography{galaxy}

\appendix

\section{Effects of locus projection}
\label{AppA}
The improvement in the estimate of $r-i$ color resulting from the locus projection depends on
the local slope of the locus. If the locus has a steep or almost vertical slope, as for stars with $g-r \sim
1.4$ (cf. figure \ref{locusfit}), the knowledge of $g-r$ color does not further constrain $r-i$
color. On the other hand, for shallow slopes the knowledge of $g-r$ color determines the intrinsic
$r-i$ to a much better accuracy than the the $r-i$ measurement alone. For most of the observed $g-r$
color range we are closer to the second regime, with the locus having a slope of
$d(r-i)/d(g-r)\sim0.3$ for $0 < g-r < 1$.

\begin{figure}
\plotone{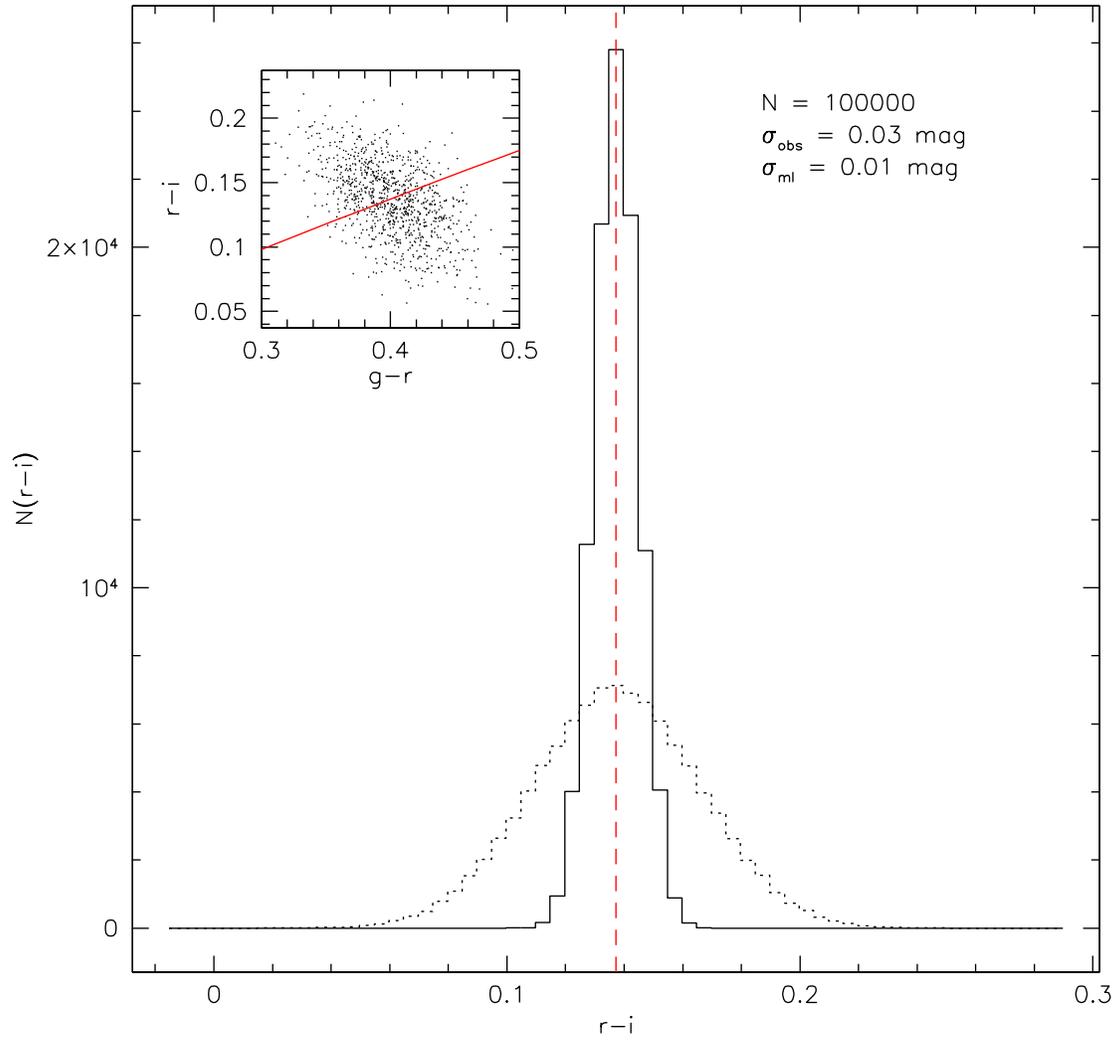}
\caption{An illustration of the reduction in the $r-i$ color scatter after applying the 
locus projection. An ensemble of $10^5$ stars with the same color ($g-r = 0.4, r-i = 0.143$) 
is subjected to SDSS photometric errors ($\sigma_{g} = \sigma_{r} = \sigma_{i} = 0.02$~mag), 
resulting in the color distribution shown in the inset. Its $r-i$ histogram (dashed line) 
has a root-mean-square scatter of $\sigma_{r-i} = 0.03$ mag.
After the colors are locus corrected (solid histogram), the scatter is reduced to $\sigma_{r-i} =
0.01$ mag. The amount of reduction of the scatter depends on the slope of the locus -- the more
horizontal the locus gets, the better is the determination of the true $r-i$ color.
\label{locusfitdemo}}
\end{figure}

To illustrate the effects of locus projection, in Fig.~\ref{locusfitdemo} we simulate an ensemble
of $10^5$ stars with the same color ($g-r = 0.4, r-i = 0.143$) subjected to photometric errors
representative of SDSS observations ($\sigma_{r-i} = \sigma_{g-r} = 0.03$~mag). Errors
introduce a scatter in observed colors as shown in the inset. Using
only the $r-i$ information we obtain the expected $\sigma_{r-i} = 0.03$~mag scatter in observed
$r-i$ color (dashed histogram). By projecting the colors to the locus, the scatter is reduced to
$\sigma_{r-i} = 0.01$~mag. As we construct density maps for stars binned in color bins
with $\Delta(r-i) = 0.05$~mag width, this is a significant reduction in scatter.

There are other benefits of locus projection. Figure \ref{locusbias} illustrates how photometric
errors can bias the determination of the number of stars in regions with  large gradients of the 
number density (in the $g-r$ vs. $r-i$ color-color diagram). The solid line shows
the true density distribution of stars as a function of $r-i$ color. In this toy model, we
used $\rho(r-i) \propto (r-i)^2$, with a sharp cutoff at $r-i = 0.24$ (solid histogram).
Scattering the stars drawn from that distribution with the $\sigma_{r-i} = \sigma_{g-r} = 0.03$~mag
photometric errors, and binning in $r-i$ color bins, produces the dashed histogram (effectively,
a convolution of the original distribution with the SDSS photometric errors). It systematically
underestimates the original density distribution by as much as $50\%$ in the region of highest
negative gradient (near the $r-i = 0.24$ cutoff), while overestimating the density in regions with
positive density gradient. The exact under or overestimation depends on the scale at which the true
density changes appreciably. If it is significantly smaller than $\sigma_{r-i}$ or $\sigma_{g-r}$,
as
is the case near the cutoff in figure \ref{locusbias}, the biases become significant. Applying the
locus projection gives a much better estimate of the original distribution (dotted histogram).

This is not a purely hypothetical example -- this type of strong density gradient is observed near
$g-r \sim 0.2$ (cf. figure \ref{locusfit}). Figure \ref{locusbias} compares the number of stars per
$r-i$ color bin for observed $r-i$ colors (dashed histogram) and the colors estimated by locus
projection (solid histogram) for the 48 million stars in our star catalog. In the regions
of maximal gradients the locus projection has a $\sim 20\%$ effect on the total number of stars per
$r-i$ bin. Not including this correction would similarly bias the density normalization of Galactic
models deduced from this sample. The bias would further propagate to luminosity function
determination\footnote{For this particular work, the discussed bias is not a problem as we make
use of only the reddest bins ($r-i > 1.0$) to measure the Galactic model parameters}.

One may further be concerned about the dependence of the location of the $r-i$ vs. $g-r$ stellar 
locus on other parameters. We do observe a slight magnitude dependence of the
locus, especially for stars with bluer $g-r$ colors. Figure \ref{rivsr} shows the dependence of the
measured $r-i$ color on the $r$ band magnitude for stars with $0.29 < g-r < 0.31$ (the "worst case 
$g-r$ color", where we find the dependence of the locus location on the magnitude to be the
largest). 
We find a weak, approximately linear dependence, with $d(r-i)_{locus}/dr \sim$ 0.007~mag/mag
(solid line). The horizontal dashed line shows the location of the locus for this bin as given by
eq.~\ref{eq.locus}. The two match for $r = 18$. We have chosen to disregard this magnitude
dependence when performing the locus projection procedure. We reason that the real dependence is not
on magnitude but on metallicity\footnote{For example, a part of the shift in locus is likely created
by the distant metal-poor stars of the halo, which are preferentially found at fainter magnitudes
in our sample.} and probably other unknown factors for which we do not have a firm handle. By
attempting to correct for those, we risk introducing additional unknown and more
complicated biases. Secondly, the magnitude dependence is relatively small and
 only affects the bluest bins.

Finally, the three panels in Fig.~\ref{binspill} show the histograms of observed $r-i$ colors for
stars which
have their locus projected colors in the $0.1 < (r-i)_e < 0.15$ color
range. The top panel shows the histogram for all stars in the sample. The 0.03~mag scatter
is comparable to SDSS photometric errors. To check for the effects discussed in
the previous paragraph, the center and bottom panels show histograms for the brightest and faintest
magnitude bin respectively. Slight magnitude dependence can be seen as a small shift of histogram
median to the left (center panel) and right (top panel). Also, the worsening of photometric
precision at the faint end is quite visible in the bottom panel, as the scatter increases to
0.08~mag.

\begin{figure}
\plotone{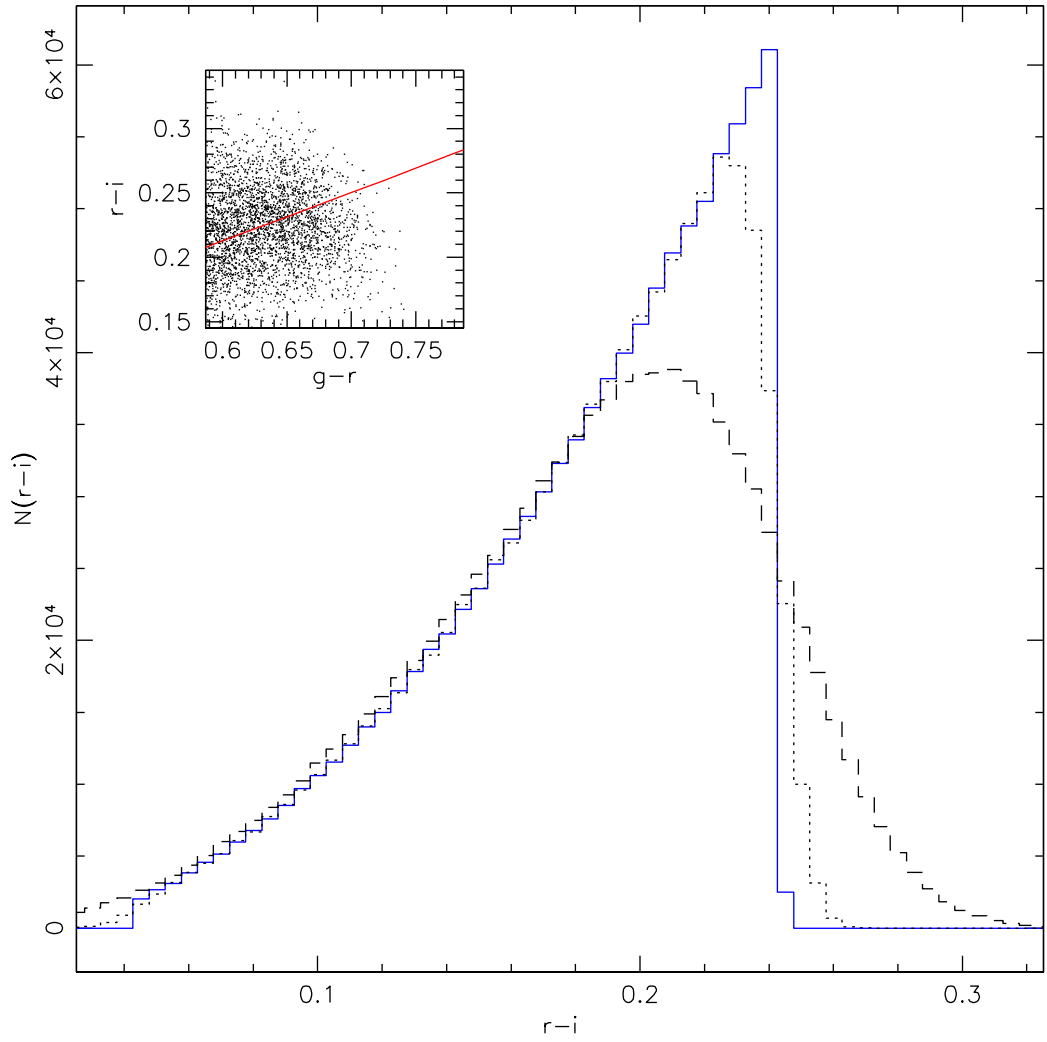}
\caption{An illustration of the bias in the determination of the number
density introduced by photometric errors. The solid histogram shows a toy model 
distribution as a function of the $r-i$ color ($\rho(r-i) \propto (r-i)^2$, with a sharp 
cutoff at $r-i = 0.24$). The dashed histogram is the convolution of the original distribution 
with SDSS photometric errors ($\sigma_{g} = \sigma_{r} = \sigma_{i} = 0.02$~mag). The measured 
density of stars with $r-i = 0.24$ is underestimated by as much as 50\%. Applying the locus 
projection gives an improved estimate of the original distribution (dotted histogram).
\label{locusfitdemo2}}
\end{figure}

\begin{figure}
\plotone{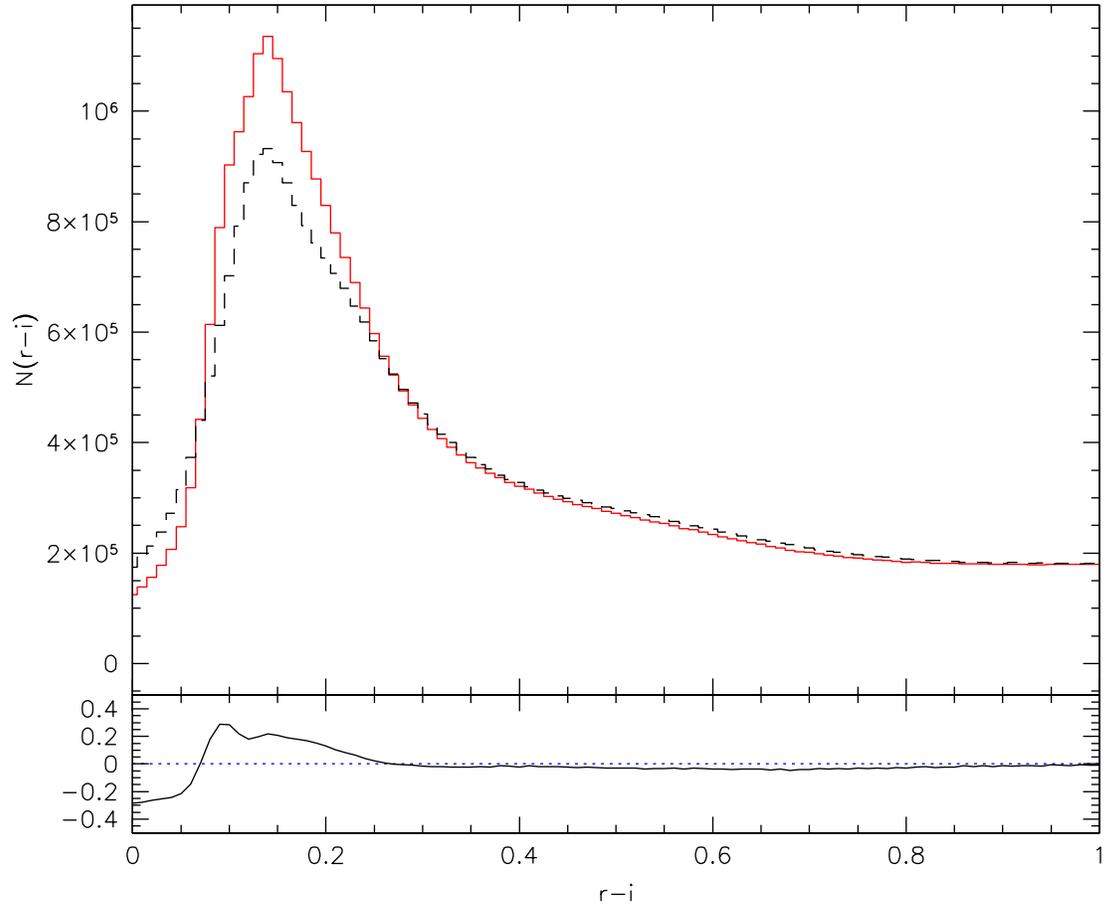}
\caption{A comparison of the number of stars per $r-i$ bin for observed $r-i$ (dashed histogram) and
the locus corrected $(r-i)_e$ colors (solid histogram) for the 48 million stars in our star catalog.
In the regions of maximal gradients the locus projection has a $\sim 20\%$ effect on the total
number of stars per $r-i$ bin. Not including this correction would similarly bias the density
normalization of Galactic models.
\label{locusbias}}
\end{figure}

\begin{figure}
\plotone{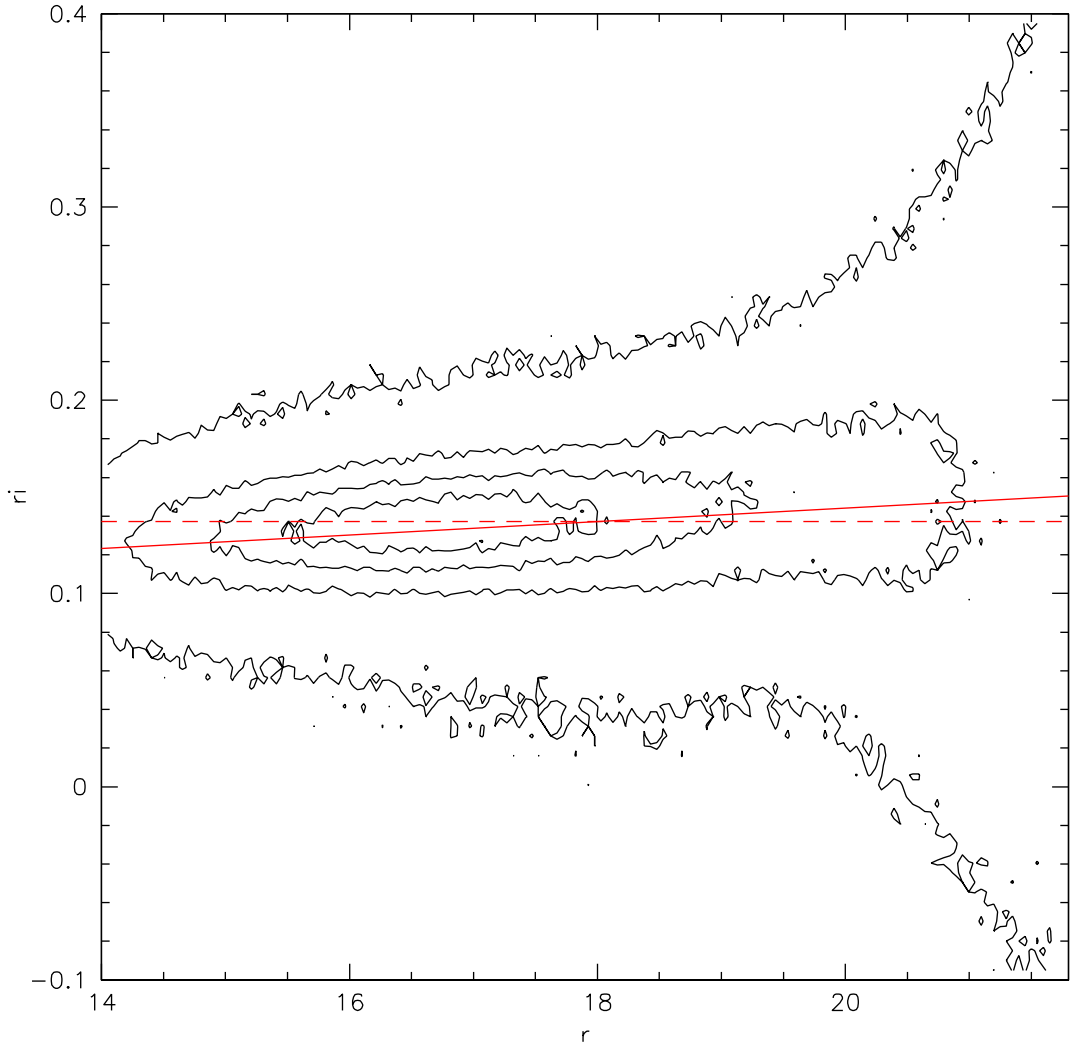}
\caption{The dependence of the measured $r-i$ color on the $r$ band magnitude for stars 
with $0.29 < g-r < 0.31$. An approximately linear dependence, with $d(r-i)/dr \sim$0.007
mag/mag is observed (solid line). The horizontal dashed line shows the location of the
locus for this bin as given by eq.~\ref{eq.locus}. The two match for $r = 18$. 
\label{rivsr}}
\end{figure}

\begin{figure}
\plotone{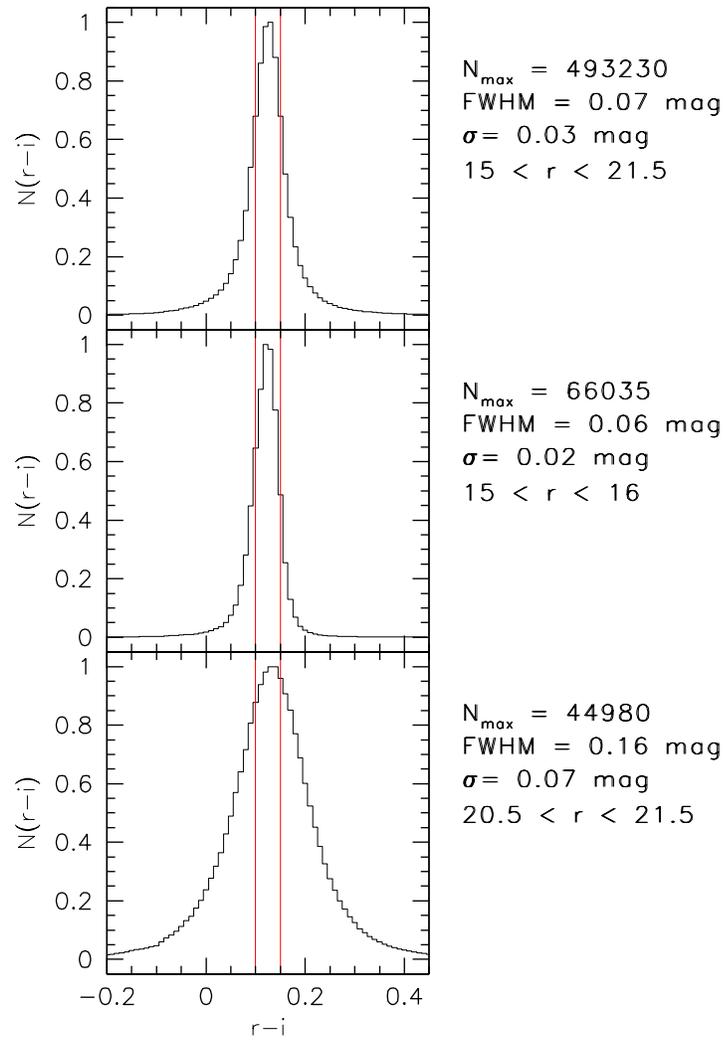}
\caption{The effect of ignoring the magnitude dependence of the locus. The histograms 
show observed $r-i$ colors of stars having locus-corrected colors in the $0.1 < (r-i)_e <
0.15$ color range. The top panel shows the histogram for all stars in the sample.
The spread of $\sigma = 0.03^{\mathrm{mag}}$ is comparable to SDSS photometric errors. 
The middle and bottom panels show histograms for the brightest and faintest magnitude bin,
respectively. A weak magnitude dependence can be seen as a small shift of histogram median 
to the left (middle panel) and right (bottom panel). 
\label{binspill}}
\end{figure}
\end{document}